\documentclass[11pt]{article}

\include{setup}

\begin{document}

\thispagestyle{empty}
\def\thefootnote{\fnsymbol{footnote}}
\setcounter{footnote}{1}
\null
\strut\hfill MPP-2009-164 \\
\vskip 0cm
\vfill
\begin{center}
  {\Large \boldmath{\bf Radiative corrections to the
      neutral-current Drell--Yan process
      in the Standard Model and its minimal
      supersymmetric extension}
\par} \vskip 2.5em
{\large
{\sc Stefan Dittmaier$^1$ and Max Huber$^2$
}\\[2ex]
{\normalsize \it 
$^1$ \it Albert-Ludwigs-Universit\"at Freiburg, Physikalisches Institut, \\
D-79104 Freiburg, Germany
}\\[1ex]
{\normalsize \it 
$^2$ Max-Planck-Institut f\"ur Physik (Werner-Heisenberg-Institut), \\
D-80805 M\"unchen, Germany
}\\[1ex]
}
\par \vskip 5em
\end{center}\par
\vskip .0cm \vfill {\bf Abstract:} \par
An adequate description of the neutral-current Drell--Yan process
at the Tevatron and the LHC, in particular, requires the inclusion 
of electroweak radiative corrections.
We extend earlier work in this direction in various ways.
{}First, we define and numerically compare different methods to
describe the Z-boson resonance including next-to-leading order electroweak
corrections; moreover, we provide explicit analytical expressions
for those. 
Second, we pay particular attention to contributions from $\ga\ga$
and $\ga$--quark collisions, which involve photons in the
initial state, and work out how their impact can be enhanced by
selection cuts.
Third, 
we supplement the $\mathcal{O}(\alpha)$ corrections
by universal electroweak effects of higher order, such as
universal two-loop contributions
from $\Delta\alpha$ and $\Delta\rho$, and the leading two-loop
corrections in the high-energy Sudakov regime as well as multi-photon
radiation off muons in the structure-function approach.
{}Finally, we present results on the complete next-to-leading order
electroweak and QCD corrections within the minimal 
supersymmetric extension of the Standard Model.

\par
\vskip 1cm
\noindent
November 2009
\par
\null
\setcounter{page}{0}
\clearpage
\def\thefootnote{\arabic{footnote}}
\setcounter{footnote}{0}

\section{Introduction}

The Drell--Yan-like production of $\PW$ and $\PZ$ bosons both provides a
standard candle for hadronic high-energy colliders as the Tevatron
and the LHC and offers good possibilities to search for extra
gauge bosons $\PW'$ and $\PZ'$ in high-energy tails of distributions
(see, e.g., \citeres{Gerber:2007xk, Buescher:2006jm} and references therein).
{}For instance, the investigation of the Z-boson resonance, which is well
known from LEP and SLC experiments, is of great importance for detector
calibration, while the analogous study of Jacobian peaks of the W~boson
in appropriate distributions even allow for precision measurements
of the W-boson mass.
Even the effective leptonic weak mixing angle might be 
measurable~\cite{Haywood:1999qg}
at the LHC with a precision competing with LEP and SLC.
On the theoretical side, all these tasks require precise predictions
with an inclusion of both strong and electroweak radiative corrections
and a careful estimate of the remaining theoretical uncertainties.

The largest corrections are due to strong interactions, mainly
described by perturbative QCD. The QCD corrections are known to
two loops, i.e.\ next-to-next-to-leading order (NNLO) for integrated
cross sections~\cite{Hamberg:1990np} and for differential
distributions~\cite{Anastasiou:2003yy}.  Including corrections up to
N$^3$LO in the soft-plus-virtual approximation~\cite{Moch:2005ky} the
remaining theoretical error from QCD for inclusive cross sections is
at the per-cent level or lower.  The next-to-leading-order (\NLO) QCD
corrections have been matched with parton showers
\cite{Frixione:2006gn} and combined with a summation of soft gluon
radiation~\cite{Arnold:1990yk}.

While QCD corrections to on- or off-shell W- and Z-boson production
with leptonic decays are very similar, electroweak corrections to the
different gauge-boson production processes differ considerably. At \NLO\
the electroweak corrections are completely known, both for
charged-current
(CC)~\cite{Baur:1998kt,Zykunov:2001mn,CarloniCalame:2006zq,Dittmaier:2001ay} and
neutral-current
(NC)~\cite{Baur:1997wa,Baur:2001ze,Zykunov:2005tc,CarloniCalame:2007cd,Arbuzov:2007db} processes.  A
tuned comparison of cross sections and differential distributions has
shown good agreement between the various
calculations~\cite{Buttar:2006zd, Gerber:2007xk, Buttar:2008jx}.
Since collinear singularities from photonic initial-state radiation
are absorbed into the parton distribution functions (PDF), similar to
the usual QCD factorization, a photon PDF delivers another source of
real electroweak corrections.  Corrections due to $\gamma q$ and
$\gamma\bar q$ collisions arise both in the CC case
(W~production)~\cite{DK_LH,Arbuzov:2007kp,Brensing:2007qm} and in the
NC case (dilepton
production)~\cite{CarloniCalame:2007cd,Arbuzov:2007kp}.  In the NC
case even a leading-order (LO) contribution is induced by $\ga\ga$
collisions~\cite{CarloniCalame:2007cd}.  Finally, the \NLO\ calculations
to the CC Drell--Yan process have been generalized to the
supersymmetric extension of the Standard Model (MSSM) in
\citere{Brensing:2007qm}.

Beyond \NLO\ electroweak corrections, multi-photon final-state
radiation has been considered both for 
W-boson~\cite{Placzek:2003zg,Brensing:2007qm}
and Z-boson production~\cite{CarloniCalame:2005vc};
more recently even multi-photon radiation off all charged particles
has been matched with the $\Oa$ corrections in the {\sc HORACE}
program in the CC~\cite{CarloniCalame:2006zq}
and NC~\cite{CarloniCalame:2007cd} cases.
Moreover, the impact of the leading higher-order effects due to
$\Delta\alpha$ and $\Delta\rho$ as well as the leading two-loop
corrections in the high-energy Sudakov regime have been 
investigated for the CC case in \citere{Brensing:2007qm}.

A proper combination of QCD and electroweak corrections is in progress
by various groups. Different procedures for this combination based
on factorization or addition, as implemented in {\sc HORACE}, 
are described in \citere{Balossini:2008cs}.
The results discussed there suggest that 
non-factorizable mixed strong--electroweak corrections, which start
at the two-loop level, are required in order to achieve per-cent accuracy
in the predictions.
For on-shell Z~production part of these
${\cal O}(\alpha\alpha_{\mathrm{s}})$ effects
have been calculated in \citere{Kotikov:2007vr}.

In this paper, we complete and extend the existing results on
radiative corrections to the 
NC Drell--Yan process
in various respects:
\begin{enumerate}
\item
We rederive the ${\cal O}(\alpha)$ electroweak corrections and
document the analytical results for the one-loop corrections explicitly.
Moreover, we define and numerically compare different treatments
of the Z-boson resonance in the presence of weak corrections.
Specifically, we discuss the 
``complex-mass scheme''~\cite{Denner:1999gp,Denner:2005fg},
the ``pole scheme''~\cite{Stuart:1991xk,Aeppli:1993cb},
and a scheme employing a simple factorization 
into the LO cross section containing the Z~resonance and a 
factor for the weak correction.
\item
We consistently include dilepton production processes involving
photons in the initial state, which proceed via the partonic processes
$\ga\ga\to l^-l^+$, $q\gamma \to \llb + q$, and
$\bar q\gamma \to \llb + \bar q$.
We even take into account the
known \NLO\ electroweak corrections~\cite{Denner:1998tb}
to the process $\ga\ga\to l^-l^+$,
which contributes to the LO signal process.
\item
Beyond \NLO\ we consider
universal two-loop contributions
from $\Delta\alpha$ and $\Delta\rho$, the leading two-loop
corrections in the high-energy Sudakov regime, and multi-photon
radiation off muons in the structure-function approach~\cite{Kuraev:1985hb}.
\item
{}Finally, we calculate the \NLO\
electroweak and QCD corrections within the MSSM.
\end{enumerate}
{}For the Standard Model (SM)
the presentation in this paper widely follows 
\citeres{Dittmaier:2001ay,Brensing:2007qm}, where the electroweak
\NLO\ corrections and the same type of effects beyond \NLO\ are
discussed for the CC Drell--Yan process. Similarly our discussion
of the \NLO\ corrections in the MSSM, presented here for the NC case,
proceeds along the same lines as in \citere{Brensing:2007qm} for 
the CC case.

The paper is organized as follows. In \refse{se:born} we set up our
conventions and give the lowest-order cross sections. Furthermore we
describe and discuss the different treatments of the \PZ-boson
resonance and the different 
input-parameter schemes considered in this
paper, as far as it is necessary for the LO process. In
\refse{sec:rad-corr-sm} the electroweak radiative corrections of
points 1.--3.\ given above as well as \NLO\ QCD corrections are
discussed. The \NLO\ corrections within the MSSM are described in
\refse{se:mssm}. Our discussion of numerical results, which is presented in
\refse{se:numres}, comprises integrated cross sections as well as differential
distributions for the LHC and integrated results for the Tevatron.
We also compare our results to results previously given in the literature
and discuss how effects of incoming photons can be enhanced.
Our conclusions are drawn in \refse{se:conclusion}. In the Appendix we
describe the factorization of QED-like collinear singularities into
the photon distribution function, 
give explicit expressions for the vertex and box corrections in the SM,
and provide details on the considered SUSY scenarios.

\section{Conventions and lowest-order cross sections}
\label{se:born}

In this section we set up our conventions for the discussion
of the various partonic processes contributing to the 
production of a charged lepton pair.  Apart from the Drell--Yan-like
process $\qqb\to
\gamma/\PZ \to \llb$ and its radiative corrections we consider the
photon-induced process $\gamma \gamma \to \llb$ and its
radiative corrections. Although the latter does not have a
\PZ~resonance, it is an irreducible background to $\qqb\to\gamma/\PZ
\to \llb$ and therefore should be included. The (electroweak) \NLO\
corrections to $\gamma \gamma \to \llb$ have been calculated in
\citere{Denner:1998tb}, and we only briefly review some of the results
given there.

The momenta of the incoming particles will be denoted with
$p_i$, $i=1,2$, and the ones of the outgoing particles with
$k_j$, $j=1,2,3$. Explicitly we assign the external momenta and
helicities $(\si_i,\tau_i,\lambda)$ according to
\begin{eqnarray}
  q(p_1,\sigma_1) + \bar q(p_2,\sigma_2) &\;\to\;&
  \Pl^-(k_1,\tau_1) + \Pl^+(k_2,\tau_2) \;\; [+\gamma/\Pg(k_3,\lambda)], 
  \label{qqllgam}\\
  \gamma(p_1,\sigma_1) + \gamma(p_2,\sigma_2) &\;\to\;&
  \Pl^-(k_1,\tau_1) + \Pl^+(k_2,\tau_2) \;\; [+\gamma(k_3,\lambda)],
\end{eqnarray}
where $q$ generically denotes the light up- and down-type quarks, $q=
\Pd,\Pu,\Ps,\Pc,\Pb$, and $l$ denotes the charged leptons
$l=\Pe,\mu,\tau$. The possible photons or gluons in the final state
deliver part of the real radiation contribution to the \NLO\ corrections.
The remaining part of the real \NLO\ corrections is induced by the 
crossed processes $q\gamma/\Pg \to \llb + q$ and
$\bar q\gamma/\Pg \to \llb + \bar q$.
The Mandelstam variables are defined by
\begin{equation}
  \hat s = (p_1+p_2)^2, \quad
  \hat t = (p_1-k_1)^2, \quad
  \hat u = (p_1-k_2)^2, \quad
  s_{l l} = (k_1+k_2)^2.
\end{equation} 
We neglect the fermion masses of the light quarks, $m_q$, and of the leptons,
$\Ml$, whenever possible, i.e.\ we keep these masses only as regulators
in the logarithmic mass singularities originating from collinear
photon emission or exchange. Obviously, we have $\hat s = s_{l l}$
for the non-radiative processes $\qqb\to \llb$ and $\gamma\gamma
\to \llb$.  At LO the Feynman diagrams shown in
\reffi{fig:borndiag} contribute to
the scattering amplitudes.
\begin{figure}
  \centering
  \input{borndiagall.tex}
  \vspace{-2em}
  \caption{Partonic lowest-order diagrams for $\Pp\Pp/\Pp\Ppbar\to\llb+X$.}
  \label{fig:borndiag}
\end{figure}
{}For $\qqb\to \llb$ the polarized Born amplitude $\M_{\qqb}^{\mathrm{LO}}$
can be written as
\begin{equation}
\label{eq:m0}
\M_{\qqb}^{\mathrm{LO},\sigma\tau} = 
-\frac{e^2}{\hat s} \sum_{V=\ga,\PZ} \gqqV^\sigma\, \gllV^\tau\,
\chi_V(\hat s)\,  \mathcal{A}^{\sigma \tau} \equiv f_{\qqb}^{\mathrm{LO},\sigma\tau} \, \mathcal{A}^{\sigma \tau},
\end{equation}
where $e$ is the electric unit charge,
$\gffV^\si$ are the chiral couplings of the fermions $f$ to the
vector bosons $V$, the functions $\chi_V(\hat s)$ describe the propagation 
of $V$, and $\mathcal{A}^{\sigma \tau}$ are ``standard matrix elements''
containing the spin information of the fermions.

The standard matrix element $\mathcal{A}^{\sigma \tau}$ for the quark
and lepton chiralities, $\sigma=\sigma_1=-\sigma_2$ and
$\tau=\tau_1=-\tau_2$, is defined as
\begin{equation}
\label{eq:sme}
\mathcal{A}^{\sigma \tau}= 
[ \bar{v}_q(p_2) \, \gamma^\mu \omega_\sigma\, u_q(p_1) ] \; 
[ \bar{u}_l(k_1) \, \gamma_\mu \omega_\tau \, v_l(k_2) ] \;,
\end{equation}
with an obvious notation for the Dirac spinors $\bar{v}_q(p_2)$, etc.,
and the chirality projectors
$\omega_\pm=\frac{1}{2}(1\pm\gamma_5)$.  Explicitly the
$\mathcal{A}^{\sigma \tau}$ are given by
\begin{equation}
\mathcal{A}^{\pm\pm} = 2\,\hat u  \;,\qquad
\mathcal{A}^{\pm\mp} = 2\,\hat t  \;.
\end{equation}
{}For a fermion $f$ with charge $Q_f$ and third component
$I^3_{\scrs\PW,f}$ of its weak isospin the
left- and right-handed couplings to $V=\ga,\PZ$ are given by
\begin{equation} 
\label{eq:ffVcoupl}
\gffA^\pm = -Q_f\,,\qquad
\gffZp = -\frac{\sw}{\cw} \Qf \,,\qquad
\gffZm=\frac{I^3_{\scrs\PW,f}-\sw^2\,\Qf}{\sw\cw}\,.  
\end{equation} 
The sine and cosine, $\sw$ and $\cw$, of the weak mixing angle are
fixed by the W- and Z-boson masses $\MW$ and $\MZ$ as described below
in more detail.

The propagator functions are defined by
\begin{equation}
\chi_\ga(\hat s) = 1, \qquad
\chi_\PZ(\hat{s}) = \frac{\hat s}{\hat{s}-\xMZ^2},
\end{equation}
where the complex quantities
\begin{equation}
\xMZ^2=\MZ^2- \ri \MZ \GZ\;, \qquad\xMW^2 = \MW^2 - \ri \MW \GW
\end{equation} 
denote the locations of the poles of the Z- and W-boson propagators (with
momentum transfer $p$) in the complex $p^2$~plane.
The gauge-boson widths $\Ga_V$ enter the propagator denominators only
after performing the Dyson summation of
all insertions of the (imaginary parts of the) gauge-boson self-energies,
i.e.\ using the above $\chi_\PZ(\hat{s})$ already goes beyond the
lowest perturbative order. It is well known that this unavoidable
mixing of perturbative orders jeopardizes the gauge invariance
of predictions, in particular in the presence of radiative corrections.%
\footnote{More details on this issue, specific examples as well as
proposed solutions can, e.g., be found in 
\citeres{Denner:2006ic,Grunewald:2000ju,Dittmaier:2002nd} and references therein.}
Before describing our solutions to this problem, we recall an important
feature of the explicit form of the propagator function.
While we have chosen a constant imaginary part in the
denominator of $\chi_\PZ(\hat s)$, the frequently used
on-shell (OS) renormalization scheme, as for instance defined in
\citere{Denner:1993kt}, naturally leads to a running width in the
denominator. In the approximation of massless decay products of the
boson $V$, the OS version of $\chi_V(\hat s)$ is
\begin{equation}
\chi_V(\hat{s})\big|_{\mathrm{OS}} = 
\frac{\hat s}{\hat{s}-M_{V,\mathrm{OS}}^2
+\ri M_{V,\mathrm{OS}}\Ga_{V,\mathrm{OS}}\,\times\,
\hat s/M_{V,\mathrm{OS}}^2\times \theta(\hat s)}.
\end{equation}
The two versions of $\chi_V(\hat{s})$ are formally equivalent
in the resonance region
if mass and width of $V$ are properly 
translated~\cite{Bardin:1988xt,Beenakker:1996kn}
\begin{equation}
M_V = \frac{M_V}{\sqrt{1+\Ga_V^2/M_V^2}}\Bigg|_{\mathrm{OS}}, 
\qquad
\Ga_V = \frac{\Ga_V}{\sqrt{1+\Ga_V^2/M_V^2}}\Bigg|_{\mathrm{OS}}. 
\label{eq:mwdef}
\end{equation}
Since the W and Z masses and widths are usually quoted in the OS scheme,
we shall perform this translation before our evaluations.
{}For the masses, the impact of this conversion typically is
$M_{\PZ,\mathrm{OS}}-\MZ\approx 34\MeV$ and
$M_{\PW,\mathrm{OS}}-\MW\approx 27\MeV$.
We perform our evaluation in the following schemes for
treating the Z-boson resonance, where at this point we describe the 
various procedures only as far as relevant in LO and give the details
for the corrections in the next section:
\begin{itemize}
\item
{\it Complex-mass scheme} (CMS):
The CMS was introduced in \citere{Denner:1999gp} for
LO calculations and generalized to \NLO\ in
\citere{Denner:2005fg}. In this approach the squared
W- and Z-boson masses are consistently identified with $\xMW^2$
and $\xMZ^2$, respectively, i.e.\ with the location of the poles
of the propagators in the complex $p^2$ plane.
This leads to complex couplings and, in particular, a complex weak
mixing angle via $\cw^2=1-\sw^2=\xMW^2/\xMZ^2$. The scheme fully respects
all relations that follow from gauge invariance
(Ward or Slavnov--Taylor identities, gauge-parameter cancellation),
because the gauge-boson masses are modified only by an analytic 
continuation. Beyond LO the complex masses are
introduced directly at the level of the Lagrangian by splitting the
real bare masses into complex renormalized masses and complex
counterterms, so that the usual perturbative calculus with Feynman
rules and counterterms works without modification. In contrast to
gauge invariance, unitarity is not respected order by order in perturbation
theory. However, spurious terms spoiling unitarity are of (N)\NLO\ in
an (N)LO calculation without any unnatural amplification, because
unitarity cancellations, which are ruled by gauge invariance, are respected.
More details of this scheme can also be found in \citere{Denner:2006ic}.

In the CMS the LO amplitude \refeq{eq:m0}
is, thus, evaluated with complex couplings
$\gffV^\pm$ and a complex Z-boson mass.
\item
{\it Pole scheme} (PS):
The PS exploits the fact that both the location $\mu_V^2$ of the 
$V$~propagator pole and its residue in amplitudes are gauge-independent
quantities~\cite{Stuart:1991xk,Sirlin:1991fd}.
The idea~\cite{Stuart:1991xk,Aeppli:1993cb}
is, thus, to first isolate the residue for the
considered resonance and subsequently to introduce a finite decay width
only in the gauge-independent resonant part. If done carefully this
procedure respects gauge invariance, but it should be kept in mind
that the resonant part of an amplitude is not uniquely determined by 
the propagator structure alone, but depends on a specific phase-space 
parameterization and in most cases also on the separation of
polarization-dependent parts.
A ``pole approximation''---in contrast to a full PS calculation
as performed in this paper---results 
from a resonant amplitude defined in the PS upon
neglecting non-resonant parts.

The LO amplitude \refeq{eq:m0} with real couplings defined via the
usual on-shell relation $\cw^2=1-\sw^2=\MW^2/\MZ^2$, but with the 
complex Z-boson mass in $\chi_\PZ(\hat s)$, represents the result
of a particular PS variant. The PS operation here first splits off
the polarization-dependent structure $\mathcal{A}^{\sigma\tau}$ and
subsequently introduces the Z-boson width in the resonant
part of the form factors via
$1/(\hat s-\MZ^2)\to1/(\hat s-\MZ^2+\ri\MZ\GZ)=\chi_\PZ(\hat s)/\hat s$,
while the non-resonant photon part is not changed.

\item
{\it Factorization scheme} (FS):
Many variants of factorizing resonant structures from amplitudes have
been suggested and used in the literature, but they all share the idea 
to separate a simple resonant factor from a potentially complicated
amplitude that does not involve resonances anymore. 
In \citere{Dittmaier:2001ay}, for instance, the virtual electroweak
correction to Drell--Yan-like W~production was factorized from the
resonant LO amplitude, so that the relative correction factor did not 
involve resonance factors anymore.%
\footnote{The relative electroweak correction defined in this way 
involves the W-boson width $\GW$ only in logarithms 
$\propto\alpha\ln(\hat s-\MW^2+\ri\MW\GW)$, which result from 
soft-photon exchange.}

{}For the present case of NC dilepton production we start from the
LO amplitude \refeq{eq:m0} with real couplings, as in the PS,
and define the relative correction factor for the weak 
(i.e.\ non-photonic) one-loop correction in the strict limit of
vanishing gauge-boson widths.
\end{itemize}
We can, thus, compare two different versions of LO cross sections
for $\qqb\to\ga/\PZ\to\llb$: one version delivered by the CMS, another by
the PS and FS, which coincide in LO.

The electromagnetic coupling $\alpha=e^2/(4\pi)$ yields an overall factor
to the LO predictions. Although the electric charge is always defined 
(renormalized) in the Thomson limit, the value for $\alpha$
can be fixed in different input-parameter schemes.
We support the following three different schemes:
\begin{itemize}
\item
{\it $\alpha(0)$-scheme:}
The fine-structure constant $\alpha(0)$ and all particle masses define
the complete input.
In this scheme, the relative corrections to the $\qqb\to\ga/\PZ\to\llb$
cross sections
sensitively depend on the light-quark masses via $\alpha\ln m_q$ terms
that enter the charge renormalization.
\item
{\it $\alpha(\MZ)$-scheme:}
The effective electromagnetic coupling  $\alpha(\MZ)$
and all particle masses define
the basic input.
Tree-level couplings are derived from $\alpha(\MZ)$,
and the relative corrections receive contributions from the
quantity $\De\alpha(\MZ)$,
which accounts for the running of the electromagnetic coupling
from scale $Q=0$ to $Q=\MZ$ (induced by light fermions) and cancels the
corresponding $\alpha\ln m_q$ terms that appear in the corrections to
the $q\bar q$ channels in the $\alpha(0)$-scheme.
\item
{\it $\GF$-scheme:}
The Fermi constant $\GF$ and all particle masses define
the basic input.
Tree-level couplings are derived from the effective coupling
$\alpha_{\GF}=\sqrt{2}\GF\MW^2(1-\MW^2/\MZ^2)/\pi$, 
and the relative corrections receive contributions from the
quantity $\De r$~\cite{Sirlin:1980nh},
which describes the radiative corrections to muon decay.
Since $\De\alpha(\MZ)$ is contained in $\De r$, there is no large
effect on the $q\bar q$ channels
induced by the running of the electromagnetic coupling in the
$\GF$-scheme either.
\end{itemize}
Since light-quark masses are perturbatively ill-defined and can only
play the role of phenomenological fit parameters, the $\alpha(\MZ)$- and
$\GF$-schemes are preferable over the $\alpha(0)$-scheme for the 
$q\bar q$ annihilation processes.
More details on the difference of the three schemes are provided
in the next section, where we deal with electroweak radiative corrections
(see also \citere{Dittmaier:2001ay}).

\begin{figure}
  \centering
  \includegraphics{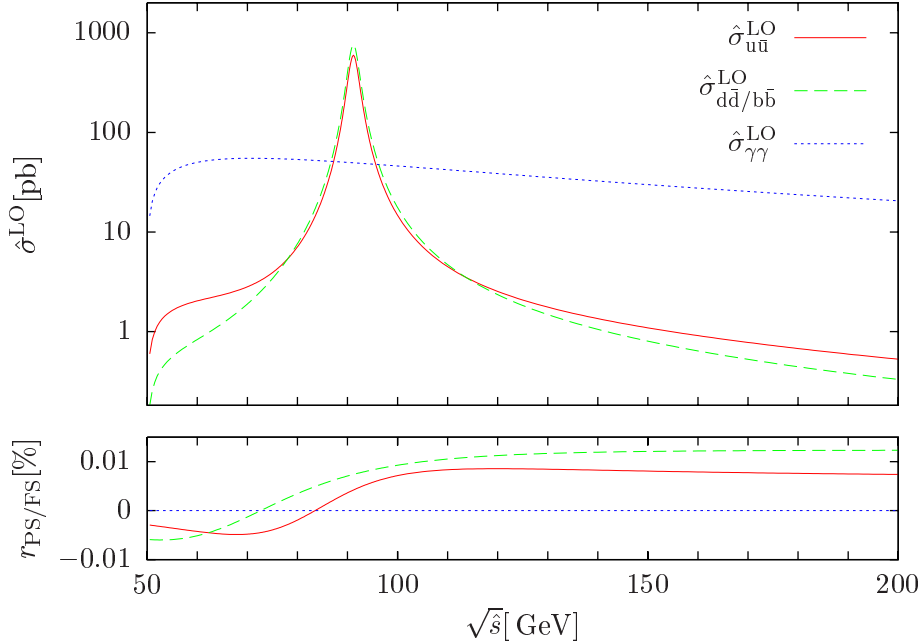}
  \caption{LO cross sections for $u\bar u/d\bar d\to\ga/\PZ\to\llb$ in
    the vicinity of the Z~resonance using the different schemes (CMS
    and PS/FS) for treating finite-width effects, employing the
    $\GF$-scheme, and the LO cross section for $\ga\ga \to \llb$.}
  \label{fig:locs}
\end{figure}

The differential LO cross section 
$\rd\hat\sigma_{\qqb}^{\mathrm{LO}}/\rd\hat\Omega$ 
is easily obtained by squaring the LO matrix element $\M_{\qqb}^{\mathrm{LO}}$,
\begin{eqnarray}
\label{eq:locs}
\biggl(\frac{\rd\hat\sigma_{\qqb}^{\mathrm{LO}}}{\rd\hat\Omega}\biggr) 
&=& \frac{1}{12} \, \frac{1}{64\pi^2\hat s} \, 
\sum_{\mathrm{pol}} 
|\M_{\qqb}^{\mathrm{LO}}|^2 \\
&=& \frac{\alpha^2}{12 \,\hat{s}^3} \bigg\{
    \,2\, \Qq^2 \Ql^2 \,(\hat{t}^2+\hat{u}^2)  
\nl
&& {}
    + 2\, \Qq \Ql\, \Re\Big[\big[ (\gqqZp \gllZp + \gqqZm \gllZm)\, \hat{u}^2
+(\gqqZp \gllZm + \gqqZm \gllZp)\, \hat{t}^2 \big]\, \chi_\PZ(\hat{s})\Big] 
\nl
&& {}
+ \Big[ (|\gqqZp|^2 |\gllZp|^2 + |\gqqZm|^2 |\gllZm|^2)\, \hat{u}^2
+(|\gqqZp|^2 |\gllZm|^2 + |\gqqZm|^2 |\gllZp|^2)\, \hat{t}^2 \Big] 
\,|\chi_\PZ(\hat{s})|^2\,\bigg\} \, .     
\nn
\end{eqnarray}
The explicit factor $1/12$ results from the
average over the quark spins and colours, and $\hat\Omega$ is the
solid angle of the outgoing $l^-$ in the partonic centre-of-mass 
frame. 
In \reffi{fig:locs} we show the integrated partonic LO cross sections 
$\hat \sigma_{\qqb}^{\mathrm{LO}} (\hat s)$ for the different
schemes (CMS and PS/FS) to treat
the finite Z~width, as obtained in the $\GF$-scheme.
We also show the relative difference $r_{\mathrm{PS/FS}}=\hat
\sigma^{\mathrm{LO}}|_{\mathrm{PS/FS}}\; /\; \hat
\sigma^{\mathrm{LO}}|_{\mathrm{CMS}} - 1$ of the results obtained in the
different schemes, which turns out to be at the $0.01$ per-cent level.

{}For completeness we state the contribution of $\gamma\gamma\to l^-
l^+$, $\rd\hat\sigma_{\gamma \gamma}^{\mathrm{LO}}/\rd\hat\Omega$, to the 
LO differential cross section,
\begin{equation}
  \label{eq:1}
  \biggl(\frac{\rd\hat\sigma_{\gamma\gamma}^{\mathrm{LO}}}{\rd\hat\Omega}\biggr)
= \frac{1}{4}\,\frac{1}{64 \pi^2 \hat{s}} \sum_{\mathrm{pol}}
|\M_{\gamma\gamma}^{\mathrm{LO}}|^2 
= \frac{\alpha^2}{2 \hat{s}}   \left( \frac{\hat t}{\hat u} + \frac{\hat u}{\hat t} \right) \, .
\end{equation}
{}For details we refer to \citere{Denner:1998tb}.
Here we just mention that we consider $\gamma\gamma\to l^- l^+$
cross sections always in the $\alpha(0)$-scheme, because the 
natural scale for the coupling of the external photons is $Q=0$.
In fact, using the $\alpha(\MZ)$- or $\GF$-scheme here would result in
large corrections containing $\alpha\ln m_q$ terms, which should be avoided.

\section{Radiative corrections to the partonic cross sections in the SM}
\label{sec:rad-corr-sm}

In this section we discuss the \NLO\ radiative corrections to
the partonic subprocesses contributing to the hadronic process
$\Pp\Pp/\Pp\Ppbar \to \Pl^-\Pl^++X$.
{}For the main contribution of $q\bar q$ annihilation, many issues 
discussed here are very similar to the case of 
$\Pep\Pem\to\ga/\PZ\to f\bar f$ as measured in the LEP and SLD
experiments, for which precision calculations
have been performed in the last two decades (see, e.g., 
\citeres{LEP1,Bardin:1999gt} and references therein).

\subsection{Survey of radiative corrections and calculational details}
\label{sec:surv-radi-corr}

\begin{figure}
  \centering
  \input{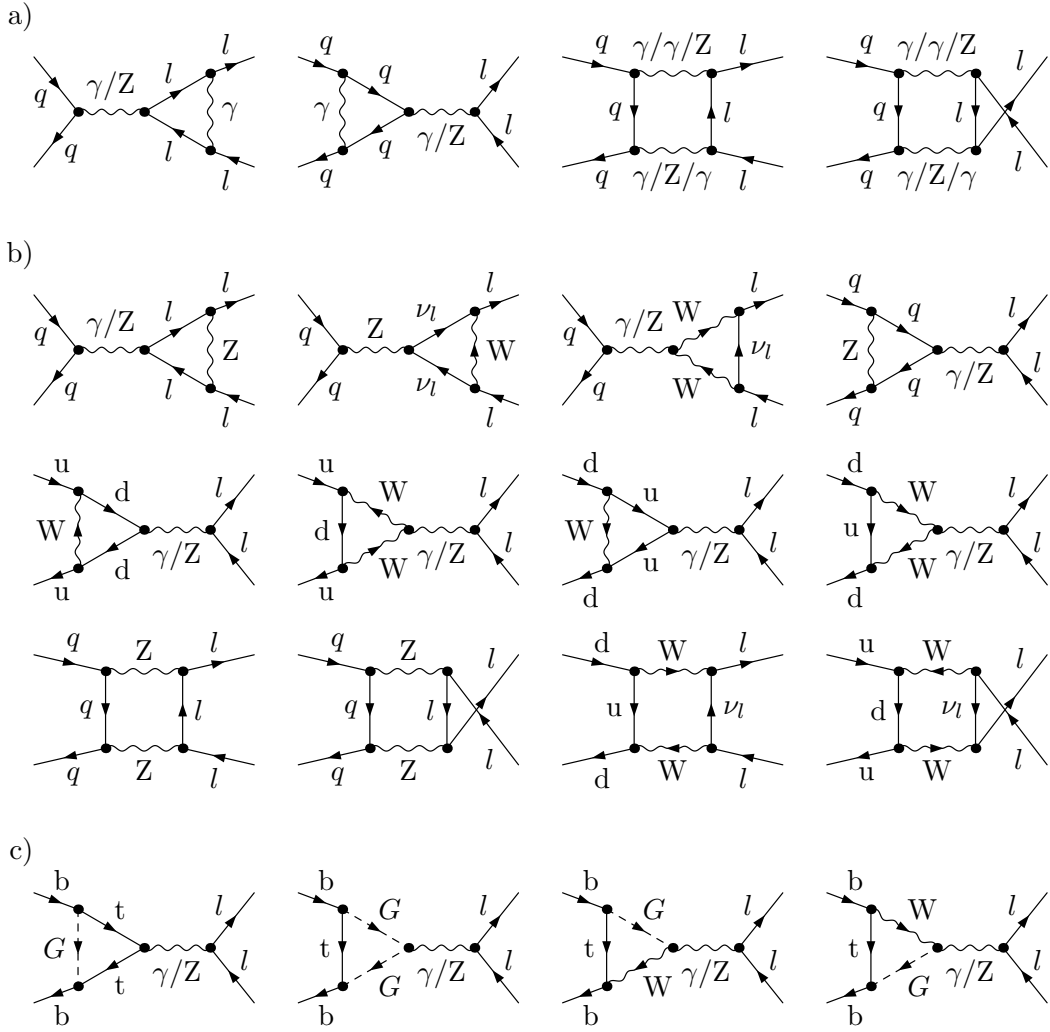}
  \vspace{-1.5em}
  \caption{Vertex and box diagrams for the electroweak virtual
    corrections to $q\bar q\to\llb$: 
    a) photonic corrections, b) weak corrections with
    light incoming quarks $q=\Pu,\Pd,\Pc,\Ps,\Pb$, and c) additional
    diagrams for incoming \Pb-quarks, where $G$ stands for would-be
    Goldstone boson fields.}
  \label{fig:vertboxdiags}
\end{figure}

The electroweak radiative \NLO\ corrections can be divided into photonic
and weak corrections. 
The photonic corrections consist of real and virtual corrections
that are induced by the emission and exchange of an additional
photon. 
Since only electrically neutral gauge bosons are involved at LO,
the photonic subset of the complete
\Oa~electroweak corrections is separately invariant under
$\mathrm{U}(1)_{\mathrm{elmg}}$ gauge transformations. 
{}For the $q\bar q$ channel this classification is, e.g., discussed
in \citere{Dittmaier:2002nd} in more detail, for the $\ga\ga$ channel
this separation was introduced in \citere{Denner:1998tb}.
{}For $q\bar q$ annihilation
the photonic corrections can be further classified into
separately $\mathrm{U}(1)_{\mathrm{elmg}}$ gauge-invariant
parts. Specifically, the photonic contributions can be split into 
initial-state corrections,
final-state corrections, and interference terms, according to their charge
proportionality to $Q_q^2$, $Q_l^2$, and $Q_q \,Q_l$, respectively. 
In this sense the photonic corrections to the $\ga\ga$ channel
are final-state corrections proportional to $Q_l^2$.
The virtual photonic corrections to the $q\bar q$ channel
are composed of the one-loop photon
exchange diagrams shown in \reffi{fig:vertboxdiags}a) and the
corresponding counterterm contributions; the counterparts for
$\ga\ga$ scattering can be found in \citere{Denner:1998tb}.
The real photonic corrections
consist of processes with single-photon emission, $\qqb\to \ga/\PZ \to \llb
+ \ga$ and $\ga \ga \to \llb + \ga$, and of the processes $q/\bar q \,
\ga \to \ga/\PZ \to \llb +q /\bar q$, which deliver 
a correction to both LO
processes $\qqb\to\llb$ and $\ga\ga\to\llb$. On the hadronic level the
photon-induced processes are, of course, suppressed due to the
smallness of the photon PDF, but on the partonic level all processes
are of the same order \Oa\ compared to the LO processes.
Since real photons effectively couple with $\alpha(0)$ and since
virtual and real photonic corrections are intrinsically linked to each
other, it is natural to identify the relative coupling of the whole
photonic correction with $\alpha(0)$, independent of the choice of the
input-parameter scheme chosen in LO. This means in $q\bar q$ 
annihilation (and in the crossing-related $q/\bar q\ga$ scattering)
and in the
$\ga\ga$ channel we scale the cross section contributions of the photonic
corrections with $\alpha(0)\alpha^2$ and $\alpha(0)^3$, respectively,
where $\al$ depends on the input-parameter scheme as discussed in 
\refse{se:born}.

The weak \Oa~corrections to the $q\bar q$ channel
comprise contributions of the transverse
parts of the photon, the $\PZ$, and the $\ga\PZ$ mixing
self-energies ($\Sigma^{\ga\ga}_{\mathrm{T}}$,
$\Sigma^{\PZ\PZ}_{\mathrm{T}}$, and $\Sigma^{\ga\PZ}_{\mathrm{T}}$), of
weak corrections to the $\ga/\PZ\,\bar q q$ and $\ga/\PZ\, \llb$
vertices, the $\PZ\PZ$ and $\PW\PW$ box diagrams, and
counterterms. The diagrams for the vertex
and box corrections are shown in \reffi{fig:vertboxdiags}b) for
incoming quarks other than b's. For incoming b-quarks, the same diagrams
as for incoming d- or s-quarks exist, but in diagrams with internal
W~bosons the b-quark turns into its massive iso-spin partner, the top-quark.
{}For this reason, in `t~Hooft--Feynman gauge there are additional versions of
those diagrams in which one or two W~bosons are replaced by would-be Goldstone
bosons; these diagrams are shown in \reffi{fig:vertboxdiags}c).
Details and explicit results on the weak corrections to the $\ga\ga$ 
channel can be found in \citere{Denner:1998tb}. In our explicit
evaluation we scale the relative weak correction with the coupling $\alpha$
as defined in the respective input-parameter scheme, i.e.\
the cross section contributions of the weak corrections scale like
$\alpha^3$ and $\alpha\alpha(0)^2$ in the $q\bar q$ and $\ga\ga$
channels, respectively.

The \NLO\ QCD corrections to $\qqb \to \llb$ are easily obtained from
the photonic initial-state corrections, \ie by setting the lepton
charge $\Ql$ to zero within the photonic corrections, and replacing
$\al(0) \, \Qq^2 \to \alpha_{\mathrm{s}}(\mu_{\mathrm{R}}) \,
C_{\mathrm{F}}$, with $C_{\mathrm{F}} = 4/3$ and
$\alpha_{\mathrm{s}}(\mu_\mathrm{R})$ representing the strong coupling
constant at renormalization scale $\mu_{\mathrm{R}}$.  {}For squared
amplitudes with an incoming gluon, $q/\bar q \, \Pg \to \ga/\PZ \to
\llb +q /\bar q$, we omit diagrams with the external photon coupling
to $l$ and replace $\al(0) \Qq^2 \to
\alpha_{\mathrm{s}}(\mu_\mathrm{R}) T_{\mathrm{F}} / 3$ with
$T_{\mathrm{F}} = 1/2$ in the respective squared amplitudes with an
incoming photon instead of a gluon.

{}For this work we have rederived the \NLO\ corrections to the $q\bar q$
channel with standard methods. More precisely, we performed two
independent calculations, with results in mutual agreement.
The one-loop diagrams and amplitudes are generated with {\sc FeynArts}
versions 1.0~\cite{Kublbeck:1990xc} and 3.2~\cite{Hahn:2000kx}.
The subsequent algebraic reductions to standard forms are done
with inhouse {\sc Mathematica} routines in one version and
with the help of {\sc FormCalc}~\cite{Hahn:1998yk} and 
{\sc FeynCalc}~\cite{Mertig:1991an} in the other.
In this reduction the appearing tensor integrals are reduced to scalar
integrals with the Passarino--Veltman algorithm \cite{Passarino:1979jh}.
The scalar integrals are evaluated using the methods and results of
\citere{'tHooft:1979xw,Beenakker:1990jr,Denner:1991qq}, where
UV divergences are treated in dimensional regularization and the 
soft and collinear singularities are
regularized by small fermion masses and an infinitesimal photon or gluon
mass $m_{\gamma/\Pg}$. 
Since the application of the CMS requires complex gauge-boson masses,
the results of \citeres{'tHooft:1979xw,Beenakker:1990jr,Denner:1991qq} 
on the loop integrals had to be generalized accordingly.%
\footnote{In detail,
the scalar two- and three-point functions with
complex masses can be explicitly found in \citeres{Denner:2005nn} and
\cite{'tHooft:1979xw}, respectively. The IR-divergent 
four-point integral of \citere{Beenakker:1990jr} is also valid
for an internal complex mass, so that only the regular four-point
functions, as e.g.\ given in \citere{Denner:1991qq} for real mass parameters,
had to be analytically continued to complex masses.
General results on singular and regular four-point integrals
with complex masses will be published elsewhere~\cite{scalints}.}
The amplitude of the
virtual correction, $\M_{\qqb}^{\virt,\,\sigma \tau}$, can be
expressed in terms of a ``form factor''
$f^{\virt,\,\sigma \tau}_{\qqb}=f^{\virt,\,\sigma \tau}_{\qqb,\phot} 
+ f^{\virt,\,\sigma \tau}_{\qqb,\weak} + f^{\virt,\,\sigma \tau}_{\qqb,\QCD}$ 
times the LO Dirac structure $\mathcal{A}^{\sigma \tau}$,
\begin{equation} 
\M_{\,\qqb}^{\virt,\,\sigma \tau} =
(f^{\virt,\,\sigma \tau}_{\qqb,\phot} + f^{\virt,\,\sigma \tau}_{\qqb,\weak} 
+ f^{\virt,\,\sigma \tau}_{\qqb,\QCD}) \,
\mathcal{A}^{\sigma \tau}.
\label{eq:m1virt}
\end{equation}
{}For self-energy and vertex corrections this is obviously possible,
since all external fermions are considered in the massless limit.
In $D\ne4$ space-time dimensions the calculation of box diagrams
actually leads to combinations of Dirac chains that are not present at
LO. However, since the box diagrams are UV~finite the
four-dimensionality of space-time can be used to reduce all Dirac
structures to the one of $\M_{\qqb}^{\mathrm{LO}}$, 
as explained in \refapp{app:virtRCs} in more detail.
{}Finally, we have rederived the photonic and QCD corrections for
massless external fermions, photons, and gluons within dimensional
regularization by making use of the results of \citere{Dittmaier:2003bc}
for translating the IR-divergent scalar integrals from mass into
dimensional regularization and by employing the statements made in
the appendix of \citere{Bredenstein:2008zb} on the structure of rational 
terms of IR origin. The results of mass and dimensional regularization for
IR divergences are in perfect agreement.

Details of our calculation of real photonic (or gluonic) corrections 
are provided in the following section, where we present our results
on the photonic and QCD corrections. 
The contributions resulting from the factorization of mass-singular
initial-state photonic or gluonic corrections are also reviewed
there.

We conclude this overview by summarizing the structure of the
hadronic cross section including the full \NLO\ corrections,
\begin{eqnarray}
\lefteqn{
\sigma^{\NLO}_{h_1h_2}(P_1,P_2) 
= \int_0^1\rd x_1\, \int_0^1\rd x_2\,
\biggl\{ \sum_{q=\Pu,\Pd,\Pc,\Ps,\Pb} 
f^{(h_1)}_q(x_1,\mu_{\mathrm{F}}^2) 
f^{(h_2)}_{\bar q}(x_2,\mu_{\mathrm{F}}^2) }&&
\nl
&& \qquad {} \times
\biggl[\int\rd\hat\sigma_{q\bar q}^{\mathrm{LO}}(x_1 P_1,x_2 P_2)
\left(1 + \delta^{\virt}_{q\bar q,\QCD} + \delta^{\virt}_{q\bar q,\phot} 
+ \delta^{\virt}_{q\bar q,\weak} \right)
\nl
&& \qquad\qquad {} 
+ \int\rd\hat\sigma^{\real}_{q\bar q,\QCD}(x_1 P_1,x_2 P_2)
+ \int\rd\hat\sigma^{\fact}_{q\bar q,\QCD}(x_1 P_1,x_2 P_2)
\nl
&& \qquad\qquad {} 
+ \int\rd\hat\sigma^{\real}_{q\bar q,\phot}(x_1 P_1,x_2 P_2)
+ \int\rd\hat\sigma^{\fact}_{q\bar q,\phot}(x_1 P_1,x_2 P_2) \biggr]
\nl
&& \quad {} +\sum_{q=\Pu,\Pd,\Pc,\Ps,\Pb} f^{(h_1)}_{\Pg}(x_1,\mu_{\mathrm{F}}^2) 
\biggl[ f^{(h_2)}_q(x_2,\mu_{\mathrm{F}}^2)
\biggl(
\int\rd\hat\sigma_{\Pg q}^{\mathrm{LO}}(x_1 P_1,x_2 P_2)
+\int\rd\hat\sigma^{\fact}_{\Pg q}(x_1 P_1,x_2 P_2) \biggr)
\nl
&& \qquad\qquad {} 
+ (q\to\bar q) \biggr]
\nl
&& \quad {} + \sum_{q=\Pu,\Pd,\Pc,\Ps,\Pb} f^{(h_1)}_\ga(x_1,\mu_{\mathrm{F}}^2) 
\biggl[ f^{(h_2)}_q(x_2,\mu_{\mathrm{F}}^2)
\biggl(
\int\rd\hat\sigma_{\ga q}^{\mathrm{LO}}(x_1 P_1,x_2 P_2)
+\int\rd\hat\sigma^{\fact}_{\ga q}(x_1 P_1,x_2 P_2) \biggr)
\nl
&& \qquad\qquad {} 
+ (q\to\bar q) \biggr]
\nl
&& \quad {} + \frac{1}{2} f^{(h_1)}_\ga(x_1,\mu_{\mathrm{F}}^2) 
f^{(h_2)}_\ga(x_2,\mu_{\mathrm{F}}^2)
\biggl[
\int\rd\hat\sigma_{\ga\ga}^{\mathrm{LO}}(x_1 P_1,x_2 P_2)
\left( 1 + \delta^{\virt}_{\ga\ga,\phot} + \delta^{\virt}_{\ga\ga,\weak} \right)
\nl
&& \qquad\qquad {} 
+ \int\rd\hat\sigma^{\real}_{\ga\ga,\phot}(x_1 P_1,x_2 P_2)
+ \int\rd\hat\sigma^{\fact}_{\ga\ga,\phot}(x_1 P_1,x_2 P_2) \biggr]
\biggr\}
\nl
&& {} + (h_1\leftrightarrow h_2).
\label{eq:nlocs}
\end{eqnarray}
Here $f^{(h_i)}_a(x,\mu_{\mathrm{F}}^2)$ are the \NLO\ PDF for finding
the parton $a$ with momentum fraction $x$ in the hadron $h_i$ with
momentum $P_i$ at the factorization scale $\mu_{\mathrm{F}}$.  The
contributions $\hat\sigma^{\fact}_{ab,\QCD/\phot}$ result from the PDF
redefinitions that describe the absorption of collinear initial-state
singularities of gluonic or photonic origin (see next section). 
The factors $\delta_{ab,X}^{\virt}$ represent the virtual corrections to the
squared LO matrix elements for the $ab$ initial state,
\begin{equation}
  \label{eq:defcorrfact}
  2\Re\{\M_{ab,X}^{\virt}(\M_{ab,X}^{\LO})^*\} 
\equiv \delta_{ab,X}^{\virt} \,|\M_{ab,X}^{\LO}|^2\, .
\end{equation}

\subsection{Photonic and QCD corrections}
\label{sec:photonic-corrections}

The issue of a gauge-invariant treatment of the photonic
and QCD corrections has been discussed in \citere{Dittmaier:2002nd}
in detail (including even massive fermions). 
{}From the arguments given there and the discussion above, it is clear
that a consistent way of evaluating the photonic and QCD corrections
is to use the complex Z-boson mass $\mu_\PZ$ wherever
it appears. Since the weak mixing angle is derived from the ratio
of the W and Z~masses, and $\MW$ does not appear elsewhere in these
corrections, the quantity $\cw$ can be treated as free parameter in the
context of photonic and QCD corrections, and $\sw$ as well as the
couplings $\gffZ^\pm$ are derived from $\cw$. Specifically, we
set $\cw$ to $\mu_\PW/\mu_\PZ$ in the CMS and to $\MW/\MZ$ in the PS and 
{}FS; the numerical difference is, however, marginal, 
as expected.

The virtual photonic corrections can be decomposed into vertex and box
contributions,
\begin{equation} 
f^{\virt,\,\si\tau}_{\qqb,\,\phot} =
f^{\mathrm{vert},\,\si\tau}_{\qqb,\,\phot}(\hat s) +
f^{\mathrm{box},\,\si\tau}_{\qqb,\,\phot}(\hat s,\hat t)\, ,
\label{eq:dvirtphot}
\end{equation} 
where the vertex part contains also the photonic contributions to the
fermionic wave-function corrections.
The vertex correction $f^{\mathrm{vert}}_{\phot}(\hat s)$
consists of an initial- and a final-state part and reads
\begin{equation}
\label{eq:delta-vert-phot}
f^{\mathrm{vert},\,\si\tau}_{\qqb,\,\phot}(\hat s) = 
   - \frac{e^2}{\hat s} \,\left[ 
     \hat F_{qqV,\phot}(\hat s) + \hat F_{llV,\phot}(\hat s) \right] \,
 \left[ \Qq\Ql
   + \gqqZ^\sigma\, \gllZ^\tau \, \chi_\PZ(\hat s)
 \right] \;,
\end{equation}
with the renormalized vertex form factor
\begin{equation}
\hat F_{ffV,\phot}(\hat s) = - \frac{Q_f^2\al}{2\pi} \biggl[
\ln\bigg(\frac{m_\ga^2}{\hat s}\bigg)\ln\bigg(\frac{m_f^2}{\hat s}\bigg)
+\ln\bigg(\frac{m_\ga^2}{\hat s}\bigg)
+\frac{1}{2}\ln\bigg(\frac{m_f^2}{\hat s}\bigg)-\frac{1}{2}\ln^2\biggl(\frac{m_f^2}{\hat s}\biggr)-\frac{2 \pi^2}{3} + 2
\biggr] \;,
\end{equation}
where irrelevant imaginary parts have been discarded.
The interference terms of the virtual photonic corrections are due to the
photonic box diagrams and can be written as
\begin{equation}
\label{eq:photbox}
f^{\mathrm{box},\,\si\tau}_{\qqb,\,\phot}(\hat s,\hat t) = 
f_{\qqb}^{\ga\ga,\,\sigma \tau}(\hat s,\hat t) + f_{\qqb}^{\PZ\ga,\,\sigma \tau}(\hat s,\hat t) \;.
\end{equation}
The correction factors $f_{\qqb}^{VV',\sigma \tau}$ are given in
\refapp{app:virtRCs}.

\begin{figure}
  \centering
  \input{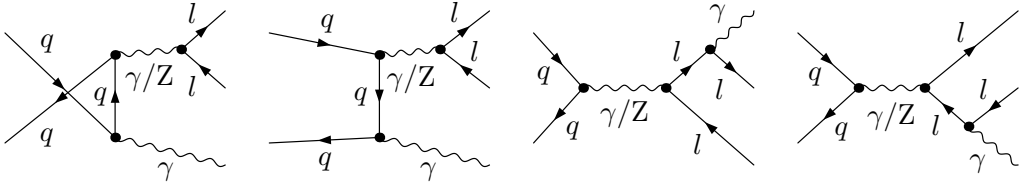}
  \vspace*{-2em}
  \caption{Diagrams for real-photon emission.}
  \label{fig:realphot}
\end{figure}

\begin{figure}
  \centering
  \input{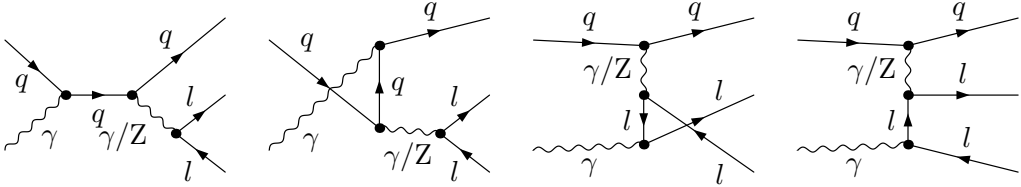}
  \vspace*{-2em}
  \caption{Diagrams for photon-induced processes with incoming quarks.}
  \label{fig:incphot}
\end{figure}

The real photonic bremsstrahlung
corrections to $\qqb\to \gamma/\PZ\to \llb$, whose
diagrams are shown in \reffis{fig:realphot} and \ref{fig:incphot}, 
are calculated using the Weyl--van-der-Waerden
spinor formalism adopting the conventions of
\citere{Dittmaier:1999nn}. 
This results in very compact expressions
for the helicity amplitudes 
$\M^{\sigma_1, \sigma_2, \tau_1, \tau_2}_{ab}(\lambda)
=\sum_{V=\ga,\PZ} \M_{ab,V}^{\sigma_1, \sigma_2, \tau_1, \tau_2}(\lambda)$,
where $ab= \qqb, q\ga, \bar q\ga$ refers to the partonic initial states and
$V$ to the exchanged bosons in the respective diagrams.
{}For real-photon emission we get
\begin{eqnarray}
\M^{-++-}_{q \bar q,V}(+) &=& \phantom{-}2\sqrt{2}\;e^3\gqqVm\,\gllVp
\sprod{p_2}{k_2}^2 \left[ \frac{Q_q}{s_{ll}-\mu_V^2} \,
\frac{\csprod{k_1}{k_2}}{\sprod{p_1}{k_3}\sprod{p_2}{k_3}}
-\frac{Q_l}{\hat s-\mu_V^2} \,
\frac{\csprod{p_1}{p_2}}{\sprod{k_1}{k_3}\sprod{k_2}{k_3}}
\right],
\nl
\M^{-+-+}_{q \bar q,V}(+) &=& -2\sqrt{2}\;e^3\gqqVm\,\gllVm \sprod{p_2}{k_1}^2
\left[ \frac{Q_q}{s_{ll}-\mu_V^2} \,
\frac{\csprod{k_1}{k_2}}{\sprod{p_1}{k_3}\sprod{p_2}{k_3}}
-\frac{Q_l}{\hat s-\mu_V^2} \,
\frac{\csprod{p_1}{p_2}}{\sprod{k_1}{k_3}\sprod{k_2}{k_3}}
\right]
\hspace{2em}
\label{eq:mqqbar}
\end{eqnarray}
in the limit of massless fermions, and we have defined $\mu_\ga=0$.
The spinor products are defined by
\begin{equation}
\langle pq\rangle=\epsilon^{AB}p_A q_B
=2\sqrt{p_0 q_0} \,\Biggl[
{\mathrm{e}}^{-\ri\phi_p}\cos\frac{\theta_p}{2}\sin\frac{\theta_q}{2}
-{\mathrm{e}}^{-\ri\phi_q}\cos\frac{\theta_q}{2}\sin\frac{\theta_p}{2}
\Biggr],
\end{equation}
where $p_A$, $q_A$ are the associated momentum spinors for the light-like
momenta
\begin{eqnarray}
p^\mu&=&p_0(1,\sin\theta_p\cos\phi_p,\sin\theta_p\sin\phi_p,\cos\theta_p),\nl
q^\mu&=&q_0(1,\sin\theta_q\cos\phi_q,\sin\theta_q\sin\phi_q,\cos\theta_q).
\end{eqnarray}
Owing to helicity conservation in the case of massless fermions all
amplitudes with $\si_1=\si_2$ or $\tau_1=\tau_2$ vanish.
The remaining six non-zero helicity amplitudes are obtained from the
amplitudes \refeq{eq:mqqbar} via a parity (P)
transformation
\begin{equation}
  \label{eq:paritytrafo}
  \M_{ab,V}^{-\sigma_1, -\sigma_2, -\tau_1, -\tau_2}(-\lambda) = \sgn(\sigma_1 \sigma_2 \tau_1 \tau_2) 
\left(\M_{ab,V}^{\sigma_1, \sigma_2, \tau_1, \tau_2}(\lambda)\right)^* 
\Big|_{g^\pm_{ffV}\leftrightarrow g^\mp_{ffV}},
\end{equation}
and a CP transformation
\begin{equation}
  \label{eq:CPtrafo}
  \M_{ab,V}^{-\sigma_2, -\sigma_1, -\tau_2, -\tau_1}(-\lambda) = - \sgn(\sigma_1 \sigma_2 \tau_1 \tau_2) 
\left(\M_{ab,V}^{\sigma_1, \sigma_2, \tau_1, \tau_2}(\lambda)\right)^* 
\Big|_{p_1 \leftrightarrow p_2 \atop k_1 \leftrightarrow k_2}.
\end{equation}
Note that in the above formulas the complex masses in the propagators
and the couplings are not complex conjugated, since P and
CP~transformations only act on the wave functions and momenta entering
the amplitudes. 

Apart from the partonic channels with a $\qqb$ pair in the initial state, we
also include the photon-induced processes whose diagrams are shown in
\reffi{fig:incphot} for incoming quarks. Of course, there are also the
corresponding channels for incoming anti-quarks. The amplitudes for
the photon-induced processes 
\beqar
 q(p_{1},\sigma_{1}) + \ga(p_{2},\lambda) &\;\to\;& 
l^-(k_1,\tau_1) + l^+(k_2,\tau_2) + q(k_3,\sigma_{1}),
\nn\\
\ga(p_{1},\lambda) + \bar q(p_{2},\sigma_{2}) &\;\to\;& 
l^-(k_1,\tau_1) + l^+(k_2,\tau_2) + \bar q (k_3,\sigma_{2}) 
\eeqar
are related to the ones for
real-photon emission by crossing symmetry,
\begin{eqnarray}
  \M^{\sigma_1, -\sigma_2, \tau_1, \tau_2}_{q \ga}(-\lambda) &=& 
- \sgn(\sigma_2)\, \M^{\sigma_1, \sigma_2, \tau_1, \tau_2}_{\qqb}(\lambda)\Big|_{ p_2\leftrightarrow -k_3  } \,,\\
  \M^{-\sigma_1, \sigma_2, \tau_1, \tau_2}_{\ga\bar q}(-\lambda) &=&  
- \sgn(\sigma_1)\, \M^{\sigma_1, \sigma_2, \tau_1, \tau_2}_{\qqb}(\lambda)\Big|_{p_1\leftrightarrow -k_3 } \,.
\end{eqnarray}
In terms of Weyl--van-der-Waerden spinors the crossing transformation
$p \to - p$ of a four-momentum $p$ is obtained by inverting the
conjugated parts only,
\begin{equation}
  p_{\dot{A}}\to - p_{\dot{A}}, \qquad p_{A}\to p_{A}\,.
\end{equation}
The contributions $\hat\sigma^{\mathrm{real}}_{q\bar q}$ and
$\hat\sigma^{\mathrm{real}}_{q/\bar q \ga}$ to the partonic cross
section are given by
\begin{equation}
  \label{eq:real-corr-cs}
  \int\rd\hat\sigma^{\mathrm{real}}_{a b} 
= \frac{N_{\mathrm{c},ab}}{4}\frac{1}{2 \hat s} \int \rd\Phi_\gamma \,
   \sum_{\mathrm{pol}} 
   \Big|\M^{\sigma_1, \sigma_2, \tau_1,
  \tau_2}_{a b}(\lambda)\Big|^2\,,
\end{equation}
where the colour factors for the different initial states are
$N_{\mathrm{c},q\bar q}=1/3$ and
$N_{\mathrm{c},q\ga}=N_{\mathrm{c},\bar q\ga}=1$.
The phase-space integral is defined by
\begin{equation}
\int \rd\Phi_\gamma =
\int\frac{\rd^3 {\bf k}_1}{(2\pi)^3 2k_{1,0}}
\int\frac{\rd^3 {\bf k}_2}{(2\pi)^3 2k_{2,0}}
\int\frac{\rd^3 {\bf k}_3}{(2\pi)^3 2k_{3,0}} \,
(2\pi)^4 \delta(p_1+p_2-k_1-k_2-k_3).
\label{eq:dGg}
\end{equation}

The phase-space integrals in the real corrections
$\hat\sigma^{\mathrm{real}}_{q\bar q}$ and
$\hat\sigma^{\mathrm{real}}_{q/\bar q \ga}$ contain logarithmic collinear
divergences in the limit of massless fermions. Moreover,
the real-photon emission integral contains a logarithmic soft singularity
because of the masslessness of the photon.
To regularize the soft and collinear singularities we
introduce small fermion masses and an infinitesimal photon mass
according to the generally known factorization properties of the
squared amplitudes in the singular phase-space regions.
This step is usually performed via
{\it phase-space slicing}, which isolates singular regions in phase space,
or via a {\it subtraction formalism}, which employs an auxiliary
function in the whole phase space in order to cancel all singularities.
In our calculation we proceed as in the treatment of hadronic
W~production as described in \citeres{Dittmaier:2001ay,Brensing:2007qm},
i.e.\ we employ three different methods:
(i) soft phase-space slicing with effective collinear 
factors~\cite{Dittmaier:2001ay},
(ii) two-cutoff phase-space slicing~\cite{Baur:1998kt}
for soft and collinear singularities,
and (iii) dipole subtraction~\cite{Dittmaier:1999mb,Dittmaier:2008md}.
Since the detailed formulas for the CC case can be transferred to the
present NC case in a straightforward way, we do not go into formal
details here, but restrict ourselves to the most important
features of the singularity structure in the final result.

The analytical results on the photonic corrections to $\ga\ga\to\llb$
can be found in \citere{Denner:1998tb}. The following discussion
of final- and initial-state singularities includes both 
$q\bar q$ and $\ga\ga$ scattering.

Two types of final-state collinear singularities arise.
{}First, there is a collinear singularity if the $l^-l^+$ system in the
final state receives a small invariant mass $M_{ll}$, e.g., via a
collinear $\ga^*\to\llb$ splitting. Since we, however,
set a lower limit on $M_{ll}$,
this singular configuration is excluded from our
region of interest.
Second, collinear photon radiation off the final-state charged leptons
is enhanced by the mass-singular factor $\alpha\ln(Q/\Ml)$
(with $Q$ denoting a typical hard scale). The nature of this
singularity is discussed in more detail in \refse{se:Multi-photon},
where an effective treatment of collinear multi-photon emission
is described.

Singularities connected to collinear splittings in the initial state
result from 
$q\to \Pg/\ga q^*$, $\bar q\to\Pg/\ga\bar q^*$
(gluon/photon bremsstrahlung in $q\bar q$ annihilation),
$\Pg/\ga\to q\bar q^*$, $\Pg/\ga\to \bar qq^*$ 
(gluon/photon splittings into $q\bar q$ pairs in $\Pg/\ga q$
and $\Pg/\ga \bar q$ scattering),
$q\to q\ga^*$, $\bar q\to \bar q\ga^*$
(forward scattering of $q$ or $\bar q$ in $\ga q/\bar q$
scattering), and
$\ga\to l^\pm l^{\mp*}$
(photon splitting into $l^+l^-$ pairs in $\ga q/\bar q$
and $\ga\ga$ scattering).
The last splitting corresponds to configurations with a charged 
lepton $l^\pm$ lost in the beam direction (proton remnant), 
i.e.\ it only contributes
if not both charged leptons are required in the event signature;
the contribution of this configuration is enhanced by the factor
$\alpha\ln(Q/\Ml)$ (again with $Q$ denoting a typical hard scale).
The other splittings lead to $l^+l^-$ pairs in the final state
with a gluon, (anti-)quark, or photon lost in the proton remnant;
the corresponding contributions are enhanced by factors
$\alpha_{\mathrm{s}}\ln(Q/\Mq)$ and $\alpha Q_q^2\ln(Q/\Mq)$ for
gluonic and photonic splittings, respectively.
These (non-perturbative) singular contributions are absorbed into
the PDF via factorization, where finite contributions to this
PDF redefinition define the factorization scheme.
In detail the LO PDF $f^{(h)}_a(x)$, describing the emission of parton $a$
out of the hadron $h$ with longitudinal momentum fraction $x$,
are split according to (see, e.g., \citere{Baur:1998kt,Diener:2005me})
\begin{eqnarray}
  \label{eq:redef-pdf}
  f^{(h)}_{q/\bar q}(x) \rightarrow f^{(h)}_{q/\bar q}(x,\mu_{\mathrm{F}}^2) 
  &-& \;\frac{\alpha_{\mathrm{s}} C_{\mathrm{F}}}{2\pi}\;\int^1_x \frac{\rd z}{z} \; 
  f^{(h)}_{q/\bar q}\left(\frac{x}{z},\mu_{\mathrm{F}}^2\right) \nl
  &&\qquad \times \left\{\ln\biggl(\frac{\mu_{\mathrm{F}}^2}{m_q^2}\biggr) \;
  \left[P_{ff}(z)\right]_+ -
  \left[P_{ff}(z)\;(2\ln(1-z)+1)\right]_+ + C^\MSbar_{ff}(z)\;
  \right\} \nl
  &-&\; \frac{\alpha_{\mathrm{s}} T_{\mathrm{F}}}{2\pi} \;\int^1_x \frac{\rd z}{z} \;
  f^{(h)}_g\left(\frac{x}{z},\mu_{\mathrm{F}}^2\right) \;
  \Big\{\ln\biggl(\frac{\mu_{\mathrm{F}}^2}{m_q^2}\biggr)
  \; P_{f\ga}(z) + C^\MSbar_{f\ga}(z) \;\Big\}\nl
  &-&\; \frac{\alpha\,Q_q^2}{2\pi} \; \int^1_x \frac{\rd z}{z}\; 
  f^{(h)}_{q/\bar q}\left(\frac{x}{z},\mu_{\mathrm{F}}^2\right) \\
  &&\qquad \times \left\{\ln\biggl(\frac{\mu_{\mathrm{F}}^2}{m_q^2}\biggr) \; 
  \left[P_{ff}(z)\right]_+ - \left[P_{ff}(z)\;(2\ln(1-z)+1)\right]_+ 
  + C^\DIS_{ff}(z)\; \right\} \nl
  &-&3 \frac{\alpha\,Q_q^2}{2\pi} \;\int^1_x \frac{\rd z}{z} \; 
  f^{(h)}_\ga\left(\frac{x}{z},\mu_{\mathrm{F}}^2\right) \; 
  \left\{\ln\biggl(\frac{\mu_{\mathrm{F}}^2}{m_q^2}\biggr) \; P_{f\gamma}(z) +
  C^\DIS_{f\gamma}(z) \;\right\}, \nl
  f^{(h)}_\ga(x) \rightarrow f^{(h)}_\ga(x,\mu_{\mathrm{F}}^2) &-&\; \frac{\alpha\,Q_q^2}{2\pi}
  \sum_{a=q,\bar{q}} \int^1_x \frac{\rd z}{z} \;
  f^{(h)}_a\left(\frac{x}{z},\mu_{\mathrm{F}}^2\right) 
\label{eq:photonPDFredef}
\\
  && \qquad \times \left\{\ln\biggl(\frac{\mu_{\mathrm{F}}^2}{m_q^2}\biggr) \;
  P_{\gamma f}(z) 
  - P_{\ga f}(z)\,(2\ln z+1)
  + C^\DIS_{\gamma f}(z) \;\right\}\nonumber
\end{eqnarray}
into \NLO\ PDF $f^{(h)}_a(x,\mu_{\mathrm{F}}^2)$, which now include parton
emission up to a scale of the order of the factorization scale 
$\mu_{\mathrm{F}}$.
The splitting functions are given by
\begin{equation}
  \label{eq:pff-pfa-paf}
  P_{ff}(z)   = \frac{1+z^2}{1-z}\,, \qquad
  P_{f \ga}(z)= z^2 + (1-z)^2\,, \qquad
  P_{\ga f}(z)= \frac{1 + (1-z)^2}{z}.
\end{equation}
The coefficient functions $C_{ij}(z)$, defining the finite parts,
coincide with the usual definition in $D$-dimensional
regularization for exactly massless partons where the $\ln m_q$ terms
appear as $1/(D-4)$ poles.
Details about the photon PDF redefinition are given in \refapp{app:photonPDF}.
{}Following standard definitions of QCD, we
distinguish the $\MSbar$ and DIS-like schemes which are formally
defined by the coefficient functions
\begin{eqnarray}
    C^{\MSbar}_{ff}&=&C^{\MSbar}_{f\ga}=C^{\MSbar}_{\ga f}=0\,, \\
    C^{\DIS}_{ff}(z) &=& \left[ \; P_{ff}(z) \; 
    \biggl( \; \ln
    \biggl( \frac{1-z}{z} \biggr) - \frac{3}{4} \; \biggr)
    + \frac{9+5z}{4} \; \right]_+ \;,\\
    C^{\DIS}_{f\gamma}(z) &=&
    P_{f\gamma}(z)\;\ln\biggl(\frac{1-z}{z}\biggr)
    - 8z^2 + 8z -1 \;,\\
    C^{\DIS}_{\gamma f}(z) &=& - C^{\mathrm{DIS}}_{ff}(z) \;. 
  \label{eq:redef-pdf-C-schema}
\end{eqnarray}
We use the MRST2004qed PDF \cite{Martin:2004dh} which consistently
include ${\cal O}(\alpha_{\mathrm{s}})$ QCD and
${\cal O}(\alpha)$ photonic corrections. These PDF include a photon
distribution function for the proton and thus allow to take into
account photon-induced partonic processes. As explained in
\citere{Diener:2005me}, the consistent use of these PDF requires the
$\MSbar$ factorization scheme for the QCD, but the DIS scheme for the
photonic corrections.%
\footnote{Note that our choice of the factorization scheme and
coefficient functions for incoming photons differs from the previously presented
results~\cite{CarloniCalame:2007cd,Arbuzov:2007kp}.
In \citere{Arbuzov:2007kp} the $\MSbar$ scheme was employed, and
the coefficient function $C^{\DIS}_{\gamma f}(z)$ of
\citere{CarloniCalame:2007cd} was fixed somewhat ad hoc.
Our redefined photon PDF is fixed in such a way that
the momentum sum rule for the total proton momentum is respected,
i.e.\ our fixation of the photon PDF follows the same logic as the 
gluon PDF redefinition in the DIS scheme for QCD factorization.
In the first preprint version of this paper we employed a different 
factorization prescription that also respected the proton momentum sum
rule, but was not in line with the standard NLO QCD conventions
for factorization schemes. Therefore, we switched to the scheme
described here (see comments at the end of \refapp{app:photonPDF}).}
The contributions $\hat\sigma^{\fact}_{ab,\QCD/\phot}$ appearing in
\refeq{eq:nlocs} result from the corrections
in the PDF replacements \refeq{eq:redef-pdf} and \refeq{eq:photonPDFredef} after these
substitutions are made in the LO prediction for the hadronic
cross section. More precisely, $\hat\sigma^{\fact}_{ab,\QCD/\phot}$
corresponds to the \NLO\ QCD/photonic contribution (i.e.\ linearized
in $\alpha_{\mathrm{s}}$ or $\alpha$) proportional to the
PDF combination 
$f^{(h_1)}_a(x_1,\mu_{\mathrm{F}}^2)f^{(h_2)}_b(x_2,\mu_{\mathrm{F}}^2)$
of the partonic $ab$ initial state.

\subsection{Weak corrections}
\label{sec:weak-corrections}

In the following we sketch the structure of the weak corrections 
and emphasize those points that
are relevant for the treatment of the resonance and for the change
from one input-parameter scheme to another.  
The correction factor $f^{\virt}_{\qqb,\weak}$, which is introduced
in Eq.~\refeq{eq:m1virt}, is decomposed
according to the splitting into self-energy, vertex, and box diagrams,
\begin{equation}
  f^{\virt,\,\si\tau}_{\qqb,\weak} = 
  f^{\mathrm{self},\,\si\tau}_{\qqb,\weak}(\hat s) + 
  f^{\mathrm{vert},\,\si\tau}_{\qqb,\weak}(\hat s) + 
  f^{\mathrm{box},\,\si\tau}_{\qqb,\weak}(\hat s,\hat t)\,.
\label{eq:dvirtew}
\end{equation} 
The self-energy corrections comprise contributions from the
$\ga\ga$, $\ga\PZ$, and $\PZ\PZ$ self-energies, the results of which
can be found in \citere{Denner:1993kt} in `t~Hooft--Feynman gauge.
Self-energy corrections to the external
fermion states are absorbed into vertex counterterms, as usually
done in on-shell renormalization schemes.
The one-loop diagrams for the weak vertex and box
corrections are shown in \reffis{fig:vertboxdiags}b) and c),
their complete expressions are provided in \refapp{app:virtRCs}.

\paragraph{(i) Complex-mass scheme}

We first describe the calculation in the CMS. 
The self-energy corrections explicitly read
\beq
\label{eq:dsevvew}
f^{\mathrm{self},\si\tau}_{\qqb,\weak}(\hat s) =
e^2 \,\frac{Q_q \,Q_l}{\hat{s}^2} 
\,\hat\Sigma^{\ga\ga}_{\mathrm{T}}(\hat s) 
+ e^2 \,\frac{\gqqZ^\sigma\,\gllZ^\tau}{(\hat s - \xMZ^2)^2} 
\,\hat\Sigma^{\PZ\PZ}_{\mathrm{T}}(\hat s) 
- e^2 \,\frac{Q_l\,\gqqZ^\sigma+Q_q\,\gllZ^\tau}{\hat{s}\,(\hat s - \xMZ^2)}
\,\hat\Sigma^{\ga\PZ}_{\mathrm{T}}(\hat s) \;,
\eeq
where $\hat\Sigma^{VV'}_{\mathrm{T}}$ denote the renormalized 
(transverse) gauge-boson self-energies,
\begin{eqnarray}
\Sigma^{\ga\ga}_{\mathrm{T}}(\hat s) &=&
\Sigma^{\ga\ga}_{\mathrm{T}}(\hat s) + \delta {\cal Z}_{\ga\ga} \, \hat{s}\;, 
\nn\\
\Sigma^{\PZ\PZ}_{\mathrm{T}}(\hat s) &=&
\Sigma^{\PZ\PZ}_{\mathrm{T}}(\hat s) - \delta \MZ^2 + \delta {\cal Z}_{\PZ\PZ} (\hat s - \xMZ^2)\;, 
\nn\\
\Sigma^{\ga\PZ}_{\mathrm{T}}(\hat s) &=&
\Sigma^{\ga\PZ}_{\mathrm{T}}(\hat s) + \frac{1}{2} \delta {\cal Z}_{\ga \PZ}\,\hat{s} +\frac{1}{2} \delta {\cal Z}_{\PZ \ga}\, (\hat{s} - \xMZ^2)\;.
\end{eqnarray}
As mentioned above, the explicit results of \citere{Denner:1993kt} on the
unrenormalized self-energies $\Sigma^{VV'}_{\mathrm{T}}$ can be used,
however, we stress that complex gauge-boson masses and couplings
have to be inserted everywhere. 
The renormalization constants $\de\MZ^2$ and 
$\delta {\cal Z}_{VV'}$ are defined in Eqs.~(4.9) and (4.10) of
\citere{Denner:2005fg} for the CMS. They are expressed in terms
of gauge-boson self-energies and consistently evaluated with complex
parameters (but real-valued momenta); 
in particular, no real part is taken in their definition,
in contrast to the usual on-shell renormalization scheme, as, e.g., 
defined in \citere{Denner:1993kt}.

The vertex corrections can be written as
\beq
\label{eq:dvertew}
f^{\mathrm{vert},\,\si\tau}_{\qqb,\weak}(\hat s) =
- e^2 \, \frac{Q_q Q_l}{\hat s} 
\left[\hat F^\sigma_{qq\ga,\weak}(\hat s) + 
\hat F^\tau_{ll\ga,\weak}(\hat s) \right]
- e^2 \,\frac{\gqqZ^\sigma \gllZ^\tau}{\hat s - \xMZ^2} 
\left[\hat F^\sigma_{qqZ,\weak}(\hat s) 
+ \hat F^\tau_{llZ,\weak}(\hat s) \right] 
\;, 
\eeq
with the renormalized vertex form factors
\beq
\hat F^\sigma_{ffV,\weak}(\hat s) = F^\sigma_{ffV,\weak}(\hat s) 
+ \delta_{ffV,\weak}^{\mathrm{ct},\,\sigma} \;.
\eeq
The explicit expressions for the unrenormalized
form factors $F_{ffV,\weak}(\hat s)$
are given in \refapp{app:virtRCs}. The subscript ``weak'' indicates
that the contributions from photon-exchange diagrams are omitted both
in the form factors and in the vertex counterterms 
$\delta_{ffV,\weak}^{\mathrm{ct},\,\si}$.
In the $\alpha(0)$-scheme, the counterterms are given by
\begin{eqnarray}
\delta^{\mathrm{ct},\si}_{ff\ga,\weak} &=& 
\frac{\delta e}{e} + \frac{1}{2}\delta {\cal Z}_{\ga \ga} + \delta {\cal Z}^\si_{f,\weak} - \frac{1}{2} \frac{\gffZ^\si}{Q_f} \delta {\cal Z}_{\PZ \ga}\;,
\nn\\ 
\delta^{\mathrm{ct},\si}_{ffZ,\weak} &=& \frac{\delta \gffZ^\si}{\gffZ^\si} + \frac{1}{2} \delta {\cal Z}_{\PZ\PZ} + \delta {\cal Z}^\si_{f,\weak} - \frac{1}{2} \frac{Q_f}{\gffZ^\si} \delta {\cal Z}_{\ga \PZ}\;,
\end{eqnarray}
with
\beq
\delta\gffZ^+ = -\frac{\sw}{\cw} Q_f \left( \frac{\delta e}{e} 
+ \frac{1}{\cw^2} \frac{\delta{\sw}}{\sw} \right)\,, \qquad
\delta \gffZ^- = \frac{I^3_{\mathrm{w},f}}{\sw \cw}\left(\frac{\delta e}{e} 
+ \frac{\sw^2-\cw^2}{\cw^2}\frac{\delta{\sw}}{\sw} \right) + \delta\gffZ^+   \;.
\hspace{1em}
\eeq
Note that the subscript ``weak'' appears only on the fermionic
wave-function renormalization constants $\delta {\cal Z}^\si_f$,
obtained from the the fermion self-energies,
because only those receive a photonic contribution.
We again emphasize the difference between the renormalization
constants in the CMS~\cite{Denner:2005fg} and the usual on-shell 
scheme~\cite{Denner:1993kt}. In the CMS, all quantities are derived
from complex masses and couplings, and no real parts are taken from the
self-energies that enter the renormalization constants. Explicit results
can be found in \citere{Denner:2005fg}.
In particular, the renormalization constant
of the weak mixing angle, $\delta\sw$, is connected to the mass
renormalization of the complex gauge-boson masses. 

The charge renormalization
constant $\delta e/e$ contains logarithms of the light-fermion masses,
inducing large corrections proportional to $\alpha\ln m_f$,
which are related to the running of the electromagnetic coupling
$\alpha(Q)$ from $Q=0$ to a high-energy scale. In order to render
these quark-mass logarithms meaningful, it is necessary to adjust
these masses to the asymptotic tail of the hadronic contribution to
the vacuum polarization 
$\Pi^{\ga\ga}(Q^2)=\Sigma^{\ga\ga}_{\mathrm{T}}(Q^2)/Q^2$
of the photon. Using $\alpha(\MZ)$, as defined in
\citere{Burkhardt:1995tt}, as input this adjustment is implicitly
incorporated, and the counterterm reads
\begin{equation}
\label{eq:dffVal}
\delta_{ffV}^{\mathrm{ct},\,\si}\Big|_{\alpha(\MZ)} =
\delta_{ffV}^{\mathrm{ct},\,\si}\Big|_{\alpha(0)} -
\frac{1}{2}\Delta\alpha(\MZ),
\end{equation}
where 
\begin{equation}
\label{eq:dalpha}
  \Delta\alpha(\MZ) = 
  \Pi^{\ga\ga}_{f\ne \Pt}(0) - \Re\{\Pi^{\ga\ga}_{f\ne \Pt}(\MZ^2)\}
\approx
\frac{\alpha(0)}{3\pi}\sum_{f\ne\mathrm{t}} N^{\mathrm{c}}_f Q_f^2
\left[\ln\biggl(\frac{\MZ^2}{m_f^2}\biggr)-\frac{5}{3}\right],
\end{equation}
with $\Pi^{\ga\ga}_{f\ne \Pt}$ denoting the photonic vacuum
polarization induced by all fermions other than the top quark (see
also \citere{Denner:1993kt}), and $N^{\mathrm{c}}_l=1$ and $N^{\mathrm{c}}_q=3$
are the colour factors for leptons and quarks, respectively.
In contrast to the $\alpha(0)$-scheme
the coun\-ter\-term 
$\delta_{ffV}^{\mathrm{ct},\,\si}\big|_{\alpha(\MZ)}$ does not involve 
light-quark masses, since all corrections of the form
$\alpha^n\ln^n(m_f^2/\MZ^2)$ are absorbed in the LO cross
section parametrized by
$\alpha(\MZ)=\alpha(0)/[1-\Delta\alpha(\MZ)]$.  In the
$\GF$-scheme, the transition from $\alpha(0)$ to $\GF$ is ruled by the
quantity $\Delta r^{(1)}$, which is deduced from muon decay,
\begin{equation}
\alpha_{\GF}\equiv\frac{\sqrt{2}\GF\MW^2(\MZ^2-\MW^2)}{\pi\MZ^2}
=\alpha(0)\left(1+\Delta r^{(1)}\right) + {\cal O}(\alpha^3).
\end{equation}
The counterterm $\delta_{ffV}^{\mathrm{ct},\,\si}$ in the $\GF$-scheme reads
\begin{equation}
\label{eq:dffVr}
\delta_{ffV}^{\mathrm{ct},\,\si}\Big|_{\GF} =
\delta_{ffV}^{\mathrm{ct},\,\si}\Big|_{\alpha(0)} 
- \frac{1}{2}\Delta r^{(1)},
\end{equation}
where the one-loop correction $\Delta r^{(1)}$ is evaluated with
complex masses and couplings in the CMS. This translation of
$\Delta r^{(1)}$ into the CMS is easily obtained upon analytical
continuation of the result given in 
\citere{Denner:1993kt} in the on-shell scheme.
Note that $\Delta r^{(1)}$ implicitly contains large contributions
from $\Delta\alpha(\MZ) \sim 6\%$ and the (one-loop) correction
$(\cw^2/\sw^2)\Delta\rho^{(1)} \sim 3\%$ induced by the
$\rho$-parameter, where $\Delta\rho^{(1)}\propto\GF\Mt^2$.
Thus, the large fermion-mass logarithms are also resummed in the
$\GF$-scheme, and the LO cross section in
$\GF$-parametrization absorbs large universal corrections induced by
the $\rho$-parameter. In \refse{se:ips} we further elaborate on 
higher-order effects induced by $\De\alpha$ and $\De\rho$.

The box correction $f^{\mathrm{box},\,\si\tau}_{\qqb}(\hat s,\hat t)$
is the only virtual correction that depends also on the scattering
angle, i.e.\ on the variables $\hat t$ and $\hat u=-\hat s-\hat t$.
The boxes are decomposed into the contributions of the $\PZ\PZ$ and
$\PW\PW$ box diagrams,
\begin{equation}
\label{eq:ewbox}
f^{\mathrm{box},\,\sigma \tau}_{\qqb,\weak}(\hat s,\hat t) = 
f_{\qqb}^{\PZ\PZ,\,\sigma \tau}(\hat s,\hat t) +
f_{\qqb}^{\PW\PW,\,\sigma \tau} (\hat s,\hat t)\;.
\end{equation}
The individual correction factors are given in
\refapp{app:virtRCs}. In \refapp{app:virtRCs} we also give the
explicit expressions for the vertex and box corrections for incoming
\Pb-quarks, where due to the large mass of the top quark additional
diagrams [see \reffi{fig:vertboxdiags}c)] have to be taken into
account.

\paragraph{(ii) Pole scheme}

As explained in \refse{se:born}, the application of the pole 
scheme~\cite{Stuart:1991xk,Aeppli:1993cb} 
starts from a fixed-order calculation without any special treatment
of the resonance. 
Specifically we calculate the weak corrections in the on-shell
renormalization scheme of \citere{Denner:1993kt}, i.e.\ in our
PS calculation all masses and couplings are real quantities and
the Z~decay width $\GZ$ only appears where it is made explicit in 
the following formulas. The input-parameter schemes are defined
as in the previous section, with $\De\alpha$ and $\Delta r^{(1)}$
derived from real input parameters.

In a second step, the resonance pole is isolated
from the non-resonant remainder and dressed by a properly Dyson-summed
Breit--Wigner propagator. The definition of a
gauge-independent residue on resonance, in general, 
involves some freedom in the more-dimensional phase space, because
the resonance location fixes only a single invariant. In our case,
for instance, two different definitions of the residue result
if we write the resonant contribution to the LO amplitude
either as $r_1(\hat s,\hat t)/(\hat s-\MZ^2)$ or
as $r_2(\hat s,\hat u)/(\hat s-\MZ^2)$ and simply set
$\hat s\to\MZ^2$ in the numerators. Off resonance ($\hat s\ne\MZ^2$)
the two versions
for the residue, $r_1(\MZ^2,\hat t)$ and $r_2(\MZ^2,\hat u)$,
obviously are not the same due to $\hat s+\hat t+\hat u=0$.
We apply
the pole scheme to the form factors $f^{\virt,\,\sigma \tau}_{\qqb,\weak}$
as defined in \refeq{eq:m1virt}, i.e.\ we single out the resonance pole
after splitting off the spin-dependent standard matrix elements
$\mathcal{A}^{\sigma \tau}$ from the amplitude.
Note that the resonant part of 
$f^{\virt,\,\sigma \tau}_{\qqb,\weak}$
comprises self-energy and vertex corrections only, which merely
depend on $\hat s$, but not on the variables $\hat t$ and $\hat u$.

{}For the vertex corrections this procedure is very simple. 
The contributions involving Z-boson exchange, 
$f^{\mathrm{vert},\PZ,\,\si\tau}_{\qqb}$,
are modified as follows,
\begin{eqnarray}
f^{\mathrm{vert},\PZ,\,\si\tau}_{\qqb,\weak}(\hat s) &=& 
- e^2 \, \frac{\gqqZ^\sigma \gllZ^\tau}{\hat s - \MZ^2} 
\left[\hat F^\sigma_{qqZ,\weak}(\hat s) +\hat F^\tau_{llZ,\weak}(\hat s)  \right] 
\nl
&\to& - e^2 \, \gqqZ^\sigma \gllZ^\tau 
\biggl[ \frac{\hat F^\sigma_{qqZ,\weak}(\MZ^2)  
    +\hat F^\tau_{llZ,\weak}(\MZ^2)}{\hat s - \MZ^2 + \ri \MZ \GZ} 
\nl
&&\qquad  {}
+ \frac{\hat F^\sigma_{qqZ,\weak}(\hat s) 
    -\hat F^\sigma_{qqZ,\weak}(\MZ^2) 
  + \hat F^\tau_{llZ,\weak}(\hat s)
    -\hat F^\tau_{llZ,\weak}(\MZ^2)}{\hat s - \MZ^2}
\biggr] \;, 
\hspace{2em}
\end{eqnarray}
while the non-resonant contributions involving photon exchange
are kept unchanged. Off resonance the introduction of the finite Z-decay
width $\GZ$ in the denominator of the vertex corrections changes
the amplitude only in ${\cal O}(\alpha^2)$ relative to LO, i.e.\ the
effect is beyond \NLO.

The treatment of the self-energy corrections is somewhat more
involved and requires the inclusion of the LO amplitude.
The sum of the LO and self-energy contributions is modified as follows,
\begin{eqnarray}
\label{eq:PSself}
f_{\qqb}^{\mathrm{LO},\,\sigma \tau}
+f^{\mathrm{self},\,\sigma \tau}_{\qqb,\weak} &=&
-e^2 \Biggl\{ \frac{ \Qq \, \Ql}{\hat{s}} 
\left[ 1-\frac{\hat \Sigma^{\ga\ga}_{\mathrm{T}}(\hat s)}{\hat s}\right] 
+ \frac{\gqqZ^\sigma\, \gllZ^\tau}{\hat{s}-M_{\PZ}^2} 
\left[ 1-\frac{\hat \Sigma^{\PZ\PZ}_{\mathrm{T}}(\hat s)}{\hat s - \MZ^2}\right]
\nl
&& \qquad
{} + \frac{Q_l\, \gqqZ^\sigma +Q_q \,\gllZ^\tau}{\hat s} 
\frac{\hat \Sigma^{\ga\PZ}_{\mathrm{T}}(\hat s)}{\hat s - \MZ^2}  
\Biggr\} 
\nl
&=&-e^2 \Biggl\{ \frac{ \Qq \, \Ql}{\hat{s}} 
\left[ 1-\frac{\hat \Sigma^{\ga\ga}_{\mathrm{T}}(\hat s)}{\hat s}\right] 
+\frac{\gqqZ^\sigma\, \gllZ^\tau}{\hat{s}-M_{\PZ}^2} 
\Biggl[1 - \frac{\hat \Sigma^{\PZ\PZ}_{\mathrm{T}}(\MZ^2)}{\hat s - \MZ^2} \nl
&&\qquad\quad {} 
- \hat\Sigma^{'\PZ\PZ}_{\mathrm{T}}(\MZ^2) -\frac{\hat \Sigma^{\PZ\PZ}_{\mathrm{T}}(\hat s) -\hat \Sigma^{\PZ\PZ}_{\mathrm{T}}(\MZ^2) -\left(\hat s-\MZ^2\right)\hat \Sigma^{'\PZ\PZ}_{\mathrm{T}}(\MZ^2)}{\hat s - \MZ^2}
\Biggr]
\nl
&&\qquad {}
+\frac{Q_l\, \gqqZ^\sigma +Q_q \,\gllZ^\tau}{\hat s} 
\left[\frac{\hat \Sigma^{\ga\PZ}_{\mathrm{T}}(\MZ^2)}{\hat s - \MZ^2} 
 + \frac{\hat \Sigma^{\ga\PZ}_{\mathrm{T}}(\hat s)-\hat \Sigma^{\ga\PZ}_{\mathrm{T}}(\MZ^2)}{\hat s - \MZ^2}  \right]\Biggr\} 
\nl
&\to& - e^2 \Biggl\{ \frac{ \Qq \, \Ql}{\hat{s}} 
\left[ 1-\frac{\hat \Sigma^{\ga\ga}_{\mathrm{T}}(\hat s)}{\hat s}\right] + \gqqZ^\sigma\, \gllZ^\tau \Biggl[\frac{1-\hat \Sigma^{'\PZ\PZ}_{\mathrm{T}}(\MZ^2)}{\hat s-\MZ^2+\ri \MZ \GZ} \nl
&&\qquad\quad {}
- \frac{\hat \Sigma^{\PZ\PZ}_{\mathrm{T}}(\hat s) -\hat \Sigma^{\PZ\PZ}_{\mathrm{T}}(\MZ^2) -\left(\hat s-\MZ^2\right)\hat \Sigma^{'\PZ\PZ}_{\mathrm{T}}(\MZ^2)}{\left(\hat s-\MZ^2\right)^2} \Biggr]
\nn\\
&&\qquad {}
+ \left(Q_l\, \gqqZ^\sigma +Q_q \,\gllZ^\tau \right) \Bigg[  
\frac{1}{\hat s - \MZ^2+ \ri \MZ \GZ} \frac{\hat \Sigma^{\ga\PZ}_{\mathrm{T}}(\MZ^2)}{\MZ^2}  \nl
&&\qquad\quad {} 
+ \frac{1}{\hat s - \MZ^2} \Bigg(\frac{\hat \Sigma^{\ga\PZ}_{\mathrm{T}}(\hat s)}{\hat s} - \frac{\hat \Sigma^{\ga\PZ}_{\mathrm{T}}(\MZ^2)}{\MZ^2} \Bigg)  \Bigg]\Biggr\} 
\end{eqnarray}
with $\hat \Sigma^{'\PZ\PZ}_{\mathrm{T}}(\hat s)= \partial\hat
\Sigma^{\PZ\PZ}_{\mathrm{T}}(\hat s)/ \partial \hat s$. Here we have
used the fact that in the on-shell renormalization scheme
the renormalized \PZ-boson self-energy fulfills
$\Re\hat\Sigma^{\PZ\PZ}_{\mathrm{T}}(\MZ^2) = 0$ and that the
resummed terms account for some imaginary parts via
$\Im\hat\Sigma^{\PZ\PZ}_{\mathrm{T}}(\MZ^2) = \MZ\GZ$, which holds
in \Oa. 
Off resonance the modification changes the amplitude only in
${\cal O}(\alpha^2)$, i.e.\ beyond \NLO.
In the resonance region ($\hat s\approx\MZ^2$) the terms involving $\GZ$ 
in the denominators do not count as \Oa\ corrections, but as LO terms.
Thus, in order to achieve \NLO\ accuracy there, $\GZ$ has to be inserted
with \NLO\ precision (or better), or the experimental value should be used.
Since the residue of the
propagator is a gauge-independent quantity, this modification can be
done in the resonant parts without spoiling gauge invariance. In
our numerical evaluation we use the experimental value.
We finally note that the result \refeq{eq:PSself} of the PS
substitution can also be obtained upon considering the
resonance region of an amplitude that results from the full
Dyson summation of the matrix propagator of the $\ga/\PZ$ system
(see, e.g., \citeres{LEP1,Bardin:1999gt}).

The weak box corrections do not become resonant, so that they are
not modified in the pole scheme.

\paragraph{(iii) Factorization scheme}

As a third option to define the weak corrections, we make use of
the fact that the {\it relative} weak corrections 
$\delta^{\mathrm{virt}}_{\qqb,\weak}$ to the differential partonic
cross sections are regular functions of $\hat s$, even in the resonance
region ($\hat s\to\MZ^2$) without introducing a finite Z~width. 
{}For the virtual photonic corrections
this is not the case because of the appearance of corrections
proportional to $\alpha\ln(\hat s-\MZ^2)$.
We, thus, can define the weak \NLO\ correction to the differential
partonic cross section in the FS scheme by
\beq
\rd\hat\sigma_{\qqb,\weak}\Big|_{\mathrm{FS}} =
\delta^{\mathrm{virt}}_{\qqb,\weak}
\Big|_{\GZ=0,\,\de\MZ=\Sigma^{\PZ\PZ}_{\mathrm{T}}(\MZ^2)}
\;\times\;
\rd\hat\sigma_{\qqb}^{\mathrm{LO}},
\eeq
where the LO cross section $\rd\hat\sigma_{\qqb}^{\mathrm{LO}}$, as given in 
\refeq{eq:locs}, contains the Z~resonance structure.
The subscripts on $\delta^{\mathrm{virt}}_{\qqb,\weak}$ indicate that
the Z~width is set to zero everywhere and that the Z-mass counterterm
is derived from the full on-shell Z-boson self-energy (i.e.\
including both real and imaginary parts), in order to avoid double
counting of the width effect already present in the LO cross section.
This simple scheme respects gauge invariance, because the LO
contribution does (see \refse{se:born}) and the relative correction
is derived from the ratio of two gauge-invariant quantities, viz.\
the weak correction and the LO contribution without any Dyson
summation.

As in the PS, the FS calculation only employs real
masses and couplings; the width $\GZ$ merely enters the LO cross
section. The input-parameter schemes are defined in complete analogy
to the PS.

\vspace{2em}

\begin{figure}
  \centering
  \includegraphics{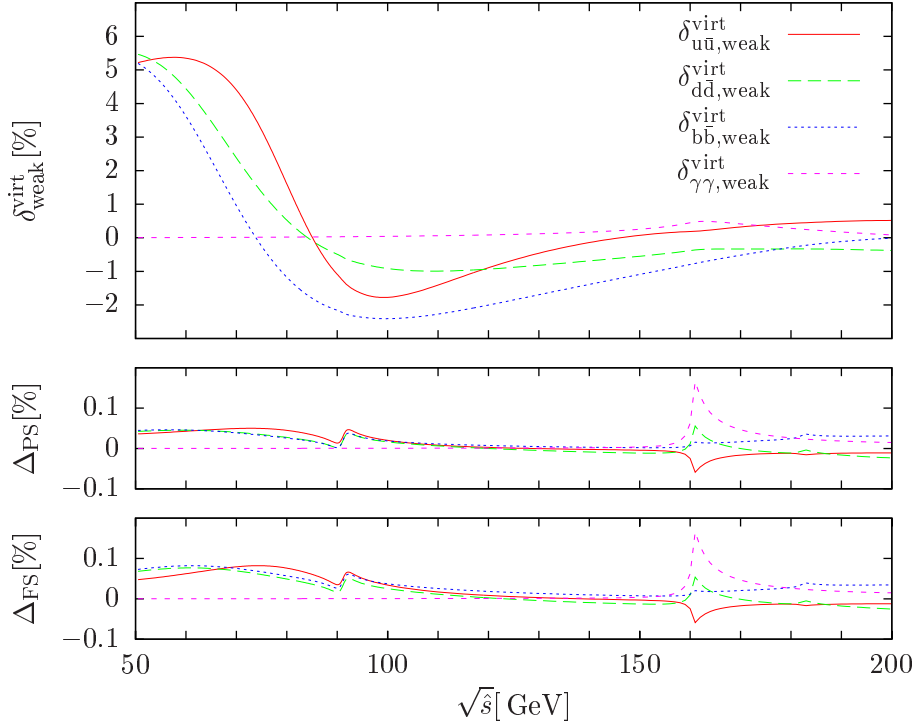}
  \caption{Weak corrections $\delta^\mathrm{virt}_{\qqb,\weak}$ and
    $\delta^\mathrm{virt}_{\ga\ga,\weak}$ to the total partonic cross
    sections for the different initial states and the
    differences $\Delta_{\mathrm{X}}$ between scheme X and the CMS for
    treating the Z~resonance.}
  \label{fig:ewcorr}
\end{figure}

In \reffi{fig:ewcorr} we show the relative weak corrections
$\delta^{\mathrm{virt}}_{\qqb,\weak}$ to the total partonic $q\bar q$
cross sections for incoming up-type, down-type, and $\Pb$-quarks, 
and $\delta^{\mathrm{virt}}_{\ga\ga,\weak}$~, the weak corrections to
$\ga\ga\to\llb$.
The difference between
$\delta^{\mathrm{virt}}_{\Pd\bar \Pd,\weak}$ and
$\delta^{\mathrm{virt}}_{\Pb\bar \Pb,\weak}$ is due to diagrams involving
W~bosons and top quarks. It turns out that the differences
\begin{equation}
  \label{eq:rel-diff-zmassschemes}
  \Delta_{\mathrm{X}}= \delta^{\mathrm{virt}}_{\qqb,\weak}\Big|_{\mathrm{X}}
           -\delta^{\mathrm{virt}}_{\qqb,\weak}\Big|_{\mathrm{CMS}}
\end{equation}
between the different schemes for treating the resonance
are below one per mille (apart from the W-pair threshold at
$\sqrt{\hat s}=2\MW$ where only the CMS delivers smooth results). 
We, therefore, conclude that all three
schemes are equally good in describing the weak corrections to the
Z~resonance.

\subsection{Higher-order electroweak effects}
\label{se:hoew} 

In this section we describe the inclusion of leading higher-order
electroweak corrections to the parton processes $q\bar q\to
\gamma/\PZ\to \llb$. First, we discuss the inclusion of the
leading universal higher-order corrections originating from the
renormalization of the electroweak couplings in the various
input-parameter schemes.
Second, we consider the leading
electroweak corrections in the Sudakov regime which are enhanced by
large logarithms $\ln^2(\hat s/\MW^2)$. Last we
discuss the inclusion of multi-photon final-state radiation, which is
treated using the structure-function approach.

\subsubsection{Leading electroweak effects and choice of couplings}
\label{se:ips}

At moderate scales the leading electroweak non-photonic 
corrections in the SM are due to the running
of the electromagnetic coupling $e^2 = 4 \pi \alpha$ from
zero-momentum transfer to the electroweak scale, and the large mass
splitting between the bottom and the top quark and the associated
breaking of the weak isospin symmetry. These leading effects
are usually quantified by $\Delta\alpha$ and $\Delta\rho$, 
respectively, and their two-loop effects can be included in a straightforward
way, as described in
\citeres{Denner:1993kt,Consoli:1989pc,Consoli:1989fg,Diener:2005me}. 
Starting from the calculation within the on-shell renormalization scheme
with the electromagnetic coupling fixed by $\alpha(0)$, i.e.\
within the $\alpha(0)$ input parameter scheme defined above, 
the corrections associated with the running of $\alpha$ are included by
the resummation of $\Delta\alpha$ via the substitution
\begin{equation}
  \label{eq:alfa0toalfaMZ}
  \alpha(0) \to \alpha(\MZ) = \frac{\alpha(0)}{1-\Delta \alpha(\MZ)}
\end{equation}
in the LO prediction, where $\Delta\alpha$ is defined in
\refeq{eq:dalpha}.
{}For $\Delta \rho$ the leading effects are taken
into account via the replacements
\begin{equation}
  \label{eq:swtoswbar}
  \sw^2 \to \swbar^2 \equiv \sw^2 + \Delta \rho \; \cw^2\;,\qquad \cw^2 \to \cwbar^2 \equiv  1 - \swbar^2 = (1 - \Delta \rho)\; \cw^2 \;.
\end{equation}
This recipe is correct up to $\Ord{\Delta \rho^2}$ and also reproduces
correctly terms of $\Ord{\Delta \alpha(\MZ)\Delta \rho}$
\cite{Consoli:1989pc,Consoli:1989fg} in processes with four light
external fermions. Note that in ${\cal O}(\Delta\rho^2)$ both one- and
two-loop corrections to $\Delta\rho$ become relevant; explicitly we
use the result
\begin{equation}
\Delta\rho = 3 x_\Pt\left[1+\rho^{(2)}\left(\MH^2/\Mt^2\right)x_\Pt\right]
\biggl[1-\frac{2\alpha_{\mathrm{s}}}{9\pi}(\pi^2+3) \biggr],
\qquad
3x_\Pt = \frac{3\sqrt{2}\GF\Mt^2}{16\pi^2} = 
\Delta\rho^{(1)}\Big|^{\GF},
\end{equation}
with the function $\rho^{(2)}$ given in Eq.~(12) of
\citere{Fleischer:1993ub}.
In the following we isolate the genuine two-loop effects induced by
$\Delta\alpha$ and $\Delta\rho$ after properly subtracting the
corresponding one-loop contributions $\Delta\alpha$ and $\Delta\rho^{(1)}$
already contained in
the full \NLO\ electroweak corrections.

The leading one- and two-loop effects of $\Delta\alpha$ and $\Delta\rho$
in the $\alpha(0)$ scheme are included in the LO cross section 
\refeq{eq:locs} upon performing the substitutions 
\refeq{eq:alfa0toalfaMZ} and \refeq{eq:swtoswbar}.
In this context,
the basic ingredients in \refeq{eq:locs} are the products 
$\alpha(0)^2 Q_q^2 Q_l^2$,
$\alpha(0)^2 Q_q Q_l\gqqZ^\si\gllZ^\tau$, and 
$\alpha(0)^2(\gqqZ^\si\gllZ^\tau)^2$ 
of the electroweak couplings defined in \refeq{eq:ffVcoupl}.
In the following we define $\bargffZ^\si$ to result from
$\gffZ^\si$ upon applying $\refeq{eq:swtoswbar}$.
Carrying out the above substitutions and keeping terms up to
two-loop order, the results for these products can be written as
\beqar
\alpha(0)^2\,Q_q^2 Q_l^2 &\;\to\;&
\alpha(\MZ)^2\, Q_q^2 Q_l^2 = 
\alpha(0)^2 \, Q_q^2 Q_l^2 
\left[ 1+2\Delta\alpha(\MZ)+3\Delta\alpha(\MZ)^2 + \dots\right],
\nn\\[.5em]
\alpha(0)^2\,Q_q Q_l\,\gqqZ^\si\gllZ^\tau &\;\to\;&
\alpha(\MZ)^2\,Q_q Q_l\,\bargqqZ^\si\bargllZ^\tau
\nn\\
&=&
\alpha(0)^2 Q_q Q_l\,\biggl\{ \gqqZ^\si\gllZ^\tau 
\left[1+2\Delta\alpha(\MZ)+3\Delta\alpha(\MZ)^2 \right]
\nn\\
&& \qquad\qquad\quad {}
+\Delta\rho \,a_{ql}^{\si\tau} + \Delta\rho^2 \,b_{ql}^{\si\tau}
+2\Delta\alpha(\MZ)\Delta\rho \,a_{ql}^{\si\tau} +\dots \biggr\},
\nn\\[.5em]
\alpha(0)^2\,(\gqqZ^\si\gllZ^\tau)^2 &\;\to\;&
\alpha(\MZ)^2\,(\bargqqZ^\si\bargllZ^\tau)^2 
\nn\\
&=&
\alpha(0)^2 \biggl\{ (\gqqZ^\si\gllZ^\tau)^2
\left[1+2\Delta\alpha(\MZ)+3\Delta\alpha(\MZ)^2 \right]
+2\Delta\rho \,a_{ql}^{\si\tau} \,\gqqZ^\si\gllZ^\tau 
\nn\\
&& \qquad\quad {}
+ \Delta\rho^2 \left( (a_{ql}^{\si\tau})^2
+2b_{ql}^{\si\tau}\,\gqqZ^\si\gllZ^\tau \right)
+4\Delta\alpha(\MZ)\Delta\rho \,a_{ql}^{\si\tau} \,\gqqZ^\si\gllZ^\tau
\nn\\
&& \qquad\quad {}
+\dots \biggr\},
\eeqar
where we have introduced the shorthands
\beq
a_{ql}^{\si\tau} =
\frac{Y_{q^\si} Y_{l^\tau}}{4\cw^2}
-\frac{\cw^2 I^3_{{\scrs\PW},{q^\si}} I^3_{{\scrs\PW},{l^\tau}}}{\sw^4},
\qquad
b_{ql}^{\si\tau} =
\frac{Y_{q^\si} Y_{l^\tau}}{4\cw^2}
+\frac{\cw^4 I^3_{{\scrs\PW},{q^\si}} I^3_{{\scrs\PW},{l^\tau}}}{\sw^6},
\eeq
with $Y_{f^\si}=2(Q_f-I^3_{{\scrs\PW},{f^\si}})$ denoting the weak 
hypercharge of fermion $f^\si$ with chirality $\si=\pm$.
Dropping the LO contribution and subtracting the relevant
one-loop terms, which are proportional to $\Delta\alpha(\MZ)$ and
$\Delta\rho^{(1)}$, the leading two-loop contributions to the coupling
combinations read
\beqar
\alpha^2\,Q_q^2 Q_l^2 \Big|^{\alpha(0)}_{\mathrm{LL}^2} &\;=\;&
3\alpha(0)^2 \, Q_q^2 Q_l^2 \, \Delta\alpha(\MZ)^2,
\nn\\[.5em]
\alpha^2\,Q_q Q_l\,\gqqZ^\si\gllZ^\tau \Big|^{\alpha(0)}_{\mathrm{LL}^2} 
&\;=\;&
\alpha(0)^2 Q_q Q_l\,\biggl\{ 3\gqqZ^\si\gllZ^\tau \,\Delta\alpha(\MZ)^2
+\left(\Delta\rho-\Delta\rho^{(1)}\Big|^{\alpha(0)}\right) \,a_{ql}^{\si\tau} 
\nn\\
&& \qquad\qquad\quad {}
+ \Delta\rho^2 \,b_{ql}^{\si\tau}
+2\Delta\alpha(\MZ)\Delta\rho \,a_{ql}^{\si\tau} \biggr\},
\nn\\[.5em]
\alpha^2\,(\gqqZ^\si\gllZ^\tau)^2 \Big|^{\alpha(0)}_{\mathrm{LL}^2} 
&\;=\;&
\alpha(0)^2 \biggl\{ 3(\gqqZ^\si\gllZ^\tau)^2 \, \Delta\alpha(\MZ)^2
+2\left(\Delta\rho-\Delta\rho^{(1)}\Big|^{\alpha(0)}\right)
\,a_{ql}^{\si\tau} \,\gqqZ^\si\gllZ^\tau 
\nn\\
&& \qquad\quad {}
+ \Delta\rho^2 \left( (a_{ql}^{\si\tau})^2
+2b_{ql}^{\si\tau}\,\gqqZ^\si\gllZ^\tau \right)
+4\Delta\alpha(\MZ)\Delta\rho \,a_{ql}^{\si\tau} \,\gqqZ^\si\gllZ^\tau
\biggr\},
\nn\\
\eeqar
where we have indicated the $\alpha(0)$ input-parameter scheme by
superscripts. In $\Delta\rho^{(1)}$ the superscript means which value of
$\alpha$ is used in its evaluation.

The transition from the $\alpha(0)$ to the $\alpha(\MZ)$ scheme is rather
easy. Since $\alpha(\MZ)$ is already used as LO coupling, only the 
replacement \refeq{eq:swtoswbar} applies, but not \refeq{eq:alfa0toalfaMZ}.
Thus, starting from the formulas in the $\alpha(0)$ scheme given above,
the terms involving $\Delta\alpha(\MZ)$ should be dropped, and we
obtain for the leading two-loop terms
\beqar
\alpha^2\,Q_q^2 Q_l^2 \Big|^{\alpha(\MZ)}_{\mathrm{LL}^2} &\;=\;& 0,
\nn\\[.5em]
\alpha^2\,Q_q Q_l\,\gqqZ^\si\gllZ^\tau \Big|^{\alpha(\MZ)}_{\mathrm{LL}^2} 
&\;=\;&
\alpha(\MZ)^2 Q_q Q_l\,\biggl\{ 
\left(\Delta\rho-\Delta\rho^{(1)}\Big|^{\alpha(\MZ)}\right) \,a_{ql}^{\si\tau} 
+ \Delta\rho^2 \,b_{ql}^{\si\tau} \biggr\},
\nn\\[.5em]
\alpha^2\,(\gqqZ^\si\gllZ^\tau)^2 \Big|^{\alpha(\MZ)}_{\mathrm{LL}^2} 
&\;=\;&
\alpha(\MZ)^2 \biggl\{ 
2\left(\Delta\rho-\Delta\rho^{(1)}\Big|^{\alpha(\MZ)}\right)
\,a_{ql}^{\si\tau} \,\gqqZ^\si\gllZ^\tau 
\nn\\
&& \qquad\qquad {}
+ \Delta\rho^2 \left( (a_{ql}^{\si\tau})^2
+2b_{ql}^{\si\tau}\,\gqqZ^\si\gllZ^\tau \right)
\biggr\},
\eeqar

In the $\GF$-scheme, $\alpha_{\GF}$ effectively involves a factor
$\alpha(\MZ)\sw^2$, so that the basic replacements read
$\alpha_{\GF}\to\alpha_{\GF}\bar\sw^2/\sw^2$ and
$\gffZ^\si\to\bargffZ^\si$.
This procedure leads to the following leading two-loop terms,
\beqar
\alpha^2\,Q_q^2 Q_l^2 \Big|^{\GF}_{\mathrm{LL}^2} &\;=\;& 
\alpha_{\GF}^2 \,Q_q^2 Q_l^2 \,\biggl\{
2\left(\Delta\rho-\Delta\rho^{(1)}\Big|^{\GF}\right) \frac{\cw^2}{\sw^2} 
+ \Delta\rho^2 \,\frac{\cw^4}{\sw^4} \biggr\},
\nn\\[.5em]
\alpha^2\,Q_q Q_l\,\gqqZ^\si\gllZ^\tau \Big|^{\GF}_{\mathrm{LL}^2} 
&\;=\;&
\alpha_{\GF}^2 Q_q Q_l\,\biggl\{ 
\left(\Delta\rho-\Delta\rho^{(1)}\Big|^{\GF}\right) \,
\biggl(a_{ql}^{\si\tau}+\frac{2\cw^2}{\sw^2}\,\gqqZ^\si\gllZ^\tau\biggr) 
\nn\\
&& \qquad\qquad\quad {}
+ \Delta\rho^2 \,\biggl( b_{ql}^{\si\tau} 
+ \frac{2\cw^2}{\sw^2}\,a_{ql}^{\si\tau}
+ \frac{\cw^4}{\sw^4}\,\gqqZ^\si\gllZ^\tau \biggr)
\biggr\},
\nn\\[.5em]
\alpha^2\,(\gqqZ^\si\gllZ^\tau)^2 \Big|^{\GF}_{\mathrm{LL}^2} 
&\;=\;&
\alpha_{\GF}^2 \biggl\{ 
2\left(\Delta\rho-\Delta\rho^{(1)}\Big|^{\GF}\right) \,
\gqqZ^\si\gllZ^\tau \,
\biggl(a_{ql}^{\si\tau}+\frac{\cw^2}{\sw^2}\,\gqqZ^\si\gllZ^\tau\biggr)
\nn\\
&& \qquad\quad {}
+ \Delta\rho^2 \biggl[ (a_{ql}^{\si\tau})^2
+2\gqqZ^\si\gllZ^\tau 
\biggl( b_{ql}^{\si\tau} +\frac{2\cw^2}{\sw^2}\,a_{ql}^{\si\tau} \biggr)
+ \frac{\cw^4}{\sw^4}\,(\gqqZ^\si\gllZ^\tau)^2 \biggr]
\biggr\}.
\nn\\
\eeqar
We recall that in the CC case~\cite{Brensing:2007qm}
the $\GF$-scheme absorbs the full $\Delta\alpha$ and $\Delta\rho$
terms into the LO prediction (at least up to two loops), because
the CC coupling factor $\alpha_{\GF}/\sw^2$ does not receive such
universal corrections. In the present NC case this absorption is
not complete, and only a numerical analysis can assess the size
of the remaining explicit universal two-loop corrections.

\begin{figure}
  \centering
  \includegraphics[width=\textwidth]{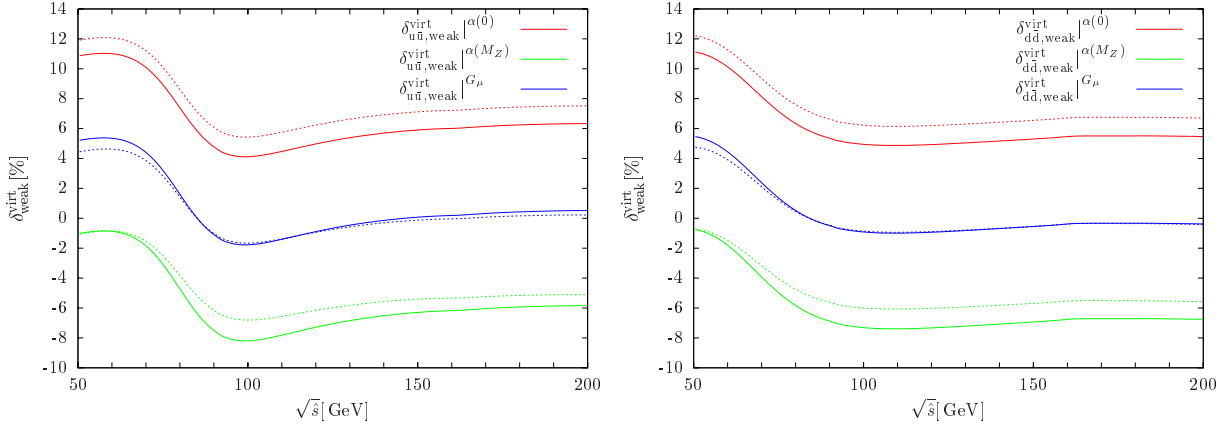}
  \caption{Weak corrections $\delta^\mathrm{virt}_{\Pu\bar\Pu,\weak}$
    and $\delta^\mathrm{virt}_{\Pd\bar\Pd,\weak}$ to the partonic cross
    sections for the different input-parameter schemes, with (dashed
    lines) and without leading higher-order corrections due to $\De\al$
    and $\De\rho$.}
  \label{fig:ewcorr-ips}
\end{figure}

In \reffi{fig:ewcorr-ips} we show the weak corrections
$\delta^\mathrm{virt}_{\Pu\bar\Pu,\weak}$ and
 $\delta^\mathrm{virt}_{\Pd\bar\Pd,\weak}$ to the partonic cross
sections for the different input-parameter schemes, including the
corresponding higher-order corrections due to $\Delta \alpha$ and
$\Delta \rho$. 
It is clearly visible that the impact of the universal two-loop
corrections is largest in the $\alpha(0)$-scheme and 
smallest in the $\GF$-scheme, as expected. 
We, therefore, conclude that the $\GF$-scheme should be the most stable
w.r.t.\ higher-order electroweak effects among the discussed input-parameter
schemes. From the above formulas it is also clear that none of the schemes
is fully optimized to absorb the effects of $\Delta\alpha$ and $\Delta\rho$
into the LO prediction as much as possible. While the $\alpha(\MZ)$
scheme is more suited for photon exchange, where no leading
$\Delta\rho$ corrections arise, the $\GF$-scheme describes Z-exchange
diagrams better, because the generic NC coupling $e/(\sw\cw)$ 
is closer to the weak gauge coupling $e/\sw$ than to $e$. 
In view of the accuracy required for hadron collider physics we do not
see, however, the necessity to switch to a mixed optimized scheme and
take the $\GF$-scheme as default in the following.

\subsubsection{Leading weak corrections in the Sudakov regime}
\label{se:sudakov}

{}For dilepton production at large lepton transverse momenta, 
the parton kinematics is restricted to the Sudakov
regime, characterized by large Mandelstam parameters $\hat{s}$,
$|\hat{t}|$, $|\hat{u}| \gg \MW^2$. The structure of electroweak
corrections beyond ${\cal O}(\alpha)$ in this high-energy regime has
been investigated in detail by several groups in recent years
(see e.g.\ \citeres{Fadin:1999bq, Ciafaloni:2000df, Hori:2000tm,
  Melles:2001dh, Beenakker:2001kf, Denner:2003wi, Jantzen:2005xi,
  Denner:2006jr} and references therein).

As described for example in \citeres{Denner:2003wi,Denner:2006jr}, the
leading electroweak logarithmic corrections, which are enhanced by
large factors $L=\ln(\hat s/\MW^2)$, can be divided into an
SU(2)$\times$U(1)-symmetric part, an electromagnetic part, and a
subleading part induced by the mass difference between $\PW$ and
\PZ~bosons.  The last part does not contribute to corrections
$\propto(\alpha L^2)^n$ and is neglected in the following.  The
leading (Sudakov) logarithms $\propto(\alpha L^2)^n$
of electromagnetic origin cancel between
virtual and real (soft) bremsstrahlung corrections; for the subleading
logarithms such cancellations should strongly depend on the observable
under consideration.  The only source of leading logarithms is,
thus, the symmetric electroweak (sew)
part, which can be characterized by comprising \PW~bosons, \PZ~bosons,
and photons of a common mass $\MW$.  
In the following we consider this type of corrections to the
$q\bar q$ annihilation channels of the light quarks,
i.e.\ $q=\Pu,\Pd,\Pc,\Ps$, which deliver the dominating contribution
to the dilepton cross section. 

The one-loop correction $\de^{(1),\si\tau}_{q\bar q, \mathrm{sew}}$ 
to the squared
amplitude, with chiralities $\si$ and $\tau$ as defined above,
can be obtained by expanding the full result for the
virtual correction $\de^{\virt,\si\tau}_{q\bar q}$ 
(given in the appendix) for large $\hat s$, $|\hat{t}|$,
$|\hat{u}| \gg \MW^2$. The explicit result can be written as
\begin{equation}
\de^{(1),\si\tau}_{q\bar q, \mathrm{sew}} =
\frac{\alpha}{2\pi}\left\{
-L^2 C^{\mathrm{sew},\si\tau}_{1,\mathrm{NC}}
+L C^{\mathrm{ad},\si\tau}_{1,\mathrm{NC}}/C^{\si\tau}_{0,\mathrm{NC}} \right\}
\label{eq:Sud1loop}
\end{equation}
with factors
\beqar
C^{\si\tau}_{0,\mathrm{NC}} &=& e^2\left( \gqqZ^\si\gllZ^\tau+Q_q Q_l \right),
\nn\\
C^{\mathrm{sew},\si\tau}_{1,\mathrm{NC}} &=&
(\gqqZ^\si)^2+Q_q^2+(\gllZ^\tau)^2+Q_l^2+\frac{\de_{\si-}+\de_{\tau-}}{2\sw^2},
\nn\\
C^{\mathrm{ad},\si\tau}_{1,\mathrm{NC}} &=&
\frac{4}{e^2} \, (C^{\si\tau}_{0,\mathrm{NC}})^2 \,
\ln\biggl(\frac{\hat{u}}{\hat{t}}\biggr)
+\frac{e^2}{\sw^4} \, \de_{\si-}\de_{\tau-} \,
\ln\biggl(\frac{-\hat{r}}{\hat{s}}\biggr) \quad
\mbox{with} \quad
\hat r = \left\{\begin{array}{ll}
\hat t \;\mbox{for}\; I^3_{{\scrs\PW},q} I^3_{{\scrs\PW},l}>0, \\
\hat u \;\mbox{for}\; I^3_{{\scrs\PW},q} I^3_{{\scrs\PW},l}<0, 
\end{array}\right.
\hspace{2em}
\eeqar
which have been introduced in Section~8.4.1 of \citere{Denner:2006jr}.
In Eq.~\refeq{eq:Sud1loop} we did not only include the leading Sudakov
logarithms $\propto \alpha L^2$, but also the related
``angular-dependent'' contributions $\propto \alpha L \ln(-\hat{t}/\hat s)$
or $ \alpha L \ln(-\hat{u}/\hat s)$.  Our explicit ${\cal O}(\alpha)$ result
is in agreement with the general results presented in
\citeres{Denner:2003wi,Denner:2006jr}, where the corresponding
corrections are also given at the two-loop level.  These ${\cal
  O}(\alpha^2)$ corrections can be obtained from the ${\cal
  O}(\alpha)$ result by an appropriate exponentiation
\cite{Melles:2001dh}.  For the leading ``sew'' corrections (including
$\alpha^2L^4$, $\alpha^2L^3\ln(-\hat{t}/\hat{s})$, and
$\alpha^2L^3 \ln(-\hat{u}/\hat s)$
terms) this exponentiation simply reads \cite{Denner:2006jr}
\begin{equation}
|\M_{\qqb}|^2 \sim |\M_{\qqb}^{\mathrm{LO}}|^2
\exp\left\{\de^{(1)}_{\qqb,\mathrm{sew}}\right\}
= |\M_{\qqb}^{\mathrm{LO}}|^2
\left( 1+\de^{(1)}_{\qqb,\mathrm{sew}}+\de^{(2)}_{\qqb,\mathrm{sew}}
+\dots\right)
\end{equation}
with
\begin{equation}
\de^{(2)}_{\qqb,\mathrm{sew}} =
\biggl(\frac{\alpha}{2\pi}\biggr)^2\biggl\{
\frac{1}{2}L^4 (C^{\mathrm{sew}}_{1,\mathrm{NC}})^2
-L^3 C^{\mathrm{sew}}_{1,\mathrm{NC}}
C^{\mathrm{ad}}_{1,\mathrm{NC}}/C_{0,\mathrm{NC}} \biggr\},
\label{eq:Sud2loop}
\end{equation}
where we have suppressed the chirality indices $\si,\tau$ in the notation.

Particularly in the case of NC fermion--antifermion
scattering processes it was observed \cite{Jantzen:2005xi} that large
cancellations take place between leading and subleading logarithms.
In view of this uncertainty, we do not include the two-loop
high-energy logarithms in our full predictions. Instead, we evaluate
the leading two-loop part $\de^{(2)}_{\qqb,\mathrm{sew}}$ as a measure of
missing electroweak corrections beyond ${\cal O}(\alpha)$ in the
high-energy Sudakov regime.

Moreover, since the electroweak high-energy logarithmic corrections
are associated with virtual soft and/or collinear weak-boson or photon
exchange, they all have counterparts in real weak-boson or photon
emission processes which can partially cancel (but not completely, see
\citere{Ciafaloni:2000df}) the large negative corrections. To which
extent the cancellation occurs depends on the experimental
possibilities to separate final states with or without weak bosons or
photons. This issue is discussed for example in
\citeres{Ciafaloni:2006qu,Baur:2006sn}. The numerical analysis
presented in \citere{Baur:2006sn} demonstrates the effect of real
weak-boson emission in the distributions in the transverse lepton
momentum $p_{\mathrm{T},\Pl}$ and in the invariant mass
$M_{ll}$ of the lepton pair.
At the LHC, at $M_{ll}=2\TeV$
the electroweak corrections are reduced from about $-11\%$ to $-8\%$
by weak-boson emission. At $p_{\mathrm{T},\Pl}=1\TeV$
the corresponding reduction from about $-10\%$ to $-3\%$ is somewhat larger.
This illustrates the sensitivity
of weak-boson emission effects to the details of experimental
event selection, in particular, how dilepton production is separated
from diboson production.

\subsubsection{Multi-photon final-state radiation}
\label{se:Multi-photon}

The emission of photons collinear to the outgoing charged lepton leads
to corrections that are enhanced by large logarithms of the form
$\alpha\ln(\Ml^2/Q^2)$ with $Q$ denoting a characteristic scale of the
process. The {\it Kinoshita--Lee--Nauenberg} (KLN) theorem 
\cite{Kinoshita:1962ur} guarantees that these logarithms cancel if
photons collinear to the lepton are treated fully
inclusively. However, since we apply a phase-space cut on the momentum
of the outgoing lepton, contributions enhanced by these logarithms
survive if the momentum of the bare lepton is considered, i.e.\ if no
photon recombination is performed. While the concept of a bare lepton
is not realistic for electrons, it is phenomenologically relevant for
muon final states.

The first-order logarithm $\alpha\ln(\Ml^2/Q^2)$ is, of course,
contained in the full ${\cal O}(\alpha)$ correction, so that $Q$ is
unambiguously fixed in this order.  However, it is desirable to
control the logarithmically enhanced corrections beyond
${\cal O}(\alpha)$. This can be done in the so-called structure-function
approach \cite{Kuraev:1985hb}, where these logarithms are derived from
the universal factorization of the related mass singularity.  The
incorporation of the mass-singular logarithms takes the form of a
convolution integral over the LO cross section
$\sigma^{\mathrm{LO}}$,
\begin{equation}
  \sigma_{\llog\FSR} =
  \int \rd\sigma^{\mathrm{LO}}(p_1,p_2;k_1,k_2)
  \int^1_0 \rd z_1 \, \Gamma_{\Pl\Pl}^{\llog}(z_1,Q^2) \,
  \Theta_{\cut}(z_1 k_1) \, 
  \int^1_0 \rd z_2 \, \Gamma_{\Pl\Pl}^{\llog}(z_2,Q^2) \,
  \Theta_{\cut}(z_2 k_2),
\label{eq:llfsr}
\end{equation}
where the step function $\Theta_{\cut}$ is equal to 1 if the event
passes the cut on the rescaled lepton momentum $z_i k_i$ and 0
otherwise.  The variables $z_i$ are the momentum fractions describing the
respective 
lepton energy loss by collinear photon emission. Note that in contrast
to the parton-shower approaches to photon radiation (see e.g.\
\citeres{Placzek:2003zg,CarloniCalame:2005vc}),
the structure-function approach neglects the photon momenta transverse
to the lepton momentum.

{}For the structure function $\Gamma_{\Pl\Pl}^{\llog}(z,Q^2)$ we take
into account terms up to ${\cal O}(\al^3)$ improved by the well-known
exponentiation of the soft-photonic parts \cite{Kuraev:1985hb};
our precise formula can also be found in Eq.~(2.21) of
\citere{Brensing:2007qm}.
Technically, we add the cross section \refeq{eq:llfsr} to the one-loop
result and subtract the LO and one-loop contributions
\beqar
\label{eq:LL1FSR}
  \sigma_{\llog^1\FSR} &=&
  \int \rd\sigma^{\mathrm{LO}}(p_1,p_2;k_1,k_2) \int^1_0 \rd z_1 \, \int^1_0 \rd z_2 \,
  \biggl[\delta(1-z_1)\,\delta(1-z_2)
\nn\\
&& \quad {}
    +\Gamma_{\Pl\Pl}^{\llog,1}(z_1,Q^2)\,\delta(1-z_2)
    +\delta(1-z_1)\,\Gamma_{\Pl\Pl}^{\llog,1}(z_2,Q^2) \biggr] \,
  \Theta_{\cut}(z_1 k_1)\, \Theta_{\cut}(z_2 k_2)
\hspace{2em}
\eeqar
contained in \refeq{eq:llfsr} in order to avoid double counting.
The one-loop contribution to the structure function reads
\newcommand{\betal}{\be_\Pl}
\begin{eqnarray}
\label{eq:LLll}
  \Gamma_{\Pl\Pl}^{\llog,1}(z,Q^2) &=&
  \frac{\betal}{4} \left(\frac{1+z^2}{1-z}\right)_+ 
\end{eqnarray}
with the variable
\begin{equation}
\betal = \frac{2\alpha(0)}{\pi}
\left[\ln\biggl(\frac{Q^2}{\Ml^2}\biggr)-1\right],
\end{equation}
which quantifies the large logarithm.
In this context it should be noted that both the full
photonic one-loop corrections (see \refse{sec:photonic-corrections})
and the multi-photon effects discussed in this section are evaluated
with $\alpha(0)$ as the photonic coupling in the corrections. 
Thus, when subtracting the one-loop part of Eq.~\refeq{eq:LL1FSR} from the
full one-loop result, the logarithmic terms $\propto\alpha(0)\ln m_l$
cancel exactly in all our considered input-parameter schemes.

The uncertainty that is connected with the choice of $Q^2$ enters in
${\cal O}(\alpha^2)$, since all $\Oa$ corrections, including constant
terms, are taken into account.  As default we choose the value
\begin{equation}
Q = \xi\sqrt{\hat s}
\label{eq:FSR_scale}
\end{equation}
with $\xi=1$.  In order to quantify the scale uncertainty, we vary
$\xi$ between $1/3$ and $3$.

\section{Radiative corrections to the partonic cross sections in the  MSSM}
\label{se:mssm}

In this section we examine the effect of corrections to $\Pp\Pp/\Pp\Ppbar
\to \ga/\PZ \to \Pl^-\Pl^+ +X$ within the supersymmetric extension of the
SM. A similar study for the case of \PW-boson production was performed in
\citere{Brensing:2007qm}. Even though Drell--Yan processes do not represent
discovery channels for supersymmetry it is important to study the
influence of SUSY on Drell--Yan processes since they will be used at
the LHC to calibrate detectors, to monitor luminosity, and to extract
information on PDF. Measurements on Drell--Yan processes will also
allow for precision tests of the SM and its extensions through
radiative corrections. If there were large corrections due to SUSY
particles all this information would be biased and therefore not very
useful to extract information about the underlying physics.

As an estimate of the impact of supersymmetric extensions of the SM we
calculate the SUSY corrections to $\Pp\Pp/\Pp\Ppbar\to \ga/\PZ \to \Pl^-\Pl^+ +X$
within the MSSM. As in \citere{Brensing:2007qm} we calculate the full
MSSM corrections and subtract the SM corrections, so that the MSSM
corrections can be added to the SM corrections without double
counting,
\begin{equation}
  \label{eq:kfactsusy}
  \delta_{\qqb,\SUSY} \equiv
  \delta_{\qqb,\MSSM} - \delta_{\qqb,\SM}\left( \MH=M_{\Ph^0}\right) \;.
\end{equation}
Note that we identify the mass $\MH$
of the SM Higgs boson with the mass $M_{\Ph^0}$ of
the lightest Higgs boson $\Ph^0$ of the MSSM for the subtraction of the SM
corrections. 

We divide the SUSY corrections into the SUSY-QCD and the SUSY electroweak
(SUSY-EW) corrections. The SUSY-QCD corrections are due to
corrections to the $q\bar q \ga/Z$ vertices as shown in
\reffi{fig:susyqcd} and the quark wave-function renormalization via
squark--gluino loops. 
\begin{figure}
  \centering
  \input{susyqcd.tex}
  \vspace*{-2em}
  \caption{Example diagram for SUSY-QCD corrections, which are due to
squark ($\tilde q$) and gluino ($\tilde g$) exchange.}
  \label{fig:susyqcd}
\vspace{0em}
  \centering
  \input{pSUSYVertins.tex}
  \vspace*{-2em}
  \caption{Example diagrams for pure SUSY vertex corrections, which involve
squark ($\tilde q$), slepton ($\tilde l$), sneutrino ($\tilde\nu_l$),
chargino ($\tilde\chi$), and neutralino ($\tilde\chi^0$) exchange.}
  \label{fig:psusyvert}
\vspace{1em}
  \centering
  \input{pSUSYBoxins.tex}
  \vspace*{-2em}
  \caption{Example diagrams for pure SUSY box corrections, which involve
squark ($\tilde q$), slepton ($\tilde l$), sneutrino ($\tilde\nu_l$),
chargino ($\tilde\chi$), and neutralino ($\tilde\chi^0$) exchange.}
  \label{fig:psusybox}
\end{figure}
To obtain the SUSY-EW corrections, we proceed as
in \citere{Brensing:2007qm} and calculate the complete electroweak
$\Ord{\al}$ corrections in the MSSM and subtract the SM
corrections. The SUSY-EW corrections can be further divided into 
Two-Higgs-Doublet-Model (THDM) and pure SUSY corrections. The THDM
corrections are due to the extension of the Higgs sector to two Higgs
doublets. It is this part of the corrections where we subtract the SM
corrections, since in the decoupling limit where the mass 
$M_{\PAn}$ of the pseudoscalar Higgs boson $\PAn$ becomes large, both
the THDM sector of the MSSM and the SM Higgs sector coincide 
if we identify the light neutral Higgs boson of
the MSSM with the SM Higgs boson. The pure SUSY corrections consist of
sfermion, neutralino, and chargino loops (see \reffis{fig:psusyvert}
and \ref{fig:psusybox}).

\begin{figure}
  \centering
  \includegraphics{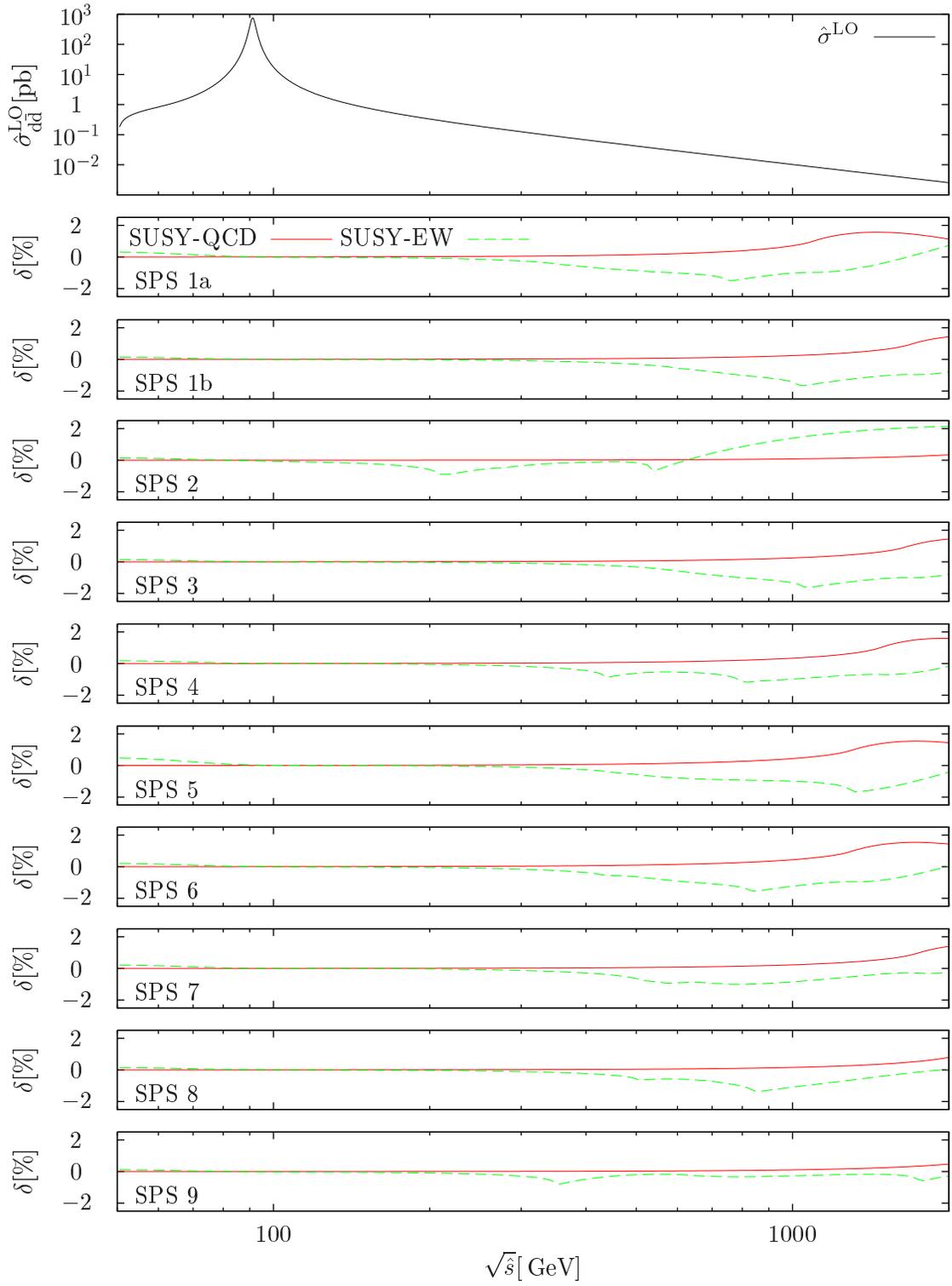}
  \caption{SUSY radiative corrections (MSSM $-$ SM) to the partonic process
    $\Pd\bar\Pd \to \llb$.}
  \label{fig:ddsusycorr}
\end{figure}

{}For the computation of the SUSY corrections we have
again performed two
independent calculations, one using the {\sc FeynArts}/{\sc
  FormCalc}/{\sc LoopTools}~\cite{Hahn:1998yk,Hahn:2000kx}
framework and the other one using {\sc FeynArts} 
and inhouse {\sc Mathematica} routines. The
calculation is done using the on-shell scheme as defined in
\citere{Denner:1993kt}. Since the LO process is a pure SM process the
renormalization of \citere{Denner:1993kt} can be applied without modification.
To treat the resonance at the \PZ-boson
peak we use the LO cross section evaluated in the FS and multiply
with a correction factor,
\begin{equation}
  \label{eq:partsusykfact}
  \hat{\sigma}_{\qqb,\SUSY} = \delta_{\qqb,\SUSY}\Big|_{\GZ=0} \times 
  \hat{\sigma}_{\qqb}^{\mathrm{LO}}\big|_{\mathrm{FS}} \;,
\end{equation}
where the relative SUSY correction $\delta_{\qqb,\SUSY}$, as defined
in \refeq{eq:kfactsusy}, can be evaluated without any special
treatment of the Z-boson resonance, i.e.\ with a zero Z-boson decay width.
We find that for the SPS benchmark scenarios \cite{Allanach:2002nj} 
(see \refapp{app:SPS}) both the SUSY-QCD and the SUSY-EW
corrections stay below $2\%$
for partonic centre-of-mass
energies up to $2 \TeV$. As an example we show in
\reffi{fig:ddsusycorr} the partonic LO cross section and radiative
corrections for $\Pd\bar\Pd$ initial-states for the different MSSM
scenarios.
%
%
%

\section{\boldmath{Numerical results for the cross section 
$\Pp\Pp/\Pp\bar{\Pp}{\to}\ga/\PZ{\to}\Pl^-\Pl^+{+}\X$}}
\label{se:numres}

In this section we describe our numerical setup and discuss the numerical
results for the proton--(anti-)proton cross section $\sigma$ of the
processes $\Pp\Pp/\Pp\Ppbar \to \ga/\PZ \to \Pl^-\Pl^+ + \X$ in the SM
and the MSSM.

\subsection{Input parameters and setup for the SM}
\label{se:SMinput}

The relevant SM input parameters are
\begin{equation}\arraycolsep 2pt
 \begin{array}[b]{lcllcllcl}
\GF & = &1.16637    \times 10^{-5} \GeV^{-2}, \\
\alpha(0) &=& 1/137.03599911    , &
\alpha(\MZ) &=&1/128.93,\quad&
\alpha_{\mathrm{s}}(\MZ) &=&0.1189,\\
M_{\PW,\mathrm{OS}} & = &80.403    \GeV, &
M_{\PZ,\mathrm{OS}} & = &91.1876    \GeV, &
M_\PH & = &115    \GeV, \\
\Gamma_{\PW,\mathrm{OS}} & = &2.141    \GeV, &
\Gamma_{\PZ,\mathrm{OS}} & = &2.4952   \GeV, \\
m_\Pe & = &0.51099892\MeV, &
m_\mu &=& 105.658369    \MeV,\quad &
m_\tau &=&1.77699    \GeV, \\
m_\Pu &=&66    \MeV, &
m_\Pc &=&1.2    \GeV, &
m_\Pt &=&174.2    \GeV, \\
m_\Pd &=&66    \MeV, &
m_\Ps &=&150    \MeV, &
m_\Pb &=&4.6    \GeV, 
 \end{array}
 \label{eq:SMpar}
\end{equation}
which essentially follow \citere{Yao:2006px}.  The masses of the light
quarks are adjusted to reproduce the hadronic contribution to the
photonic vacuum polarization of \citere{Jegerlehner:2001ca}. The CKM
matrix is set to unity. We keep finite light-quark masses in closed
fermion loops, their numerical impact is, however, extremely small in
the $\al(\MZ)$- and $\GF$-schemes. The $\Ord{\al}$-improved
MRST2004qed set of PDF \cite{Martin:2004dh} is used throughout. 
If not stated otherwise, the
QCD and QED factorization scales are identified and set to the
\PZ-boson mass \MZ.

\subsection{Phase-space cuts and event selection}
\label{se:cuts}

{}For the experimental identification of the processes
$\Pp\Pp/\Pp\Ppbar \to \ga/\PZ \to
\llb + X$ we impose the set of phase-space cuts
\begin{equation}
  \label{eq:phs-spc-cuts}
  M_{ll} > 50 \GeV, 
  \qquad p_{\mathrm{T},l^\pm} > 25 \GeV,
  \qquad |y_{l^\pm}| < 2.5,
\end{equation}
where $M_{ll}$ is the invariant mass of the dilepton system,
$p_{\mathrm{T},l^\pm}$ are the transverse momenta and $y_{l^\pm}$
the rapidities of the respective charged leptons. 
The cuts are not collinear safe with
respect to the lepton momenta, so that observables in general receive
corrections that involve large lepton-mass logarithms of the form $\al
\ln(m_l/\MZ)$. This is due to the fact that photons within a small
collinear cone around the momenta of the leptons are not treated
inclusively, \ie the cuts assume perfect isolation of photons from the
leptons. While this is (more or less) achievable for muon final
states, it is not realistic for electrons.  In order to be closer to
the experimental situation for electrons, the following photon
recombination procedure is applied:
\begin{enumerate}
\item Photons with a rapidity $|y_\gamma| > 3$, which are close to
  the beams, are considered part of the proton remnant and are not
  recombined with the (anti-)lepton.
\item If the photon survived the first step, and if the resolution
  $R_{l^\pm\gamma} = \sqrt{(y_{l^\pm}-y_\gamma)^2 +
    \phi_{l^\pm\gamma}^2}$ is smaller than 0.1 (with
  $\phi_{l^\pm\gamma}$ denoting the angles between the (anti-)lepton
  and the photon in the transverse plane), then the photon is
  recombined with the (anti-)lepton, \ie the momenta of the photon
  and of the (anti-)lepton $l^\pm$ are added and associated with the
  momentum of $l^\pm$, and the photon is discarded.
\item Finally, all events are discarded in which the resulting
  momentum of the (anti-)lepton does not pass the cuts given in
  (\ref{eq:phs-spc-cuts}).
\end{enumerate}
The same recombination procedure was also used in 
\citere{Brensing:2007qm} for single-W production.

While the electroweak corrections differ for final-state electrons and
muons without photon recombination, the corrections become universal
in the presence of photon recombination, since the lepton-mass
logarithms cancel in this case, in accordance with the KLN theorem.
Numerical results are presented for photon recombination and for bare
muons.

\subsection{SM predictions for cross sections at the LHC and the Tevatron}

In \reftas{ta:hadcs-lhc-mll} and \ref{ta:hadcs-tev-mll} we present
the integrated LO cross section together with the electroweak and QCD
correction factors $\delta_{ab}$ for the LHC with a centre-of-mass
energy $\sqrt{s} = 14 \TeV$ and for the Tevatron with $\sqrt{s} = 1.96
\TeV$. 
%
\begin{table}
  \begin{center}
      {\renewcommand{\arraystretch}{1.5}                                                                                       \renewcommand{\tabcolsep}{0.1285cm}                                                                                            \begin{tabular}{c|cccccc}		
\multicolumn{7}{c}{$\Pp\Pp\to\Pl^+\Pl^- + \X$                                                                          at $\sqrt{s}=14\TeV$}		
\\  \hline$M_{\Pl \Pl}/\mathrm{GeV}$                                                                                  & 50--$\infty$ & 100--$\infty$ & 200--$\infty$                                                                             & 500--$\infty$ & 1000--$\infty$ & 2000--$\infty$\\                                                                       \hline		
$\sigma_0/\mathrm{pb} $ & $ 738.733(6) $ & $ 32.7236(3) $ & $ 1.48479(1) $ & $ 0.0809420(6) $ & $ 0.00679953(3) $ & $ 0.000303744(1) $ \\
$\sigma_0|_\mathrm{FS/PS}/\mathrm{pb} $ & $ 738.773(6) $ & $ 32.7268(3) $ & $ 1.48492(1) $ & $ 0.0809489(6) $ & $ 0.00680008(3) $ & $ 0.000303767(1) $ \\
\rule{0pt}{5ex}$\delta_{\ga\ga,0}/\mathrm{\%} $ & $ 0.17 $ & $ 1.15 $ & $ 4.30 $ & $ 4.92 $ & $ 5.21 $ & $ 6.17 $ \\
\rule{0pt}{5ex}$\delta^{\mathrm{rec}}_{\Pq\bar{\Pq},\mathrm{phot}}/\mathrm{\%} $ & $ -1.81 $ & $ -4.71 $ & $ -2.92 $ & $ -3.36 $ & $ -4.24 $ & $ -5.66 $ \\
$\delta^{\mu^+\mu^-}_{\Pq\bar{\Pq},\mathrm{phot}}/\mathrm{\%} $ & $ -3.34 $ & $ -8.85 $ & $ -5.72 $ & $ -7.05 $ & $ -9.02 $ & $ -12.08 $ \\
\rule{0pt}{5ex}$\delta^{\mu^+\mu^-}_{\mathrm{multi-}\ga}/\mathrm{\%} $ & $ 0.073_{-0.024}^{+0.027} $ & $ 0.49_{-0.15}^{+0.18} $ & $ 0.17_{-0.05}^{+0.06} $ & $ 0.23_{-0.06}^{+0.07}$ & $ 0.33_{-0.08}^{+0.09} $ & $ 0.54_{-0.12}^{+0.13} $ \\
\rule{0pt}{5ex}$\delta_{\Pq\bar{\Pq},\mathrm{weak}}/\mathrm{\%} $ & $ -0.71 $ & $ -1.02 $ & $ -0.14 $ & $ -2.38 $ & $ -5.87 $ & $ -11.12 $ \\
\rule{0pt}{5ex}$\delta_{\mathrm{h.o.weak}}/\mathrm{\%} $ & $ 0.030 $ & $ 0.012 $ & $ -0.23 $ & $ -0.29 $ & $ -0.31 $ & $ -0.32 $ \\
\rule{0pt}{5ex}$\delta^{(2)}_{\mathrm{Sudakov}}/\mathrm{\%} $ & $ -0.00046 $ & $ -0.0067 $ & $ -0.035 $ & $ 0.23 $ & $ 1.14 $ & $ 3.38 $ \\
\rule{0pt}{5ex}$\delta_{\Pq/\bar{\Pq}\ga,\mathrm{phot}}/\mathrm{\%} $ & $ -0.11 $ & $ -0.21 $ & $ 0.38 $ & $ 1.53 $ & $ 1.91 $ & $ 2.34 $ \\
\rule{0pt}{5ex}$\delta^{\mathrm{rec}}_{\ga\ga,\mathrm{phot}}/\mathrm{\%} $ & $ -0.0060 $ & $ -0.032 $ & $ -0.11 $ & $ -0.14 $ & $ -0.16 $ & $ -0.23 $ \\
$\delta^{\mu^+\mu^-}_{\ga\ga,\mathrm{phot}}/\mathrm{\%} $ & $ -0.011 $ & $ -0.058 $ & $ -0.22 $ & $ -0.30 $ & $ -0.39 $ & $ -0.59 $ \\
\rule{0pt}{5ex}$\delta_{\ga\ga,\mathrm{weak}}/\mathrm{\%} $ & $ 0.000045 $ & $ 0.00056 $ & $ -0.025 $ & $ -0.14 $ & $ -0.31 $ & $ -0.64 $ \\
\rule{0pt}{5ex}$\delta_{\mathrm{QCD}}/\mathrm{\%} $ & $ 4.0(1) $ & $ 13.90(6) $ & $ 26.10(3) $ & $ 21.29(2) $ & $ 8.65(1) $ & $ -11.93(1) $ \\
\end{tabular}}		
 
  \end{center}
  \caption{Integrated LO cross section and relative correction 
    factors at the LHC for different values of the invariant mass 
    cut $M_{ll}$.}  
  \label{ta:hadcs-lhc-mll}
\end{table}
\begin{table}
  \begin{center}
      {\renewcommand{\arraystretch}{1.5}                                                                                       \renewcommand{\tabcolsep}{0.11cm}                                                                                            \begin{tabular}{c|cccccc}		

\multicolumn{7}{c}{$\Pp\Ppbar\to\Pl^+\Pl^- + \X$                                                                       at $\sqrt{s}=1.96\TeV$}		

\\  \hline$M_{\Pl \Pl}/\mathrm{GeV}$                                                                                  & 50--$\infty$ & 100--$\infty$ & 150--$\infty$                                                                             & 200--$\infty$ & 400--$\infty$ & 600--$\infty$\\                                                                         \hline		

$\sigma_0/\mathrm{pb} $ & $ 142.7878(7) $ & $ 6.62280(3) $ & $ 0.824114(3) $ & $ 0.294199(1) $ & $ 0.01775063(5) $ & $ 0.001778465(5) $ \\

$\sigma_0|_\mathrm{FS/PS}/\mathrm{pb} $ & $ 142.7948(7) $ & $ 6.62338(3) $ & $ 0.824183(3) $ & $ 0.294222(1) $ & $ 0.01775188(5) $ & $ 0.001778585(5) $ \\

\rule{0pt}{5ex}$\delta_{\ga\ga,0}/\mathrm{\%} $ & $ 0.15 $ & $ 0.72 $ & $ 1.54 $ & $ 1.44 $ & $ 0.83 $ & $ 0.57 $ \\

\rule{0pt}{5ex}$\delta^{\mathrm{rec}}_{\Pq\bar{\Pq},\mathrm{phot}}/\mathrm{\%} $ & $ -1.85 $ & $ -4.87 $ & $ -3.65 $ & $ -3.83 $ & $ -5.16 $ & $ -6.56 $ \\

$\delta^{\mu^+\mu^-}_{\Pq\bar{\Pq},\mathrm{phot}}/\mathrm{\%} $ & $ -3.44 $ & $ -8.93 $ & $ -6.46 $ & $ -6.86 $ & $ -9.56 $ & $ -12.42 $ \\

\rule{0pt}{5ex}$\delta^{\mu^+\mu^-}_{\mathrm{multi-}\ga}/\mathrm{\%} $ & $ 0.082_{-0.026}^{+0.032} $ & $ 0.48^{+0.18}_{-0.15} $ & $ 0.19_{-0.06}^{+0.07} $ & $ 0.20_{-0.06}^{+0.07} $ & $ 0.34_{-0.09}^{+0.10} $ & $ 0.55_{-0.14}^{+0.15} $ \\

\rule{0pt}{5ex}$\delta_{\Pq\bar{\Pq},\mathrm{weak}}/\mathrm{\%} $ & $ -0.70 $ & $ -1.01 $ & $ 0.12 $ & $ 0.15 $ & $ -1.25 $ & $ -2.60 $ \\

\rule{0pt}{5ex}$\delta_{\mathrm{h.o.weak}}/\mathrm{\%} $ & $ 0.036 $ & $ -0.00094 $ & $ -0.23 $ & $ -0.29 $ & $ -0.35 $ & $ -0.36 $ \\

\rule{0pt}{5ex}$\delta^{(2)}_{\mathrm{Sudakov}}/\mathrm{\%} $ & $ -0.00014 $ & $ -0.00044 $ & $ 0.012 $ & $ 0.047 $ & $ 0.35 $ & $ 0.78 $ \\

\rule{0pt}{5ex}$\delta_{\Pq/\bar{\Pq}\ga,\mathrm{phot}}/\mathrm{\%} $ & $ -0.070 $ & $ -0.14 $ & $ -0.16 $ & $ -0.063 $ & $ 0.090 $ & $ 0.15 $ \\

\rule{0pt}{5ex}$\delta^{\mathrm{rec}}_{\ga\ga,\mathrm{phot}}/\mathrm{\%} $ & $ -0.0059 $ & $ -0.024 $ & $ -0.054 $ & $ -0.052 $ & $ -0.035 $ & $ -0.029 $ \\

$\delta^{\mu^+\mu^-}_{\ga\ga,\mathrm{phot}}/\mathrm{\%} $ & $ -0.010 $ & $ -0.043 $ & $ -0.098 $ & $ -0.098 $ & $ -0.072 $ & $ -0.061 $ \\

\rule{0pt}{5ex}$\delta_{\ga\ga,\mathrm{weak}}/\mathrm{\%} $ & $ 0.000056 $ & $ 0.00081 $ & $ 0.0023 $ & $ -0.0038 $ & $ -0.012 $ & $ -0.014 $ \\

\rule{0pt}{5ex}$\delta_{\mathrm{QCD}}/\mathrm{\%} $ & $ 14.19(7) $ & $ 18.07(4) $ & $ 19.15(1) $ & $ 17.72(1) $ & $ 9.47(1) $ & $ 1.48(1) $ \\

\end{tabular}}

  \end{center}
  \caption{Integrated LO cross section and relative correction factors 
    at the Tevatron for different values of the invariant mass cut $M_{ll}$.}  
  \label{ta:hadcs-tev-mll}
\end{table}
The subscripts $a,\,b$ of the correction factors $\delta_{ab}$
denote the various partonic initial states for $\Pp\Pp/\Pp\Ppbar$
collisions. The results are given for six different ranges of the
dilepton invariant mass $M_{ll}$.

By definition, our LO cross section $\sigma^{\LO}$ includes only
contributions from $\Pq\bar\Pq$ initial states, i.e.\ we consistently
treat all effects from photons in the initial state as corrections.
To show the effect of the different treatments of the \PZ-boson
resonance (see \refse{sec:weak-corrections}) we give results for the
LO cross section in the FS/PS schemes $\sigma^{\LO}|_{\mathrm{FS/PS}}$, which
differ from the CMS only in the sub-permille range
($<0.01\%$).  The $\ga\ga$-induced contribution to the LO cross
section is given separately by the factor $\delta_{\ga\ga,0}$. Apart
from the suppression by the photon PDF, the partonic process $\ga\ga
\to \llb$ does not involve a \PZ-boson resonance and therefore is
strongly suppressed for low invariant mass $M_{ll}$. However, at the
LHC for higher $M_{ll}$ the $\ga\ga$-induced contribution reaches up
to $5{-}6\% $ in our default setup.  The $\Ord{\al}$ corrections
$\delta_{\ga\ga,\phot}$ and $\delta_{\ga\ga,\weak}$ have very small
effect on the integrated cross section at both the LHC and the
Tevatron.  In \refse{se:photonIS} we shall pay particular attention to
the question how an enhancement of effects of incoming photons may be
achieved, a question that is interesting for a possible empirical fit
of the photon PDF.

{}For the photonic corrections we give results for bare muons
($\delta^{\mu^+\mu^-}_{\Pq\bar{\Pq},\mathrm{phot}}$) and with the
recombination procedure described in the previous section
($\delta^{\mathrm{rec}}_{\Pq\bar{\Pq},\mathrm{phot}}$), where large
logarithms $\propto \alpha \ln\left(\frac{\Ml}{\MZ}\right)$ cancel, so
that the resulting corrections are smaller. The effect of
higher-order final-state bremsstrahlung beyond $\Ord{\al}$, as
described in Section \ref{se:Multi-photon}, is small for the integrated  
cross section, as $\delta^{\mu^+\mu^-}_{\mathrm{multi-}\ga}$ never
exceeds the $0.1\%$ level. However, as discussed below they become
relevant for the invariant-mass distribution around the resonance.
The correction $\delta^{\mu^+\mu^-}_{\mathrm{multi-}\ga}$
is given for the central scale choice $Q=
\sqrt{\hat s}$ with an uncertainty estimate obtained from varying the
scale $Q$ between $Q=3 \sqrt{\hat s}$ (upper number) and
$Q=\sqrt{\hat s}/3$ (lower number).
Although the
$\Pq/\bar{\Pq}\ga$-induced photonic processes can be considered as
being part of the $\Ord{\al}$ photonic corrections to the
$\Pq\bar{\Pq}$-induced LO process, we do not include them in 
$\delta_{\Pq\bar{\Pq},\mathrm{phot}}$, but give them separately by
$\delta_{\Pq/\bar{\Pq}\ga,\mathrm{phot}}$. They are small for all
considered $M_{ll}$ ranges in our default setup, as expected from the
suppression by a factor $\alpha$ and by the photon PDF.

Our results on the weak correction are given by
$\delta_{\Pq\bar{\Pq},\mathrm{weak}}$. For low $M_{ll}$ the
corrections to the integrated cross section are of the order of a
per cent. For the LHC at high invariant mass the weak corrections are
enhanced due to large Sudakov logarithms, eventually getting of the
same order as the photonic and QCD corrections. The smallness of the
higher-order weak effects $\delta_{\mathrm{h.o.weak}}$ and the leading
two-loop Sudakov logarithms $\delta^{(2)}_{\mathrm{Sudakov}}$, as
described in \refses{se:ips} and \ref{se:sudakov}, points towards
the stability of the results concerning higher orders in $\al$,
especially in the resonance region. 
{}Following the attitude of \citere{Brensing:2007qm} we consider the
size of $\delta^{(2)}_{\mathrm{Sudakov}}$ as a measure for the
missing EW higher-order effects beyond NLO. For the LHC this
estimate indicates a corresponding uncertainty at the level of $1{-}3\%$ for
invariant masses in the range of $1{-}2\TeV$. For the Tevatron
$\delta^{(2)}_{\mathrm{Sudakov}}$ does not even reach the per-cent level
up to invariant masses of $\sim600\GeV$, which suggests that
EW effects beyond NLO do not significantly contribute to the
theoretical uncertainty for Tevatron measurements.

The NLO QCD corrections $\delta_{\QCD}$ are evaluated for a fixed scale
$\mu_{\mathrm{R}} = \mu_{\mathrm{F}} = \MZ$ and vary strongly depending on the size of the
cut on the dilepton invariant mass. The statistical error is somewhat
larger for the QCD corrections since very large cancellations take place
between the $\Pq\Pqbar$ and the $\Pq g/\Pqbar g$ induced channels.

\subsection{SM predictions for distributions at the LHC}

{}For brevity we restrict our investigation of distributions for the NC 
Drell--Yan process to the
situation at the LHC. Already the results for integrated cross sections
indicate that the relative corrections in the Z~resonance region at the LHC and 
Tevatron are qualitatively very similar.

\begin{figure}
  \centering
  \includegraphics[width=0.99\textwidth,clip]{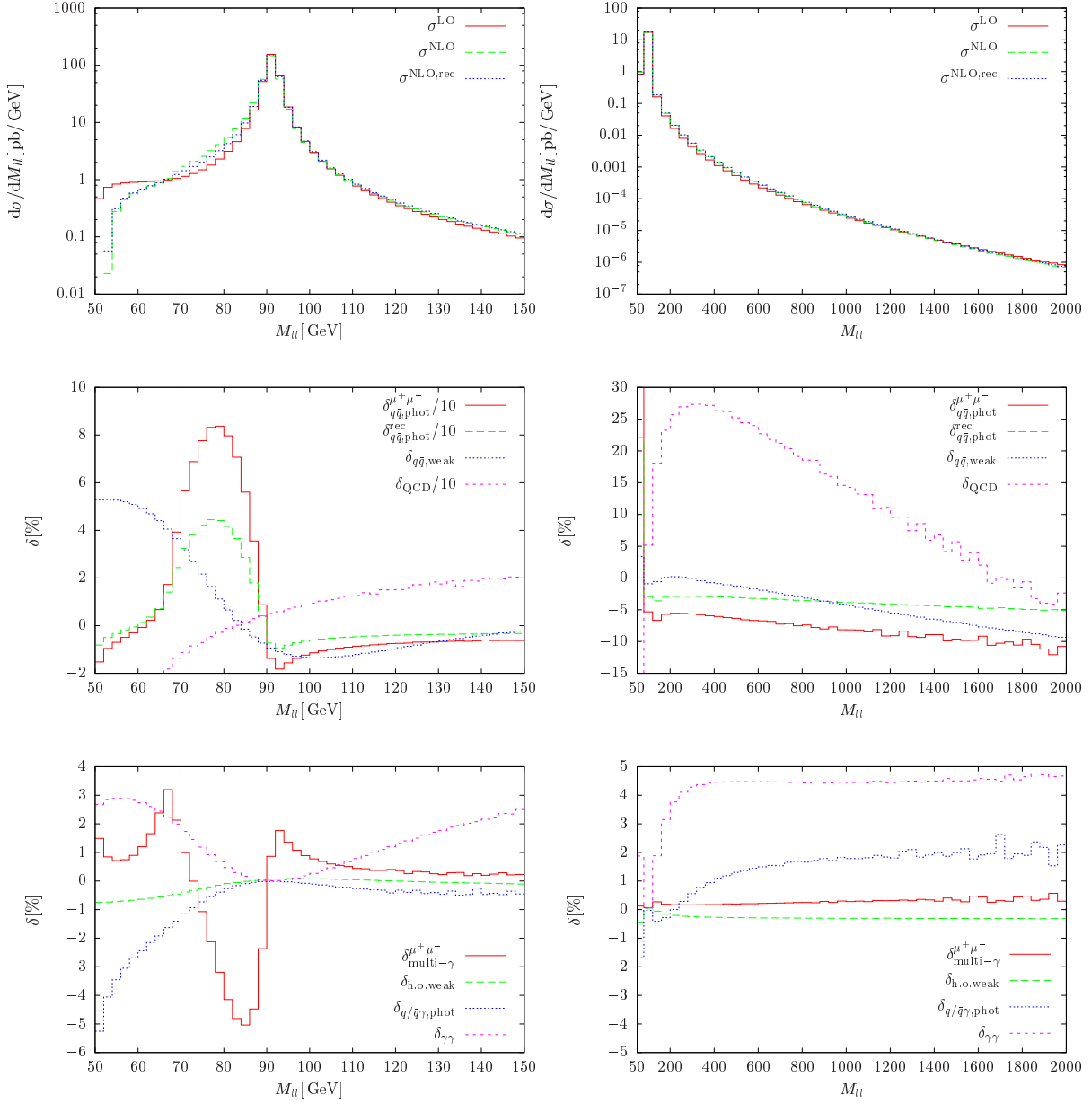}
  \caption{Dilepton invariant-mass distribution and correction
    factors at the LHC in the resonance region (left) and the 
    high-invariant-mass region (right).}
  \label{fig:hist-lhc-1}
\end{figure}
\begin{figure}
  \centering
  \includegraphics[width=0.99\textwidth,clip]{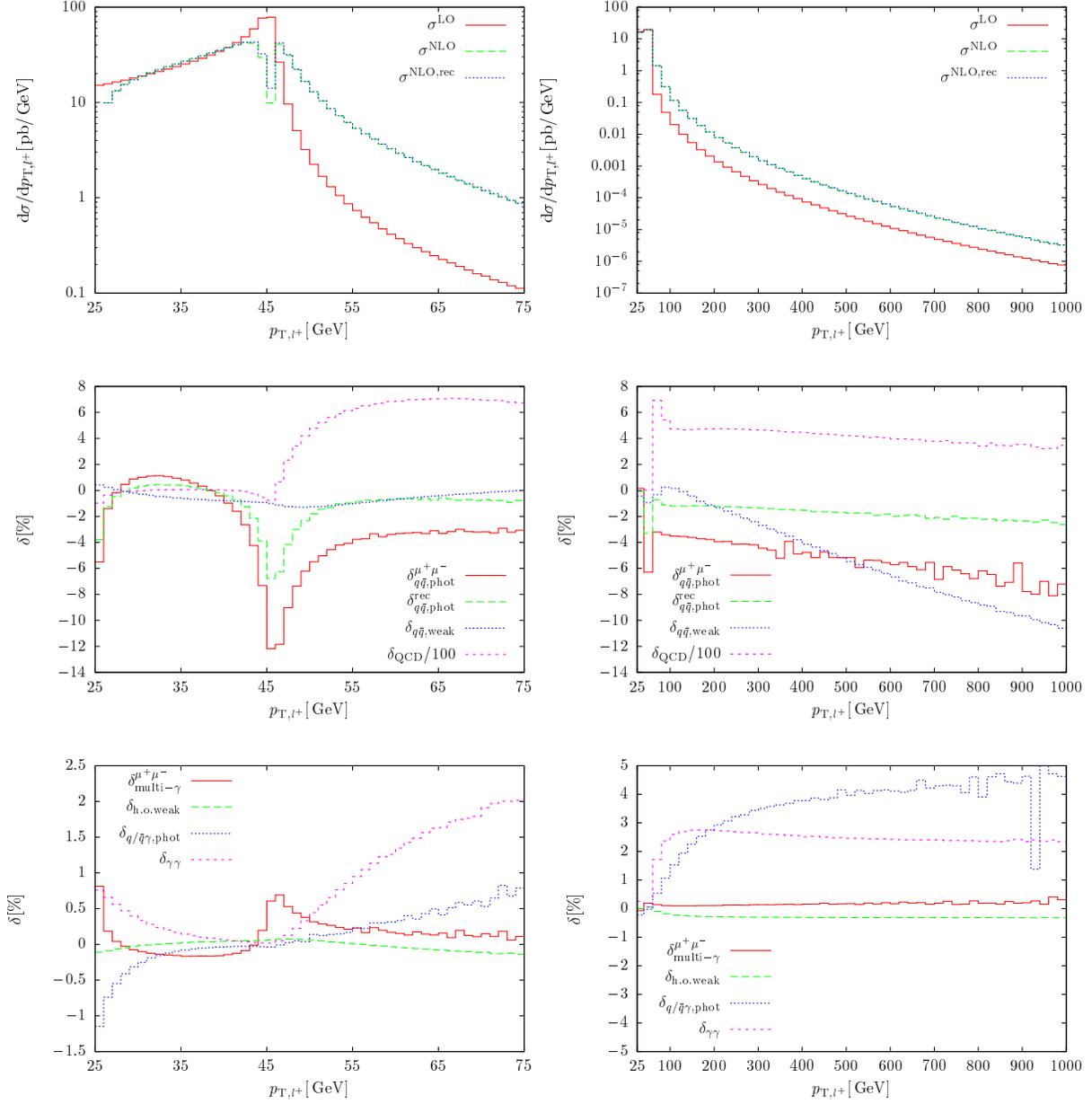}
  \caption{Lepton $p_{\mathrm{T},l^{-}}$-distribution and correction
    factors at the LHC in the resonance region (left) and the 
    high-$p_{\mathrm{T}}$ region (right).}
  \label{fig:hist-lhc-2}
\end{figure}
\begin{figure}
  \centering
  \includegraphics[width=0.99\textwidth,clip]{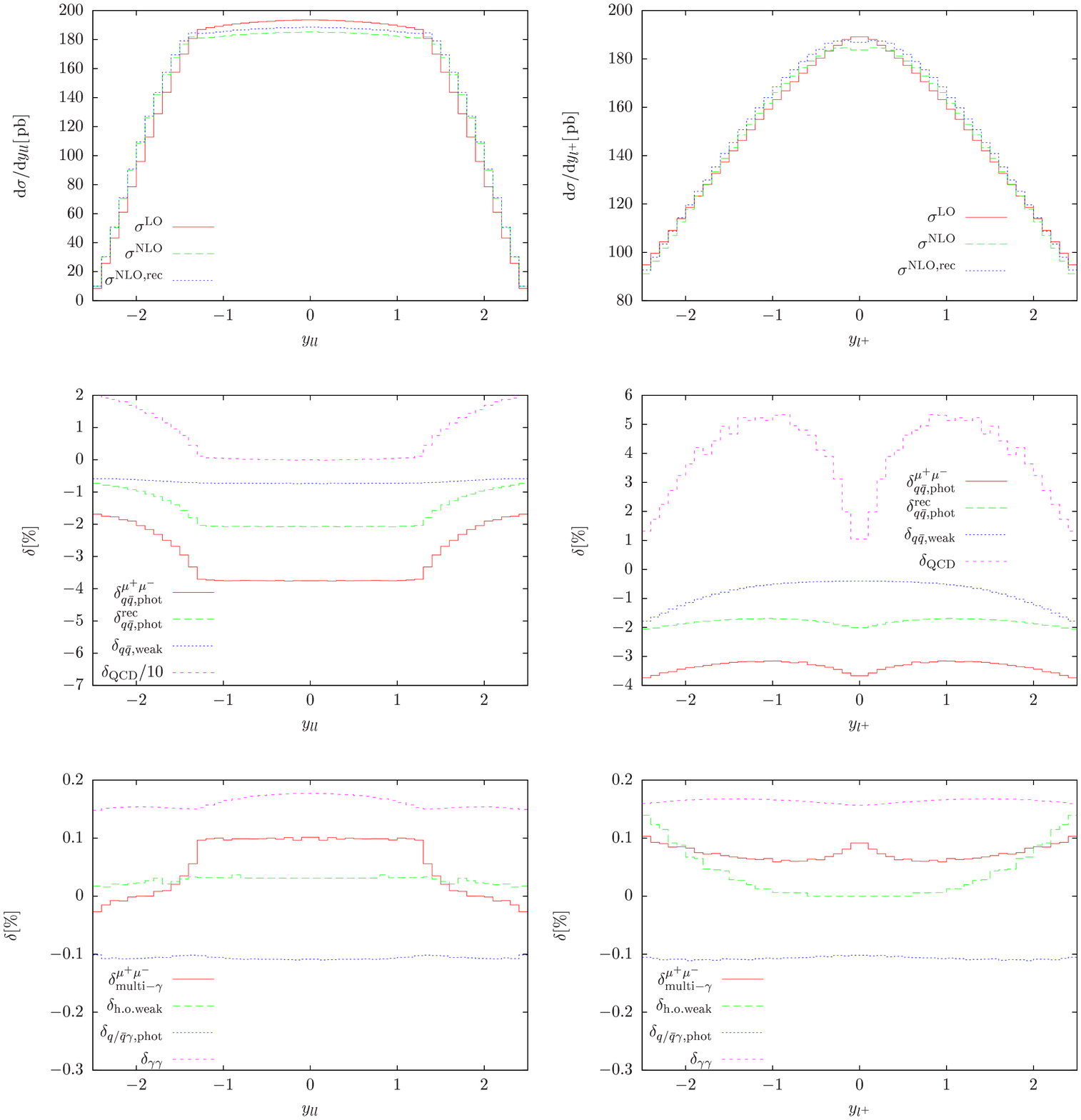}
  \caption{Dilepton (left) and lepton (right) rapidity distributions
    and correction factors at the LHC.}
  \label{fig:hist-lhc-3}
\end{figure}
{}Figures~\ref{fig:hist-lhc-1}, \ref{fig:hist-lhc-2}, and
\ref{fig:hist-lhc-3} show differential distributions and correction
factors at the LHC. The uppermost plots show absolute
distributions, followed below by the relative NLO corrections 
(QCD, photonic, and weak corrections)
normalized to the $\qqb$ LO result.
Note that for some
distributions the correction factors are rescaled. The plots at the bottom
show the higher-order electroweak and photon-induced corrections. The
distributions for $\sigma^{\NLO}$ and $\sigma^{\NLO,\rec}$ are our
best estimates for a $\mu^+\mu^-$ final state and a recombined final
state, respectively, and include all corrections shown in 
\reffis{fig:hist-lhc-1}, \ref{fig:hist-lhc-2}, and \ref{fig:hist-lhc-3}.
 
In Figure \ref{fig:hist-lhc-1} we present the dilepton invariant-mass
distribution $\rd \sigma / \rd M_{ll}$ in the resonance and the
high-invariant-mass region at the LHC. The $M_{ll}$ distribution on
the left shows the well-known large corrections due to the final-state
photon bremsstrahlung which significantly change the shape of the
resonance peak, since events belonging to the \PZ~pole are shifted to
smaller values of the invariant mass $M_{ll}$. Using photon
recombination these corrections are somewhat smaller but still in the
range of $40{-}45 \%$ at maximum. In the high-energy tail of the
distribution, where the leptons are an important background for new
physics searches, the photonic corrections rise in size up to the
order of $-10\% $. At $2 \TeV$ they are of the 
same generic size as the QCD and
the genuinely weak corrections, which are enhanced due to large
Sudakov logarithms at high energies. At the lower end of the
invariant-mass distribution the QCD corrections reach $\approx -170
\%$ which demonstrates that the inclusion of QCD corrections to NLO
only is insufficient there. In order to properly describe this end of the
distribution, which is determined by the phase-space cuts, most
probably QCD resummations are necessary---a task that is beyond the
scope of this paper.  The genuinely weak NLO corrections amount to
some per cent in the resonance region and tend to negative values for
increasing $M_{ll}$, reaching about $-10\%$ at $M_{ll}=2\TeV$. This
effect is mainly due to the well-known EW Sudakov logarithms.  The
multi-photon final-state and photon-induced corrections around the
resonance region are in the range of some per cent and thus comparable
in size to the weak NLO corrections. In particular, the higher-order
multi-photon final-state radiation reduces the effect of
bremsstrahlung at resonance. The effect of universal weak corrections
beyond NLO is very small over the whole $M_{ll}$ range. The
photon-induced corrections are strongly suppressed at the \PZ~pole,
but reach the level of a few per cent away from the pole. As we
observed for the integrated cross section, in the high-energy range
the $\ga\ga$-induced processes contribute with $\sim5\%$ to
$\sigma^\NLO$ in our default setup,
where not only the $\ga\ga$ LO contribution but also the corresponding
photonic and weak corrections are included in the plots.

The lepton $p_{\mathrm{T}}$-distribution $\rd \sigma / \rd
p_{\mathrm{T},l^{-}}$ is shown in \reffi{fig:hist-lhc-2}. The
distribution has the well-known Jacobian peak at $p_{\mathrm{T},l^{-}} \approx
\MZ / 2$. The EW corrections to the
$p_{\mathrm{T},l^{-}}$~distribution are similar in shape to the CC
case~\cite{Baur:1998kt,Zykunov:2001mn,CarloniCalame:2006zq,Dittmaier:2001ay}.
The photonic corrections, which are dominated by final-state radiation,
distort the shape of the peak and are particularly sensitive to the
fact whether photons are recombined with the outgoing leptons or not.
The weak corrections are qualitatively similar to the $M_{ll}$ distribution,
i.e.\ they are at the per-cent level and grow negative for increasing
transverse momenta.
As for the dilepton invariant-mass distribution, close to the
lower cut on $p_{\mathrm{T},l^{-}}$ the QCD corrections become negative 
and very large. In contrast to the $M_{ll}$ distribution, where the NLO QCD
corrections show a moderate size away from the lower end point, 
the NLO QCD corrections to the
$p_{\mathrm{T}}$~distribution are insufficient to describe the spectrum
at all---an effect that is well-known in the literature.
The reason for the dramatic rise of the QCD corrections for
$p_{\mathrm{T},l^{-}}\gsim\MZ/2$ lies in the fact that in LO the 
spectrum receives contributions from resonant Z~bosons only for 
$p_{\mathrm{T},l^{-}}<\MZ/2$, but in NLO resonant Z~bosons also feed
events into the range of larger $p_{\mathrm{T}}$ via the recoil
of the Z~boson against an emitted hard jet in the real corrections.
A proper description of the transition between the two regions
of Z and Z+jet production requires careful QCD 
resummations~\cite{Arnold:1990yk}.
At $p_{\mathrm{T},l^{-}} \approx \MZ / 2$ the NLO QCD
corrections are of the order of $-75\%$ and together with the negative
EW corrections this leads to the dip we observe for
$\sigma^\NLO$, and at high energies grow huge, reaching the
level of several hundred per cent.
The photon-induced
corrections are again small at the peak related to the \PZ-boson
resonance, but reach the level of a few per cent at very low and at high
$p_{\mathrm{T},l^{-}}$. The weak corrections beyond NLO are
suppressed in the whole considered $p_{\mathrm{T}}$ region. 
The photonic corrections beyond
$\Oa$ reduce the size of the NLO photonic corrections for very low 
$p_{\mathrm{T},l^{-}}$ and at the resonance peak, but do not exceed
$1\%$ in size.

{}Figure~\ref{fig:hist-lhc-3} shows the dilepton and the lepton
rapidity distributions $\rd \sigma / \rd y_{ll}$ and $\rd \sigma / \rd
y_{\Pl^-}$, respectively. The dilepton rapidity $y_{ll}$ is defined by
\begin{equation}
  \label{eq:yll-def}
  y_{ll}= \frac{1}{2} \ln \left(\frac{k_{ll}^0+k_{ll}^3}{k_{ll}^0-k_{ll}^3} \right),
 \qquad k_{ll}=k_1+k_2 .
\end{equation}
All NLO QCD and EW corrections to the $y_{ll}$ distribution are at the 
level of few per cent, indicating the perturbative stability of this
observable. This is confirmed by the extremely small size (below
$0.2\%$) of the higher-order EW effects shown in the lower left
plot of \reffi{fig:hist-lhc-3}.
Qualitatively these statements also hold true for the distribution in
the rapidity of the lepton, where the QCD corrections reach the
$5\%$~level.

\subsection{Enhancing effects of incoming photons by cuts}
\label{se:photonIS}

\begin{sloppypar}
In this section we study how the effect of photons in the
initial state can be further enhanced by choosing different
phase-space cuts. If the impact of incoming photons can be
significantly extracted, dilepton production lends itself as
a possible candidate for fitting the photon PDF.
In this discussion, however, it is essential 
to include also QCD and photonic corrections, which are potentially
large and especially sensitive to kinematical cuts.
We consider the following three ``$\ga\ga$ scenarios'', which are
defined by cuts in addition to our default setup, in order to enhance
the effect of incoming photons: 
\begin{itemize}
\item[(a)]  $p_{\mathrm{T},l^\pm} < M_{ll}/4$. \\
This cut is motivated by the consideration that $\ga\ga$ fusion
involves $t$- and $u$-channel diagrams, while the $\qqb$ annihilation
proceeds via $s$-channel diagrams only, i.e.\ $\ga\ga$ fusion prefers
a small value of $\sin\theta^*$, with
$\theta^*$ denoting the scattering angle of the lepton in the partonic 
centre-of-mass frame,
while $\qqb$ annihilation favours intermediate angles.
In LO we have $M_{ll}=\sqrt{\hat s}$ and
$p_{\mathrm{T},l^\pm} = \frac{1}{2}\sqrt{\hat s}\sin\theta^*
= \frac{1}{2}M_{ll}\sin\theta^*$, i.e.\ the above cut translates into
$\sin\theta^* < \frac{1}{2}$.
\item[(b)]  $p_{\mathrm{T},l^\pm} < 50 \GeV$. \\
{}Following the same considerations this cut translates into the condition
$\sin\theta^* < 100 \GeV/M_{ll}$ for the LO contribution.
{}For increasing invariant masses $M_{ll}$ only smaller and smaller
scattering angles are included.
\item[(c)]  $p_{\mathrm{T},l^\pm} < 100 \GeV$. \\
This case is similar to (b), but with the more relaxed LO condition
$\sin\theta^* < 200 \GeV/M_{ll}$.
\end{itemize}
\end{sloppypar}

\begin{figure}
$\ga\ga$ scenario (a): \\[.5em]
  \includegraphics[width=0.99\textwidth,clip]{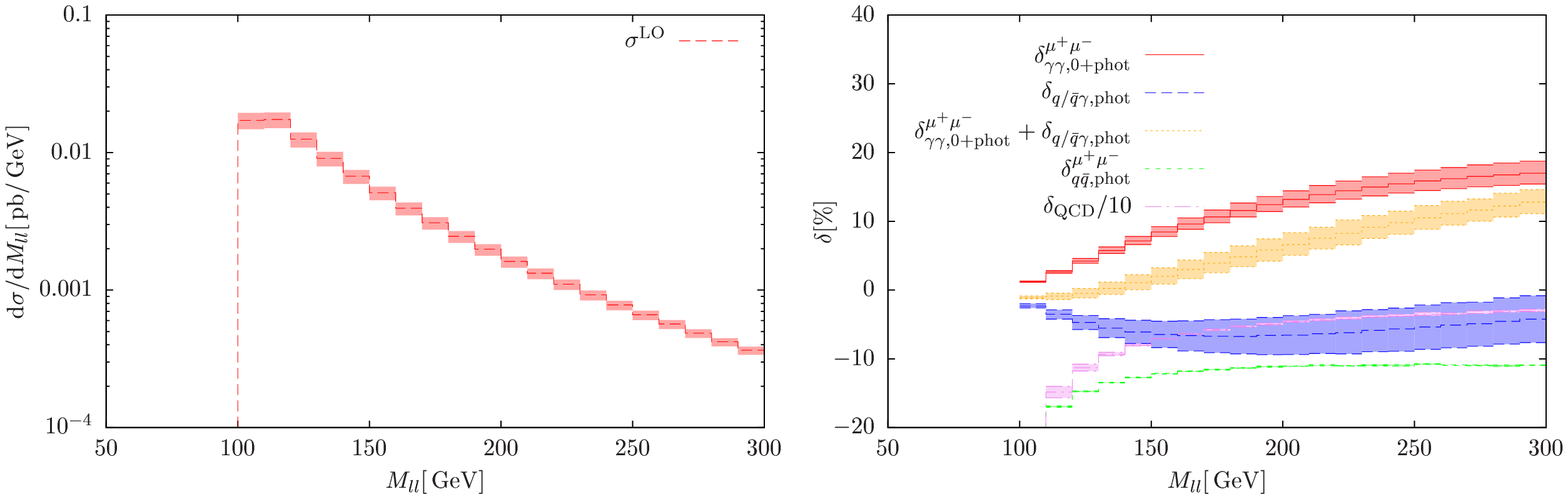}
\\[.5em]
$\ga\ga$ scenario (b): \\[.5em]
  \includegraphics[width=0.99\textwidth,clip]{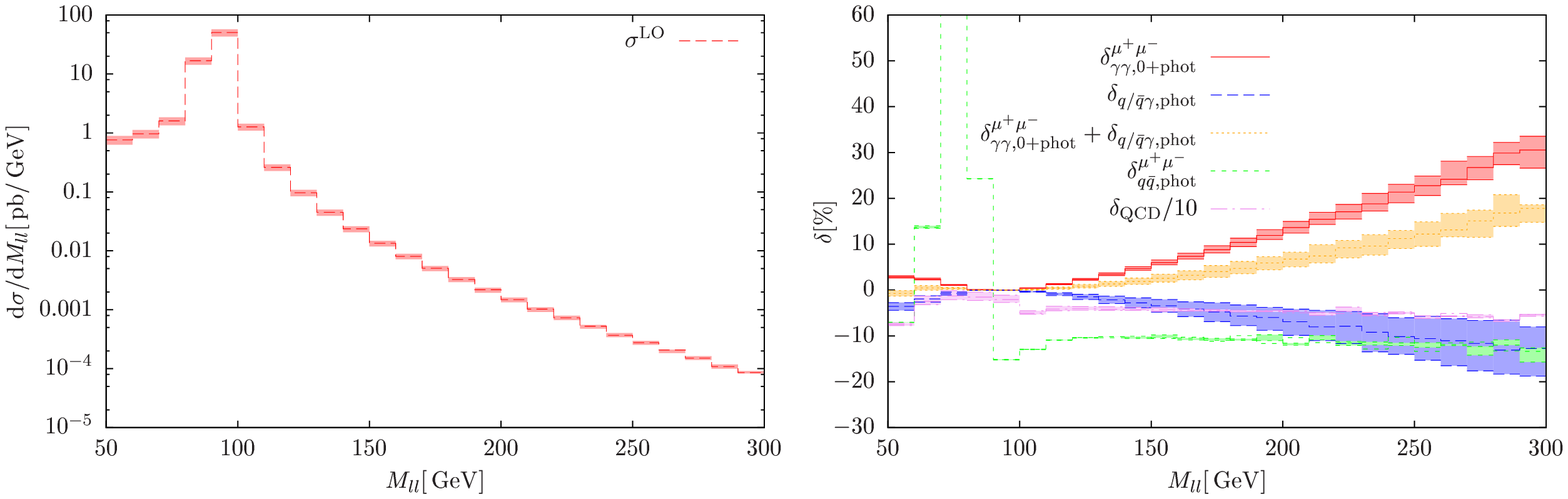}
\\[.5em]
$\ga\ga$ scenario (c): \\[.5em]
  \includegraphics[width=0.99\textwidth,clip]{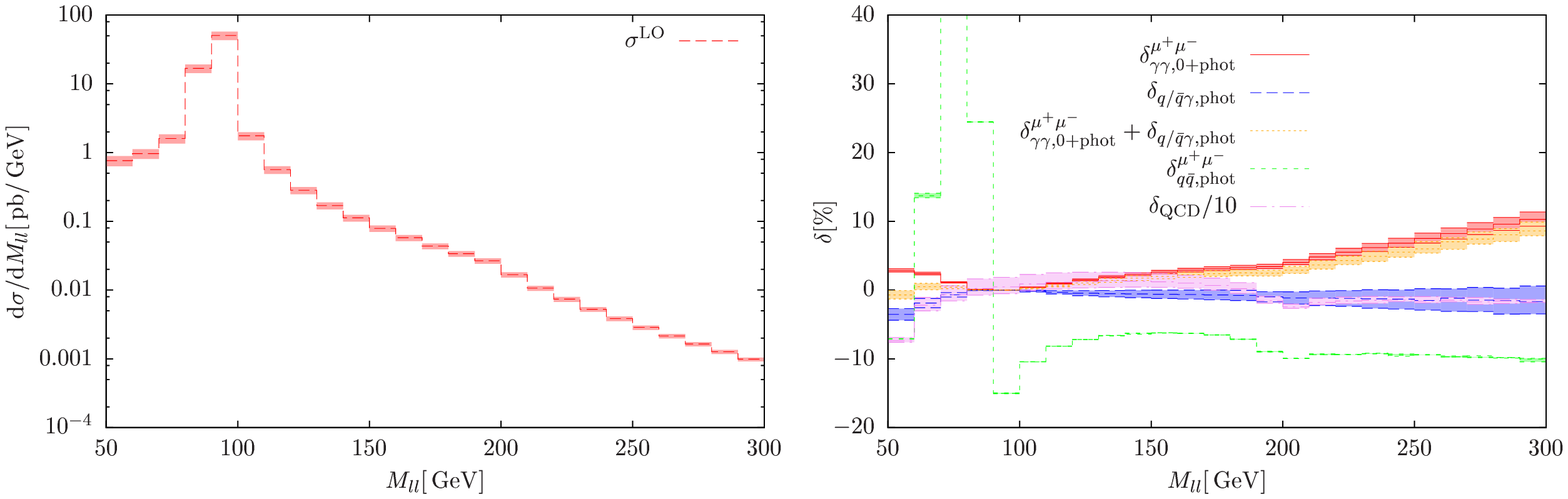}
  \caption{Dilepton invariant-mass distribution and correction
    factors at the LHC for the various $\ga\ga$ scenarios (a), (b), and (c),
    as described in the text. Within the bands the scale
    $\mu=\mu_{\mathrm{F}}=\mu_{\mathrm{R}}$ is varied in the range 
    $\MZ/2<\mu<2\MZ$.}
  \label{fig:gaga}
\end{figure}

In \refta{ta:gaga} we present
the integrated LO cross section and the photon-induced processes as
well as the photonic and QCD corrections for the three scenarios.
\begin{table}
$\ga\ga$ scenario (a): \\[-.5em]
\centerline{{\renewcommand{\arraystretch}{1.3}\renewcommand{\tabcolsep}{0.2cm} \begin{tabular*}{0.95\textwidth}{@{\extracolsep{\fill}}c|*{5}{c}}
\multicolumn{6}{c}{$\Pp\Pp\to\Pl^+\Pl^- + \X$ at $\sqrt{s}=14\TeV$}\\ \hline$M_{\Pl \Pl}/\mathrm{GeV}$                                       &  50--$\infty$ &   100--$\infty$ & 150--$\infty$ & 200--$\infty$ & 250--$\infty$ \\ \hline
$\sigma^{\mathrm{LO}}/\mathrm{pb} $  & \multicolumn{2}{c}{ $ 0.91070(6) $ } &  $ 0.28175(1) $  &  $ 0.116082(5) $  &  $ 0.058598(2) $ \\ 
$\delta_{\ga\ga,0}/\mathrm{\%} $  & \multicolumn{2}{c}{ $ 6.90 $ } &  $ 13.45 $  &  $ 16.93 $  &  $ 18.48 $ \\ 
$\delta^{\mathrm{rec}}_{\ga\ga,\mathrm{phot}}/\mathrm{\%} $  & \multicolumn{2}{c}{ $ -0.39 $ } &  $ -0.63 $  &  $ -0.71 $  &  $ -0.74 $ \\ 
$\delta^{\mu^+\mu^-}_{\ga\ga,\mathrm{phot}}/\mathrm{\%} $  & \multicolumn{2}{c}{ $ -0.65 $ } &  $ -1.06 $  &  $ -1.25 $  &  $ -1.34 $ \\ 
$\delta_{\Pq/\bar{\Pq}\ga,\mathrm{phot}}/\mathrm{\%} $  & \multicolumn{2}{c}{ $ -4.55(1) $ } &  $ -5.79(2) $  &  $ -4.68(2) $  &  $ -3.12(3) $ \\ 
$\delta^{\mathrm{rec}}_{\Pq\bar{\Pq},\mathrm{phot}}/\mathrm{\%} $  & \multicolumn{2}{c}{ $ -10.06(2) $ } &  $ -7.64 $  &  $ -7.28(1) $  &  $ -7.23 $ \\ 
$\delta^{\mu^+\mu^-}_{\Pq\bar{\Pq},\mathrm{phot}}/\mathrm{\%} $  & \multicolumn{2}{c}{ $ -15.28(2) $ } &  $ -11.49 $  &  $ -11.14 $  &  $ -11.28 $ \\ 
$\delta_{\mathrm{QCD}}/\mathrm{\%} $  & \multicolumn{2}{c}{ $ -121.2(5) $ } &  $ -51.0(1) $  &  $ -35.8(1) $  &  $ -29.1(1) $ \\ 
\end{tabular*}}
}
\\[1em]
$\ga\ga$ scenario (b): \\[-.5em]
\centerline{{\renewcommand{\arraystretch}{1.3}\renewcommand{\tabcolsep}{0.2cm} \begin{tabular*}{0.95\textwidth}{@{\extracolsep{\fill}}c|*{5}{c}}
\multicolumn{6}{c}{$\Pp\Pp\to\Pl^+\Pl^- + \X$ at $\sqrt{s}=14\TeV$}\\ \hline$M_{\Pl \Pl}/\mathrm{GeV}$                                       &  50--$\infty$ &   100--$\infty$ & 150--$\infty$ & 200--$\infty$ & 250--$\infty$ \\ \hline
$\sigma^{\mathrm{LO}}/\mathrm{pb} $ & $ 723.28(1) $ & $ 17.2883(5) $ & $ 0.37205(2) $ & $ 0.052388(3) $ & $ 0.011037(1) $ \\
$\delta_{\ga\ga,0}/\mathrm{\%} $ & $ 0.15 $ & $ 1.08 $ & $ 9.98 $ & $ 20.39 $ & $ 31.62 $ \\
$\delta^{\mathrm{rec}}_{\ga\ga,\mathrm{phot}}/\mathrm{\%} $ & $ -0.0059 $ & $ -0.046 $ & $ -0.46 $ & $ -0.99 $ & $ -1.60 $ \\
$\delta^{\mu^+\mu^-}_{\ga\ga,\mathrm{phot}}/\mathrm{\%} $ & $ -0.011 $ & $ -0.081 $ & $ -0.78 $ & $ -1.65 $ & $ -2.65 $ \\
$\delta_{\Pq/\bar{\Pq}\ga,\mathrm{phot}}/\mathrm{\%} $ & $ -0.13 $ & $ -0.66 $ & $ -4.92(2) $ & $ -8.86(7) $ & $ -12.8(1) $ \\
$\delta^{\mathrm{rec}}_{\Pq\bar{\Pq},\mathrm{phot}}/\mathrm{\%} $ & $ -1.91 $ & $ -6.79 $ & $ -6.97 $ & $ -8.09(1) $ & $ -9.14(2) $ \\
$\delta^{\mu^+\mu^-}_{\Pq\bar{\Pq},\mathrm{phot}}/\mathrm{\%} $ & $ -3.46 $ & $ -12.33 $ & $ -10.69 $ & $ -11.75(1) $ & $ -12.89(2) $ \\
$\delta_{\mathrm{QCD}}/\mathrm{\%} $ & $ -20.4(1) $ & $ -47.9(1) $ & $ -45.5(2) $ & $ -51.1(2) $ & $ -58.2(6) $ \\
\end{tabular*}}
}
\\[1em]
$\ga\ga$ scenario (c): \\[-.5em]
\centerline{{\renewcommand{\arraystretch}{1.3}\renewcommand{\tabcolsep}{0.2cm} \begin{tabular*}{0.95\textwidth}{@{\extracolsep{\fill}}c|*{5}{c}}
\multicolumn{6}{c}{$\Pp\Pp\to\Pl^+\Pl^- + \X$ at $\sqrt{s}=14\TeV$}\\ \hline$M_{\Pl \Pl}/\mathrm{GeV}$                                       &  50--$\infty$ &   100--$\infty$ & 150--$\infty$ & 200--$\infty$ & 250--$\infty$ \\ \hline
$\sigma^{\mathrm{LO}}/\mathrm{pb} $ & $ 737.827(6) $ & $ 31.8101(3) $ & $ 2.97905(5) $ & $ 0.57044(1) $ & $ 0.130466(6) $ \\
$\delta_{\ga\ga,0}/\mathrm{\%} $ & $ 0.17 $ & $ 1.11 $ & $ 3.80 $ & $ 6.68 $ & $ 11.21 $ \\
$\delta^{\mathrm{rec}}_{\ga\ga,\mathrm{phot}}/\mathrm{\%} $ & $ -0.0060 $ & $ -0.033 $ & $ -0.12 $ & $ -0.24 $ & $ -0.41 $ \\
$\delta^{\mu^+\mu^-}_{\ga\ga,\mathrm{phot}}/\mathrm{\%} $ & $ -0.011 $ & $ -0.059 $ & $ -0.23 $ & $ -0.45 $ & $ -0.76 $ \\
$\delta_{\Pq/\bar{\Pq}\ga,\mathrm{phot}}/\mathrm{\%} $ & $ -0.11 $ & $ -0.31 $ & $ -0.81 $ & $ -1.27(1) $ & $ -1.50(1) $ \\
$\delta^{\mathrm{rec}}_{\Pq\bar{\Pq},\mathrm{phot}}/\mathrm{\%} $ & $ -1.81 $ & $ -4.85 $ & $ -3.87 $ & $ -5.29 $ & $ -5.95 $ \\
$\delta^{\mu^+\mu^-}_{\Pq\bar{\Pq},\mathrm{phot}}/\mathrm{\%} $ & $ -3.34 $ & $ -9.11 $ & $ -7.30 $ & $ -9.67 $ & $ -10.03 $ \\
$\delta_{\mathrm{QCD}}/\mathrm{\%} $ & $ 3.0(1) $ & $ 9.30(6) $ & $ 1.46(6) $ & $ -19.2(1) $ & $ -17.5(1) $ \\
\end{tabular*}}
}
\caption{LO cross section from $\qqb$ annihilation together with
the relative impact from $\ga\ga$ and $q/\bar q\ga$ initial states,
as well as from photonic and QCD corrections, evaluated in the
three different $\ga\ga$ scenarios (a), (b), and (c) as described in the text.}
\label{ta:gaga}
\end{table}
The corresponding dilepton invariant-mass distributions and
the same types of corrections are shown in \reffi{fig:gaga}, where
the bands around the central lines correspond to
a variation of the renormalization and factorization scale in the
range $\MZ/2<\mu_{\mathrm{R}}=\mu_{\mathrm{F}}<2\MZ$.

The Z-resonance region is clearly dominated by $\qqb$ annihilation and
therefore not suited to access effects from photonic initial states.
Scenario (a) excludes this region by the applied cut completely, since
the lower cut on $p_{\mathrm{T},l^\pm}$ implies $M_{ll}>100\GeV$, so
that the first two columns of numbers in \refta{ta:gaga} are equal.
Phenomenologically these two columns are useless, since
``edge-of-phase-space effects'' render QCD corrections unphysically
large there.  Scenarios (b) and (c) fully include the Z~peak in the
first column of numbers in \refta{ta:gaga}, and a significant
difference between these two scenarios is visible only in the QCD
corrections.  This is also trivially visible in \reffi{fig:gaga}.  The
interesting region for studying photonic initial states is the one of
higher invariant masses $M_{ll}$, viz.\ $M_{ll}\gsim150\GeV$. In
scenarios (a) and (b) already for $M_{ll}\gsim150{-}200\GeV$ the
impact of $\ga\ga$ initial states reaches the order of 10--20\%; in
case (c) it is still 4--7\%.  In scenarios (a) and (b) it should,
however, be realized that the effect of $\ga\ga$ initial states is
systematically reduced by some per cent by the contribution of $q/\bar
q\ga$ initial states. The correlation between $\ga\ga$ and $q/\bar
q\ga$ initial states is also visible in the fact that the sum of the
two shows a somewhat smaller sensitivity to the variation of the
factorization scale than the two individual contributions, as seen in
the distributions of \reffi{fig:gaga}. In view of the smallness of the
integrated cross sections, which are of the order of $0.01\pba$ to
$1\pba$ depending on the $M_{ll}$ range and the chosen scenario, it is
not clear whether a fit of the photon PDF within a reasonable accuracy
will be feasible. This possibility has to be analyzed in a dedicated
study that carefully takes into account all experimental and
theoretical uncertainties, in particular, from higher-order QCD
effects. The QCD corrections, which are given in NLO in
\refta{ta:gaga} and \reffi{fig:gaga}, are of the order of $-30\%$ to
$-50\%$ and thus quite large; the fact that they are negative,
however, helps to further enhance the impact of photonic in initial
states. The photonic corrections to $\qqb$ initial states are about
$-10\%$ and thus go into the same direction as well, but the main
uncertainty will certainly come from QCD corrections.  The
photonic corrections, which are included in
$\delta^{\mu^+\mu^-}_{\ga\ga,0+\phot}$, and the weak (not explicitly
shown here) corrections to the $\ga\ga$ initial states will not play a
role in this context.  In view of the overall size of the
cross sections and the sizes of the $\ga\ga$ contributions and the QCD
corrections, $\ga\ga$ scenario (a) seems the most promising to access
the photon initial states.

\subsection{Comparison to SM results of other groups}
\label{se:comp}

In order to make contact to results previously presented in the 
literature~\cite{Baur:1997wa,Baur:2001ze,CarloniCalame:2007cd,Arbuzov:2007db},
we have added our results on integrated cross sections and NLO EW
corrections to the tuned comparison shown in Table~2 of
\citere{Buttar:2008jx}. In detail, for this comparison
we conformed our input to the setup of ``bare cuts'' described there.
Table~\ref{tab:cmp-SANC-HORACE-ZGRAD}, which shows our results
together with the ones obtained with the 
{\sc HORACE}~\cite{CarloniCalame:2007cd}, 
{\sc SANC}~\cite{Arbuzov:2007db},
and {\sc ZGRAD2}~\cite{Baur:1997wa,Baur:2001ze}
programs, reveals good agreement between the various
calculations.
\begin{table}
  \centering
  %
%
\begin{tabular}{|c|l|l|l|} \hline
\multicolumn{4}{|c|}{\bf LHC, $\Pp \Ppbar \to Z,\gamma \to e^+ e^-$} \\ \hline
             & LO [pb]     & NLO [pb]   & $\delta$ [\%] \\ \hline
{\sc HORACE} & 739.34(3)   & 742.29(4)  & 0.40(1)       \\
{\sc SANC}   & 739.3408(3) & 743.072(7) & 0.504(1)       \\ 
{\sc ZGRAD2} & 737.8(7)    & 743.0(7)   & 0.71(9)       \\
our results   & 739.343(1)   & 742.68(1)  & 0.451(1)      \\
\hline 
\multicolumn{4}{|c|}{\bf LHC, $\Pp \Ppbar \to Z,\gamma \to \mu^+ \mu^-$} \\ \hline
             & LO [pb]     & NLO [pb]   & $\delta$ [\%] \\ \hline
{\sc HORACE} & 739.33(3)   & 762.20(3)  & 3.09(1)       \\
{\sc SANC}   & 739.3355(3) & 762.645(3) & 3.1527(4)     \\ 
{\sc ZGRAD2} &  740(1)     &  764(1)    & 3.2(2)        \\
our results  & 739.343(1)   & 762.21(1)  & 3.092(1)       \\
\hline
\end{tabular}

  \caption{Extension of the tuned comparison shown in
  Table~2 of \citere{Buttar:2008jx} for ``bare cuts''.}
  \label{tab:cmp-SANC-HORACE-ZGRAD}
\end{table}
The remaining differences, which are at the $0.1\%$ level,
are phenomenologically irrelevant and should be due to slightly different
settings in the programs, such as the treatment of small fermion masses.

\begin{figure}
  \vspace{1em}
  \centering
  \includegraphics[width=0.495\textwidth,clip]{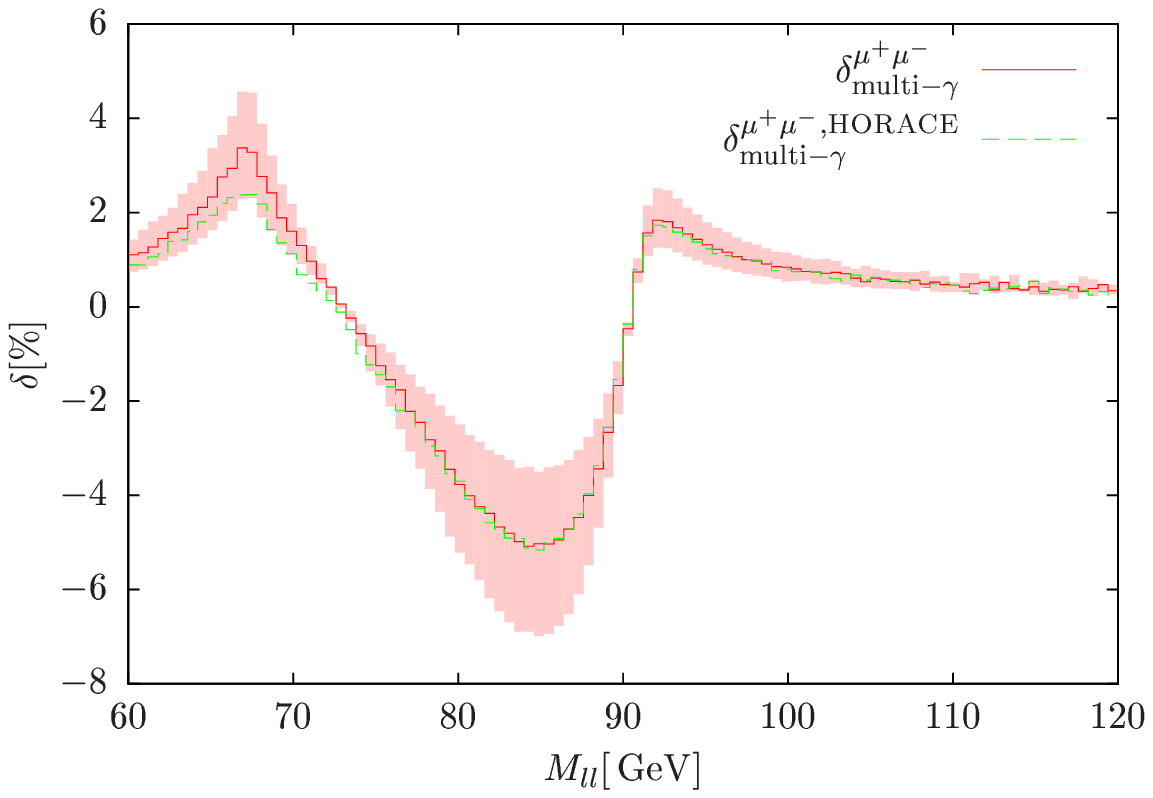}
  \caption{Comparison of the multi-photon corrections for the
    dilepton invariant-mass distribution to results obtained with
    {\sc HORACE}.}
  \label{fig:cmp-HORACE-mult-minv}
  \vspace{1em}
  \includegraphics[width=0.99\textwidth,clip]{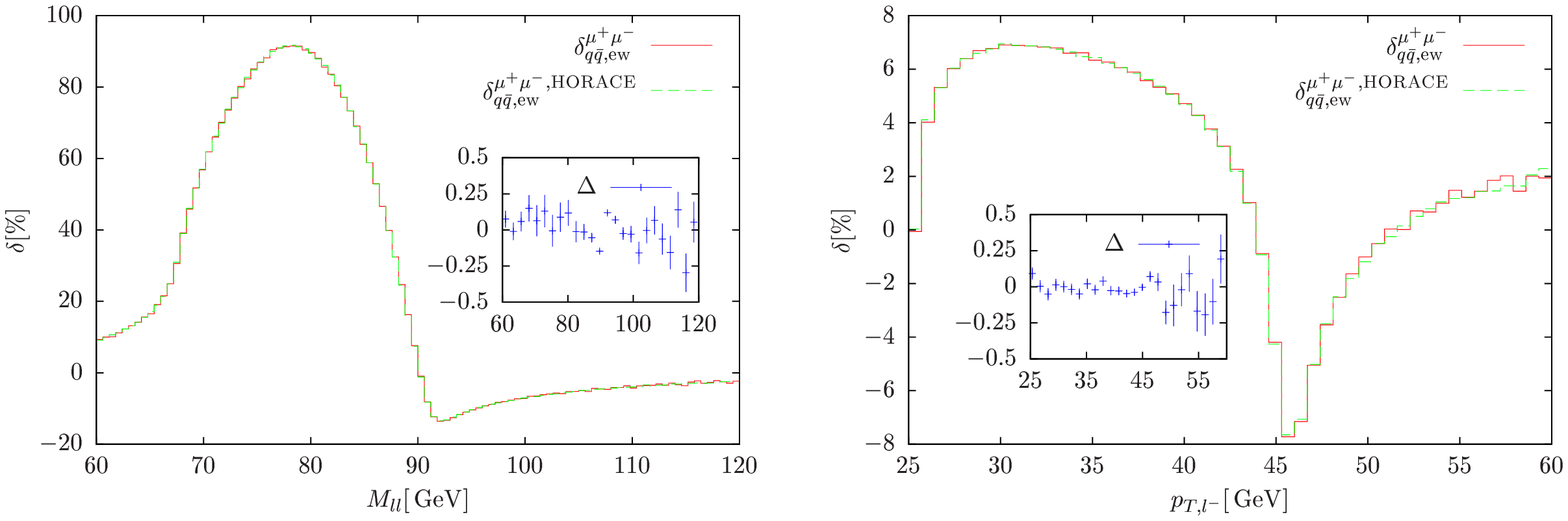}
  \caption{Comparison of the NLO EW corrections for the dilepton
    invariant-mass (left) and the $p_{\mathrm{T},l^-}$ (right)
    distributions to results obtained with {\sc HORACE}, with $\De$
    showing the difference of the two results on the relative correction
    $\de$ in the inset, where the {\sc HORACE} integration error defines
    the error bars.}
  \label{fig:cmp-HORACE-minv-ptl}
  \vspace{1em}
  \includegraphics[width=0.99\textwidth,clip]{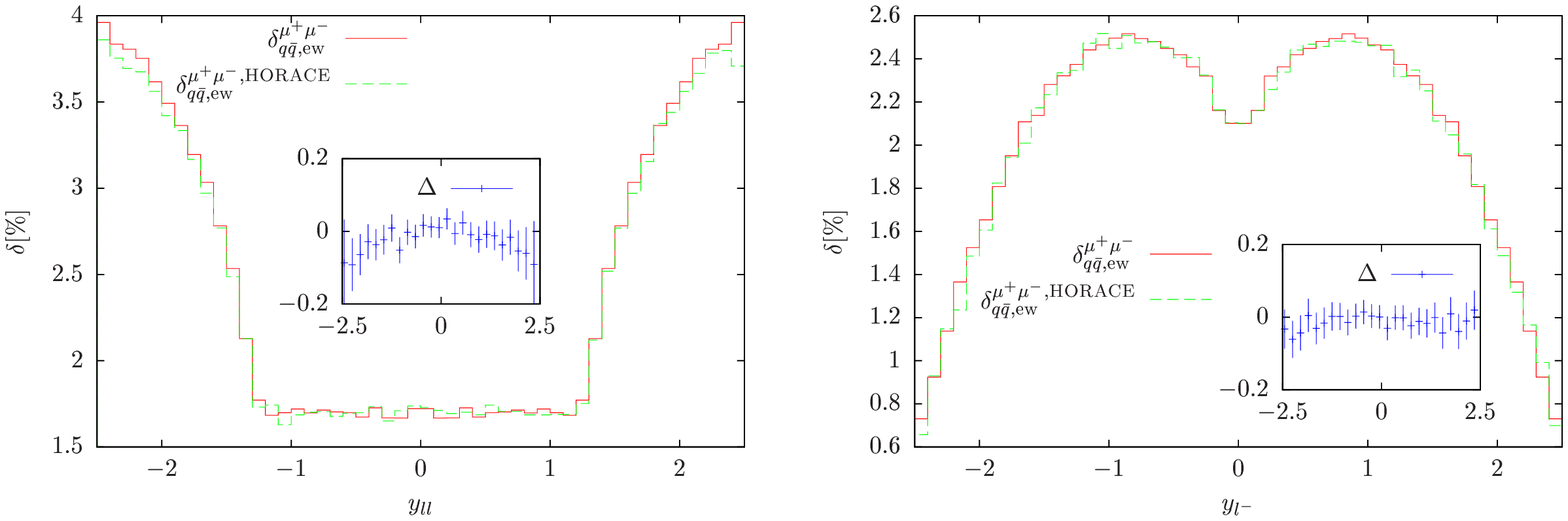}
  \caption{As in \reffi{fig:cmp-HORACE-minv-ptl}, but for the dilepton
    (left) and lepton rapidity (right) distributions.}
  \label{fig:cmp-HORACE-yll-yl}
\end{figure}

In order to demonstrate the agreement of our results on distributions
with previously published results, a comparison of the genuine NLO EW
and multi-photon corrections for various distributions to results
obtained with {\sc HORACE} is shown in
\reffis{fig:cmp-HORACE-mult-minv},~\ref{fig:cmp-HORACE-minv-ptl}, and
\ref{fig:cmp-HORACE-yll-yl}.
In this comparison the {\sc HORACE} results and the complete
numerical setup and input are taken over from \citere{CarloniCalame:2007cd}.
There is very good agreement for the genuine ${\cal O}(\alpha)$ corrections,
as it should be, because these corrections are defined exactly in the same
way in the two calculations. 
Even the multi-photon corrections perfectly agree, although
{\sc HORACE} employs a parton-shower approach for their modelling,
while we use structure functions for collinear multi-photon radiation.
The band defining our result in \reffi{fig:cmp-HORACE-mult-minv}
indicates the effect from varying 
the QED scale \refeq{eq:FSR_scale} by a factor 3 up and down.
A similar comparison in the case of the CC
Drell--Yan process was performed in \citere{Brensing:2007qm}, revealing
agreement between the two approaches at a somewhat lower level of accuracy.

\subsection{Numerical results on supersymmetric corrections in the MSSM}

Our discussion closely follows the one presented in 
\citere{Brensing:2007qm} for the CC case of single-W production.
We choose the SM input parameters and the setup of the calculation 
(input-parameter scheme, PDF, cuts, etc.) as described in 
\refse{se:SMinput} and
study the dependence of the corrections on the SUSY breaking
parameters by considering all the SPS benchmark
scenarios~\cite{Allanach:2002nj}. 
Both for the CC and NC case,
the generic suppression of the genuine SUSY corrections is
insensitive to a specific scenario.
We therefore refrain from further restricting the SPS coverage
by taking into account recent experimental bounds in favour of
a broader scope in the SUSY parameter space.
The SPS points are defined by the low-energy SUSY
breaking parameters which determine the spectrum and the couplings.
{}For the ten benchmark scenarios under consideration in 
\citere{Brensing:2007qm} and in this work, this
input~\cite{SPShomepage} is also tabulated in \refapp{app:SPS}.

Dependent SUSY parameters, such as Higgs, chargino, neutralino, or
sfermion masses, are calculated from the SPS input using tree-level
relations. Since the impact of the fermion masses of the first two
generations is negligible, these masses are set to zero in the
calculation of the corresponding sfermion mass matrices.  Following
this approach the SUSY corrections do not depend on the lepton 
generations in the partonic process $\Pq\bar \Pq\to \Pl^-\Pl^+ $, 
i.e.\ the SUSY corrections presented below are valid both for 
outgoing electrons and muons.

In Table \ref{ta:hadcs-lhc-mll-susy} we list our results for the
SUSY corrections within the MSSM at the LHC. 
\begin{table}
  \begin{center}
      {\renewcommand{\arraystretch}{1.5}                                                                                       \renewcommand{\tabcolsep}{0.175cm}                                                                                            \begin{tabular}{cc|cccccc}		
\multicolumn{8}{c}{$\Pp\Pp\to\Pl^+\Pl^- + \X$                                                                          at $\sqrt{s}=14\TeV$}		
\\  \hline \multicolumn{2}{c}{$M_{\Pl \Pl}/\mathrm{GeV}$} & 50--$\infty$ & 100--$\infty$ & 200--$\infty$ & 500--$\infty$ & 1000--$\infty$ & 2000--$\infty$\\  \hline		
SPS1a & $\delta_{\mathrm{SUSY}-\mathrm{EW}}/\%	 $ & $ 0.0094(3)  $ & $ -0.0041(1)  $ & $ -0.053  $ & $ -0.43  $ & $ -0.33  $ & $ 0.73	$ \\	   
SPS1a & $\delta_{\mathrm{SUSY}-\mathrm{QCD}}/\% 	 $ & $ 0.0060  $ & $ 0.012  $ & $ 0.062  $ & $ 0.34  $ & $ 1.19  $ & $ 0.61	$ \\ \hline
SPS1b & $\delta_{\mathrm{SUSY}-\mathrm{EW}}/\%	 $ & $ 0.0076(1)  $ & $ 0.0021  $ & $ -0.018  $ & $ -0.31  $ & $ -0.67  $ & $ -0.20	$ \\	   
SPS1b & $\delta_{\mathrm{SUSY}-\mathrm{QCD}}/\%	 $ & $ 0.0025  $ & $ 0.0049  $ & $ 0.025  $ & $ 0.13  $ & $ 0.48  $ & $ 1.38	$ \\ \hline
SPS2  & $\delta_{\mathrm{SUSY}-\mathrm{EW}}/\%	 $ & $ -0.046  $ & $ -0.12  $ & $ -0.40  $ & $ 0.30  $ & $ 1.60  $ & $ 1.94	$ \\	   
SPS2  & $\delta_{\mathrm{SUSY}-\mathrm{QCD}}/\%	 $ & $ 0.00093  $ & $ 0.0018  $ & $ 0.0091  $ & $ 0.045  $ & $ 0.15  $ & $ 0.58	$ \\ \hline
SPS3  & $\delta_{\mathrm{SUSY}-\mathrm{EW}}/\%	 $ & $ 0.0046(1)  $ & $ -0.00072(7)  $ & $ -0.021  $ & $ -0.32  $ & $ -0.66  $ & $ -0.20	$ \\	   
SPS3  & $\delta_{\mathrm{SUSY}-\mathrm{QCD}}/\%	 $ & $ 0.0026  $ & $ 0.0050  $ & $ 0.026  $ & $ 0.14  $ & $ 0.50  $ & $ 1.37	$ \\ \hline
SPS4  & $\delta_{\mathrm{SUSY}-\mathrm{EW}}/\%	 $ & $ 0.013  $ & $ 0.0005(1)  $ & $ -0.061  $ & $ -0.24  $ & $ -0.24  $ & $ 0.27	$ \\	   
SPS4  & $\delta_{\mathrm{SUSY}-\mathrm{QCD}}/\%	 $ & $ 0.0034  $ & $ 0.0066  $ & $ 0.035  $ & $ 0.18  $ & $ 0.68  $ & $ 1.30	$ \\ \hline
SPS5  & $\delta_{\mathrm{SUSY}-\mathrm{EW}}/\%	 $ & $ 0.025  $ & $ 0.013  $ & $ 0.063  $ & $ -0.23  $ & $ -0.51  $ & $ 0.11	$ \\	   
SPS5  & $\delta_{\mathrm{SUSY}-\mathrm{QCD}}/\%	 $ & $ 0.0042  $ & $ 0.0083  $ & $ 0.043  $ & $ 0.23  $ & $ 0.85  $ & $ 1.05	$ \\ \hline
SPS6  & $\delta_{\mathrm{SUSY}-\mathrm{EW}}/\%	 $ & $ 0.010  $ & $ 0.0004(2)  $ & $ -0.042  $ & $ -0.42  $ & $ -0.44  $ & $ 0.32	$ \\	   
SPS6  & $\delta_{\mathrm{SUSY}-\mathrm{QCD}}/\%	 $ & $ 0.0042  $ & $ 0.0082  $ & $ 0.043  $ & $ 0.23  $ & $ 0.85  $ & $ 1.05	$ \\ \hline
SPS7  & $\delta_{\mathrm{SUSY}-\mathrm{EW}}/\%	 $ & $ 0.0094(2)  $ & $ 0.0002(2)  $ & $ -0.049  $ & $ -0.42  $ & $ -0.032  $ & $ 0.23	$ \\	   
SPS7  & $\delta_{\mathrm{SUSY}-\mathrm{QCD}}/\%	 $ & $ 0.0023  $ & $ 0.0046  $ & $ 0.024  $ & $ 0.12  $ & $ 0.45  $ & $ 1.40	$ \\ \hline
SPS8  & $\delta_{\mathrm{SUSY}-\mathrm{EW}}/\%	 $ & $ 0.012  $ & $ 0.0039  $ & $ -0.035  $ & $ -0.34  $ & $ -0.17  $ & $ 0.22	$ \\	   
SPS8  & $\delta_{\mathrm{SUSY}-\mathrm{QCD}}/\%	 $ & $ 0.0017  $ & $ 0.0033  $ & $ 0.017  $ & $ 0.088  $ & $ 0.31  $ & $ 1.26	$ \\ \hline
SPS9  & $\delta_{\mathrm{SUSY}-\mathrm{EW}}/\%	 $ & $ -0.015  $ & $ -0.026  $ & $ -0.11  $ & $ 0.029  $ & $ 0.095  $ & $ -0.0059(1)	$ \\	   
SPS9  & $\delta_{\mathrm{SUSY}-\mathrm{QCD}}/\%	 $ & $ 0.0012  $ & $ 0.0023  $ & $ 0.012  $ & $ 0.059  $ & $ 0.20  $ & $ 0.82	$ \\ \hline
\end{tabular}}		
 
  \end{center}
  \caption{
    Relative SUSY-QCD and SUSY-EW correction factors at the LHC for 
    different values of the invariant-mass cut $M_{ll}$ for the SPS 
    benchmark scenarios. The corresponding LO cross sections can be 
    found in Table~\ref{ta:hadcs-lhc-mll}.}  
  \label{ta:hadcs-lhc-mll-susy}
\end{table}
\begin{table}
  \begin{center}
      {\renewcommand{\arraystretch}{1.5}                                                                                       \renewcommand{\tabcolsep}{0.2cm}                                                                                            \begin{tabular}{cc|cccccc}		
\multicolumn{8}{c}{$\Pp\Ppbar\to\Pl^+\Pl^- + \X$ at $\sqrt{s}=1.96\TeV$}\\  \hline 
\multicolumn{2}{c}{$M_{\Pl \Pl}/\mathrm{GeV}$}  & 50--$\infty$ & 100--$\infty$ & 150--$\infty$ & 200--$\infty$ & 400--$\infty$ & 600--$\infty$\\  \hline		
SPS1a & $\delta_{\mathrm{SUSY}-\mathrm{EW}}/\%	 $ & $ 0.0022(2)  $ & $ 0.0034(1)  $ & $ 0.052  $ & $ 0.032  $ & $ -0.14  $ & $ -0.26	$ \\	   
SPS1a & $\delta_{\mathrm{SUSY}-\mathrm{QCD}}/\% 	 $ & $ 0.0059  $ & $ 0.011  $ & $ 0.029  $ & $ 0.047  $ & $ 0.15  $ & $ 0.30	$ \\ \hline
SPS1b & $\delta_{\mathrm{SUSY}-\mathrm{EW}}/\%	 $ & $ 0.0043(1)  $ & $ 0.0050  $ & $ 0.027  $ & $ 0.021  $ & $ -0.064  $ & $ -0.21	$ \\	   
SPS1b & $\delta_{\mathrm{SUSY}-\mathrm{QCD}}/\%	 $ & $ 0.0025  $ & $ 0.0046  $ & $ 0.012  $ & $ 0.019  $ & $ 0.059  $ & $ 0.11	$ \\ \hline
SPS2  & $\delta_{\mathrm{SUSY}-\mathrm{EW}}/\%	 $ & $ -0.051  $ & $ -0.12  $ & $ -0.30  $ & $ -0.40  $ & $ 0.048  $ & $ 0.41	$ \\	   
SPS2  & $\delta_{\mathrm{SUSY}-\mathrm{QCD}}/\%	 $ & $ 0.00092  $ & $ 0.0017  $ & $ 0.0043  $ & $ 0.0070  $ & $ 0.021  $ & $ 0.040	$ \\ \hline
SPS3  & $\delta_{\mathrm{SUSY}-\mathrm{EW}}/\%	 $ & $ 0.0017(1)  $ & $ 0.0027  $ & $ 0.026  $ & $ 0.018  $ & $ -0.074  $ & $ -0.20	$ \\	   
SPS3  & $\delta_{\mathrm{SUSY}-\mathrm{QCD}}/\%	 $ & $ 0.0026  $ & $ 0.0048  $ & $ 0.012  $ & $ 0.020  $ & $ 0.061  $ & $ 0.12	$ \\ \hline
SPS4  & $\delta_{\mathrm{SUSY}-\mathrm{EW}}/\%	 $ & $ 0.0076(1)  $ & $ 0.0030  $ & $ 0.015  $ & $ -0.010  $ & $ -0.21  $ & $ -0.059	$ \\	   
SPS4  & $\delta_{\mathrm{SUSY}-\mathrm{QCD}}/\%	 $ & $ 0.0034  $ & $ 0.0063  $ & $ 0.016  $ & $ 0.026  $ & $ 0.081  $ & $ 0.16	$ \\ \hline
SPS5  & $\delta_{\mathrm{SUSY}-\mathrm{EW}}/\%	 $ & $ 0.020  $ & $ 0.024  $ & $ 0.13  $ & $ 0.14  $ & $ 0.037  $ & $ -0.0027(1)	$ \\	   
SPS5  & $\delta_{\mathrm{SUSY}-\mathrm{QCD}}/\%	 $ & $ 0.0042  $ & $ 0.0078  $ & $ 0.020  $ & $ 0.033  $ & $ 0.10  $ & $ 0.20	$ \\ \hline
SPS6  & $\delta_{\mathrm{SUSY}-\mathrm{EW}}/\%	 $ & $ 0.0048(1)  $ & $ 0.0044  $ & $ 0.034  $ & $ 0.019  $ & $ -0.14  $ & $ -0.27	$ \\	   
SPS6  & $\delta_{\mathrm{SUSY}-\mathrm{QCD}}/\%	 $ & $ 0.0042  $ & $ 0.0078  $ & $ 0.020  $ & $ 0.033  $ & $ 0.10  $ & $ 0.20	$ \\ \hline
SPS7  & $\delta_{\mathrm{SUSY}-\mathrm{EW}}/\%	 $ & $ 0.0054(1)  $ & $ 0.0040  $ & $ 0.025  $ & $ -0.0017  $ & $ -0.21  $ & $ -0.30	$ \\	   
SPS7  & $\delta_{\mathrm{SUSY}-\mathrm{QCD}}/\%	 $ & $ 0.0024  $ & $ 0.0044  $ & $ 0.011  $ & $ 0.018  $ & $ 0.056  $ & $ 0.11	$ \\ \hline
SPS8  & $\delta_{\mathrm{SUSY}-\mathrm{EW}}/\%	 $ & $ 0.0084(1)  $ & $ 0.0067  $ & $ 0.021  $ & $ 0.0059  $ & $ -0.15  $ & $ -0.19	$ \\	   
SPS8  & $\delta_{\mathrm{SUSY}-\mathrm{QCD}}/\%	 $ & $ 0.0017  $ & $ 0.0032  $ & $ 0.0082  $ & $ 0.013  $ & $ 0.040  $ & $ 0.077	$ \\ \hline
SPS9  & $\delta_{\mathrm{SUSY}-\mathrm{EW}}/\%	 $ & $ -0.017  $ & $ -0.022  $ & $ -0.028  $ & $ -0.067  $ & $ -0.014  $ & $ 0.11	$ \\	   
SPS9  & $\delta_{\mathrm{SUSY}-\mathrm{QCD}}/\%	 $ & $ 0.0012  $ & $ 0.0022  $ & $ 0.0056  $ & $ 0.0090  $ & $ 0.027  $ & $ 0.052	$ \\ \hline
\end{tabular}}		
 
  \end{center}
  \caption{
    Relative SUSY-QCD and SUSY-EW correction factors at the Tevatron for 
    different values of the invariant-mass cut $M_{ll}$ for the SPS 
    benchmark scenarios. The corresponding LO cross sections can be 
    found in Table~\ref{ta:hadcs-tev-mll}.}  
  \label{ta:hadcs-tev-mll-susy}
\end{table}
\begin{figure}
  \centering
  \includegraphics[width=0.99\textwidth,clip]{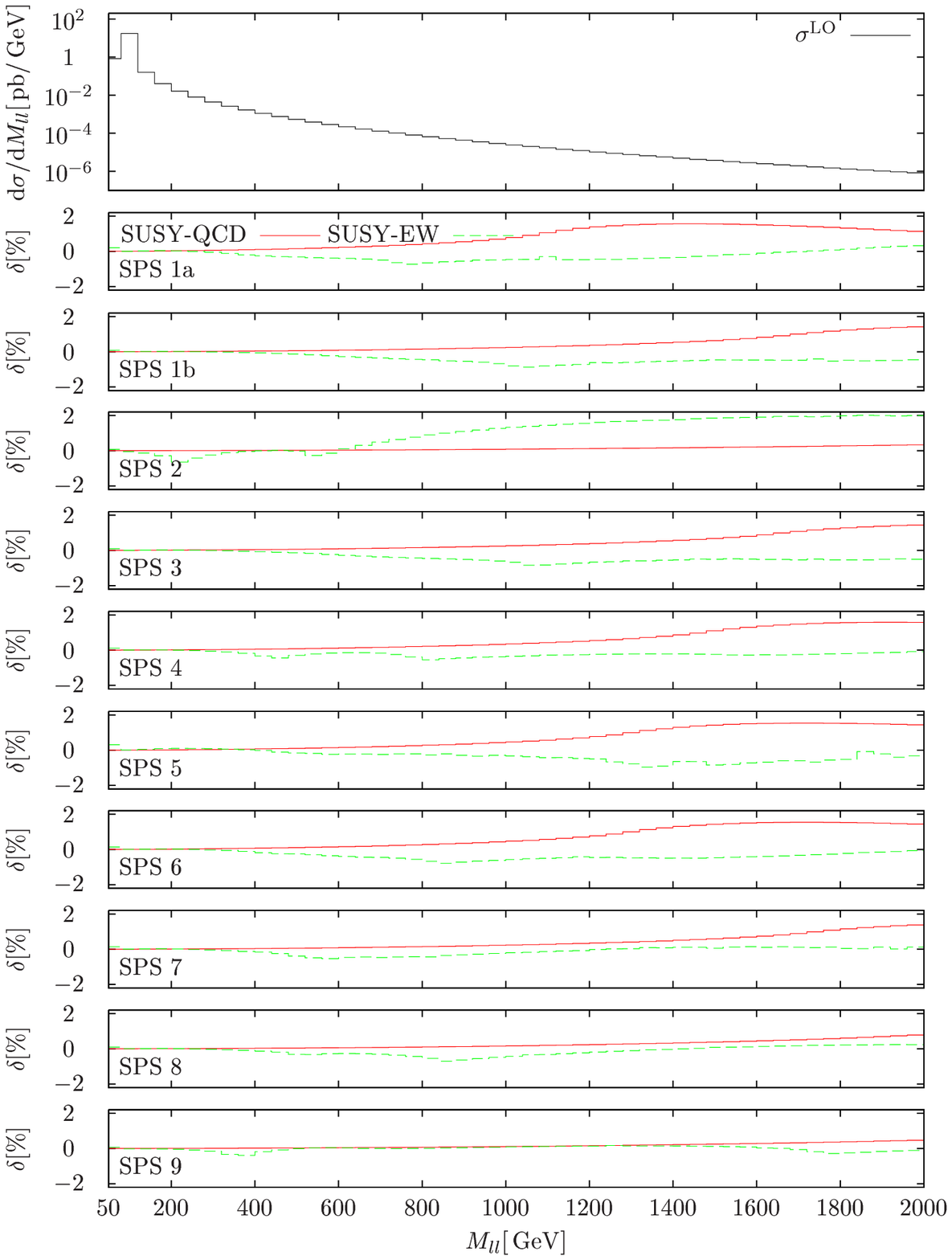}
  \caption{Dilepton invariant-mass distribution and relative SUSY-QCD
    and SUSY-EW correction factors at the LHC for the SPS benchmark
    scenarios.}
  \label{fig:hist3-susy-lhc}
\end{figure}
The corresponding LO cross sections can be found in \refta{ta:hadcs-lhc-mll}.
We give results for
SUSY-QCD and SUSY-EW corrections separately as described in
\refse{se:mssm}. As expected and similar to the CC 
case~\cite{Brensing:2007qm}, the corrections for low invariant dilepton
mass ranges are negligible at the level of $0.1\%$ or below. 
Only for very high $M_{ll}$ and only for a few scenarios the corrections
reach the level of $1{-}2\%$. Similar to the CC case the maximum is
reached for the SPS2 scenario where the gauginos are particularly light
and the squarks and sleptons are so heavy that their negative
contribution becomes effective only at higher invariant mass. 
Table~\ref{ta:hadcs-tev-mll-susy} essentially shows the same features
for the situation at Tevatron (for LO numbers see \refta{ta:hadcs-tev-mll}), 
where the SUSY corrections for the
highest reachable invariant masses are even smaller compared to the
LHC.

In \reffi{fig:hist3-susy-lhc}
we show the invariant-mass distribution $\rd \sigma / \rd M_{ll}$ 
for the LHC.
As already observed for the integrated cross section, the
distributions receive per-cent corrections
only for an invariant mass in the TeV range.
The maximum correction is again found
for SPS2 scenario where the SUSY-EW corrections reach the $2\%$ level.
The SUSY-QCD corrections reach $1\%$ for all but the SPS2, SPS8
and SPS9 scenarios but never exceed $2\%$ for all scenarios in the
considered $M_{ll}$ range.

\section{Conclusions}
\label{se:conclusion}

Neutral-current dilepton production represents one of the most
important processes at hadron colliders, such as the Tevatron and
the LHC. On the one hand, the process acts as a standard candle
that is, e.g, indispensable for detector calibration and sensitive
to the parton distribution functions of the proton; on the other,
it delivers background to many new-physics searches, such as for
new $\PZ'$ bosons. Predictions for this process, thus, ask for the
highest possible precision, i.e.\ both QCD and electroweak
corrections have to be included as far as possible.

In this paper we have recalculated and further discussed the
next-to-leading order corrections in the Standard Model,
where we have compared different methods to include radiative corrections
to the Z-boson resonance in a gauge-invariant way.
This discussion goes beyond previous work, but our numerical results 
confirm results existing in the literature.
The relevant formulas are listed explicitly and can be used by other
groups. 
We consistently include channels with incoming photons, which starts
already with the leading-order contribution of the
$\ga\ga\to\llb$ process. We also include the electroweak corrections
to this process and channels induced by $\ga q/\ga\bar q$ initial
states, i.e.\ all contributions that are formally of electroweak
next-to-leading order.

Beyond next-to-leading order we consider multi-photon radiation off
the final-state leptons in the structure-function approach, 
universal electroweak effects, and two-loop electroweak Sudakov
logarithms at high energies, so that our predictions are of
state-of-the-art precision in view of electroweak corrections.
On the side of the QCD corrections, we include next-to-leading order 
corrections only, so that further improvements via QCD resummations
or interfacing parton showers are desirable.
 
Our discussion of numerical results comprises a survey
of corrections to integrated and differential cross sections, which
shows the impact of the various types of corrections in detail.
In this context we pay particular attention to effects from
incoming photons, because it is not yet clear from results in the
literature whether these effects are phenomenologically important
or swamped by other uncertainties as, e.g., originating from QCD
effects. Our results suggest that effects from $\ga\ga$ initial
states should be significant in the invariant-mass distribution
of the dilepton pair and that these effects can be enhanced to the
level of $10{-}20\%$ by appropriate cuts. This study is, thus, 
particularly interesting for a possible future fit of the photon
distribution function that is part of the DGLAP evolution in the
presence of electroweak corrections.

{}Finally, we have presented results on radiative corrections
within the minimal supersymmetric extension of the Standard Model,
which have not yet been known in the literature. Similar to known
results on charged-current W-boson production, the difference to
the pure Standard Model corrections is small for viable
supersymmetry scenarios. In detail, the supersymmetry corrections
are below the $0.1\%$ level and thus irrelevant near the Z-boson 
resonance and grow only to about $1{-}2\%$ in the TeV range for the
dilepton invariant mass. These results confirm the role
of dilepton production as a standard candle at the Tevatron and
the LHC.

\section*{Acknowledgments}

We would like to thank Ansgar Denner, Tobias Kasprzik and 
Alexander M\"uck for many interesting and helpful
discussions. Moreover, we are grateful to the {\sc HORACE} team
for sending us explicit numbers for the comparison of differential
distributions shown in \refse{se:comp}.
This work is supported in part by the European
Community's Marie-Curie Research Training Network HEPTOOLS under
contract MRTN-CT-2006-035505.

\begin{appendix}

\section{Factorization of the photon PDF}
\label{app:photonPDF}

Since the issue of redefining the photon PDF in order to
absorb the collinear divergence arising from the splitting 
$q\to q\ga^*$ of incoming quarks (or anti-quarks)
seems not to be completely settled
in the literature yet, we here give a brief account on the
analytical structure of this collinear singularity both in
dimensional and mass regularization and derive the corresponding
factorization formula for the photon PDF in the $\MSbar$ and
$\DIS$ factorization schemes. To this end, we can make use of
existing results, e.g., available in papers on dipole subtraction
to deal with this collinear singularity.

In mass regularization this type of singularity is treated in
Section~5 of \citere{Dittmaier:2008md} in detail. After extracting
the singularity from the full phase-space integral by simple subtraction,
the singular integration is performed analytically with a small
quark mass $m_q$ and added back to the result for the cross section.
By construction this procedure can be performed with a
``spectator'' particle in the initial or final state, where its choice
is arbitrary for this type of splitting. Writing the (unpolarized)
partonic cross-section part that is to be re-added generically as
\def\cHsub{{\cal H}}
\beq
\sigma^{\sub}_{qa\to qX}(p_q,p_a) =
\frac{Q_q^2\alpha}{2\pi} \int_0^1\rd x\,
\cHsub(Q^2,x) \,
\sigma_{\gamma a\to X}(k_\ga=xp_q,p_a),
\eeq
where $\sigma_{\gamma a\to X}$ is the partonic cross section of the 
related process with an incoming photon instead of the $q\to q\ga^*$
splitting. The momenta
$p_q$ and $p_a$ correspond to the incoming quark $q$ and the
other massless particle $a$, respectively, and the functions $\cHsub$ read
(cf.\ Eqs.~(5.19) and (5.27) of \citere{Dittmaier:2008md})
\beqar
\cHsub^{qq,a}(s,x) &=& \ln\biggl(\frac{s(1-x)^2}{x^2 m_q^2}\biggr)
P_{\ga f}(x) - 2\frac{1-x}{x}, \qquad
s=(p_q+p_a)^2,
\label{eq:massrega}
\\
\cHsub^{qq}_j(P^2,x) &=& \ln\biggl(\frac{-P^2(1-x)}{x^3 m_q^2}\biggr)
P_{\ga f}(x) - 2\frac{1-x}{x}, \qquad
P^2=(\tilde p_j-k_\ga)^2,
\label{eq:massregj}
\eeqar
where $\cHsub^{qq,a}$ refers to the situation with the incoming
particle $a$ as spectator and $\cHsub^{qq}_j$ to the case with a
massless final-state spectator $j$ of momentum $\tilde p_j$.

The respective results in dimensional regularization can be deduced
from the original Catani--Seymour paper~\cite{Catani:1996vz},
where the dipole subtraction method was introduced for massless QCD.
The case with an initial-state spectator is described in Section~5.5
there, and the case with a final-state spectator in Section~5.3.
Starting from Eq.~(5.152) in the former case and from Eq.~(5.73)
in the latter, and identifying the QCD coupling factor
$C_{\mathrm{F}}\alpha_{\mathrm{s}}$ with the QED factor
$Q_q^2\alpha$, we find
\beqar
\cHsub^{qq,a}(s,x) &=& \frac{1}{\Gamma(1-\epsilon)}
\biggl(\frac{4\pi\mu^2}{s}\biggr)^\epsilon \biggl[
	-\frac{1}{\epsilon}P_{\ga f}(x)+2P_{\ga f}(x)\ln(1-x)+x \biggr],
\label{eq:dimrega}
\\
\cHsub^{qq}_j(P^2,x) &=& \frac{1}{\Gamma(1-\epsilon)}
\biggl(\frac{4\pi\mu^2}{-P^2}\biggr)^\epsilon \biggl[
        -\frac{1}{\epsilon}P_{\ga f}(x)
        +P_{\ga f}(x)\ln\biggl(\frac{1-x}{x}\biggr)+x \biggr]
\label{eq:dimregj}
\eeqar
in $D=4-2\epsilon$ dimensions. Note that in the last relation
we had to translate the kinematical variable $(\tilde p_k p_a)$
of \citere{Catani:1996vz} into our kinematics, which is accomplished
by the replacement $(\tilde p_k p_a) \to 
(\tilde p_j p_q) = (\tilde p_j k_\ga)/x = -P^2/(2x)$.
Comparing Eq.~\refeq{eq:massrega} with Eq.~\refeq{eq:dimrega},
or Eq.~\refeq{eq:massregj} with Eq.~\refeq{eq:dimregj},
we see the following correspondence between the collinear
divergence in mass and dimensional regularization,
\beq
\frac{(4\pi\mu^2)^\epsilon}{\Gamma(1-\epsilon)}\,
\frac{1}{\epsilon} \, P_{\ga f}(x) 
\quad \leftrightarrow \quad
\left( \ln m_q^2+2\ln x+1\right) P_{\ga f}(x),
\label{eq:dim2massreg}
\eeq
which is obtained in either case, i.e.\ with a spectator in the
initial or final state, as it should be.
The correspondence \refeq{eq:dim2massreg} can be used to translate
a result for the collinear singularity of the initial-state splitting
$q\to q\ga^*$ from dimensional to mass regularization, or vice versa,
in any procedure to treat the collinear divergence, i.e.\  it is
universally valid also in other subtraction procedures or in methods
employing phase-space slicing.
Applying this correspondence to the photon PDF redefinition in
dimensional regularization,
\beq
  f^{(h)}_\ga(x) \rightarrow f^{(h)}_\ga(x,\mu_{\mathrm{F}}^2) -\; \frac{\alpha\,Q_q^2}{2\pi}
  \sum_{a=q,\bar{q}} \int^1_x \frac{\rd z}{z} \;
  f^{(h)}_a\left(\frac{x}{z},\mu_{\mathrm{F}}^2\right) 
\left\{
-\frac{(4\pi)^\epsilon}{\Gamma(1-\epsilon)}\,
\biggl(\frac{\mu^2}{\mu_{\mathrm{F}}^2}\biggr)^\epsilon \,
\frac{1}{\epsilon} \, P_{\gamma f}(z) 
  + C^{\mathrm{FS}}_{\gamma f}(z) \right\}
\eeq
we obtain the result \refeq{eq:photonPDFredef} for this redefinition
in the mass regularization scheme, where the coefficient function
$C^{\mathrm{FS}}_{\gamma f}(z)$ is given by
Eq.~\refeq{eq:redef-pdf-C-schema} for the the two considered
factorization schemes (FS), independent of the chosen regularization.
Since the $\MSbar$ scheme merely rearranges the divergent terms 
in dimensional regularization
(plus some trivial universal constants), the coefficient function
obviously vanishes, $C^{\MSbar}_{\gamma f}(z)\equiv0$.
The fixation of $C^{\mathrm{DIS}}_{\gamma f}(z)$ in the DIS scheme
is less trivial.
Similar to the gluon PDF redefinition in the DIS scheme for NLO QCD
corrections, the choice \refeq{eq:redef-pdf-C-schema} ensures that
the whole PDF redefinition does not change the proton momentum
(see e.g.\ Eq.~(6.9) of \citere{Catani:1996vz}).

Finally, we mention a subtle point in the fixation of the
coefficient functions in the DIS scheme.
The DIS factorization scheme is defined in such a way that the DIS structure
function $F_2$ from electron--proton scattering does not receive any 
corrections. This condition uniquely fixes the (anti-)quark PDF
redefinition. The gluon PDF redefinition is performed in such a
way that the total proton momentum remains unchanged after the
PDF redefinition. The simplest choice obviously is to subtract
the same $z$-dependent finite part from the gluon PDF that was added
to the (anti-)quark PDFs; this is expressed by the analogon of
Eq.~\refeq{eq:redef-pdf-C-schema} in NLO QCD. In NLO QED the role
of the gluon is taken over by the photon, which finally leads to
Eq.~\refeq{eq:redef-pdf-C-schema}.
Obviously, employing the sum rule alone as criterion to fix 
$C^{\mathrm{DIS}}_{\gamma f}(z)$ is not unique, but involves
some convention, since an integral over $z$ does not fix the $z$
dependence. 
A different choice would result in mass regularization
if the coefficient functions $C^{\mathrm{DIS}}_{ab}(z)$ were just
defined to quantify the finite parts in addition to the divergent
contributions that are proportional to $\ln m_q^2 P_{ab}(z)$.
The result for the redefined photon PDF would still satisfy
the proton momentum sum rule, but the $z$~dependence of the
cross section with the incoming photon would be somewhat different.
In the first preprint version of this paper such a choice was in
fact made, and we found results on the contributions from 
$q\ga$ scattering that differ from the results of this paper at the
level of up to a few per cent in the off-shell tail of the Z-boson
resonance; for observables that are dominated by resonant Z~bosons
effects due to $q\ga$ scattering are negligible in either scheme.
The ``correct'' choice, i.e.\ the one that is in line with the
standard definitions made in NLO QCD, is obtained upon first
translating the mass-regularized divergence from the
$q\to q\ga^*$ splitting into dimensional
regularization via the correspondence \refeq{eq:dim2massreg}
and defining the remaining part as coefficient functions,
as expressed in Eq.~\refeq{eq:photonPDFredef}.

\section{Vertex and box corrections}
\label{app:virtRCs}

In this section we state explicitly the expressions for the vertex
form factors $F^\sigma_{ffV,\weak},\,V=\ga,\,\PZ$ \refeq{eq:dvertew},
and the photonic \refeq{eq:photbox} and weak box diagrams
\refeq{eq:ewbox}. The occuring scalar integrals $B_0$, $C_0$, and
$D_0$ depend on their arguments as follows,
\begin{eqnarray}
&& B_0(p_1^2,m_0,m_1)=
\frac{(2\pi\mu)^{4-D}}{\ri\pi^2}\int\rd^D q\,
\frac{1}{[q^2-m_0^2+\ri0][(q+p_1)^2-m_1^2+\ri0]},\\[.5em]
&& C_0(p_1^2,(p_2-p_1)^2,p_2^2,m_0,m_1,m_2)=
\frac{(2\pi\mu)^{4-D}}{\ri\pi^2}\int\rd^D q
\nonumber\\[.2em]
&& \hspace{1em}
\times\frac{1}{[q^2-m_0^2+\ri0]
[(q+p_1)^2-m_1^2+\ri0] [(q+p_2)^2-m_2^2+\ri0]}, \\[.5em]
&& 
D_0(p_1^2,(p_2-p_1)^2,(p_3-p_2)^2,p_3^2,
p_2^2,(p_3-p_1)^2,m_0,m_1,m_2,m_3)=
\frac{(2\pi\mu)^{4-D}}{\ri\pi^2}\int\rd^D q
\nonumber\\[.2em]
&& \hspace{1em}
\times\frac{1}{[q^2-m_0^2+\ri0] [(q+p_1)^2-m_1^2+\ri0]
[(q+p_2)^2-m_2^2+\ri0] [(q+p_3)^2-m_3^2+\ri0]},
\hspace{3em}
\end{eqnarray}
with $D$ denoting the number of space-time dimensions.
The evaluation of these scalar integrals with real or complex masses
has been briefly described in \refse{sec:rad-corr-sm}.
{}For the form factors for the weak vertex corrections we obtain
\begin{eqnarray}
{}F^+_{ff\ga/\PZ,\weak}(\hat s) &=&  
-\frac{\alpha Q_f^2 \sw^2}{4 \pi \cw^2}\left\{
2-2 \frac{( \xMZ^2+2 \hat s)}{\hat s}\BoooZ 
+ \frac{3\hat s+ 2\xMZ^2}{\hat s} \Bosoo \right. 
\nl
&&\left.
{}+ \frac{2 (\xMZ^2+\hat s)^2}{\hat s} \CoZo \right\} \,,
\\
{}F^-_{ff\ga,\weak}(\hat s)  &=& 
+\frac{\alpha}{8 \pi} \Biggl\{ \frac{1}{\hat{s} \sw^2 Q_f}
\bigg[ 2 Q_f  (2 \hat{s}+\xMW^2) \BoooW
\nl
&& {}
+ (2 I^3_{\mathrm{w},f}-Q_f)\Big(2 \hat{s}+(3 \hat{s}+2 \xMW^2)  \Bosoo+ 2 (\hat{s}+\xMW^2)^2 \CoWo \Big) 
\nl
&& {}
- 2  I^3_{\mathrm{w},f} \Big((\hat{s}+2 \xMW^2) \BoWW- 2  \xMW^2 (2\hat{s}+\xMW^2)\CoWoW \Big)\bigg] 
\nl
&& {}+ \frac{(I^3_{\mathrm{w},f}-Q_f \sw^2)^2}{\hat{s} \cw^2 \sw^2}
\bigg[ -4 \hat{s}+ 4(2 \hat{s}+\xMZ^2)  \BoooZ -2  (3 \hat{s}+2 \xMZ^2) \Bosoo 
\nl
&& {}
-4  (\hat{s}+\xMZ^2)^2 \CoZo\bigg] \Biggr\}\,, 
\\
{}F^-_{ff\PZ,\weak}(\hat s)  &=&  
+\frac{\alpha}{8 \pi} \Biggl\{ 
\frac{1}{\hat{s} \sw^2(I^3_{\mathrm{w},f}-Q_f \sw^2)}\bigg[ 2 (I^3_{\mathrm{w},f}-Q_f \sw^2)(2 \hat{s}+\xMW^2) \BoooW 
\nl
&& {} +(I^3_{\mathrm{w},f}\cw^2 - I^3_{\mathrm{w},f} \sw^2+Q_f \sw^2)\Big( 2 \hat{s} + (3 \hat{s}+2 \xMW^2) \Bosoo 
\nl
&& {} +2 (\hat{s}+\xMW^2)^2  \CoWo \Big) 
\nl
&& {} -2 \cw^2 I^3_{\mathrm{w},f} \Big( (\hat{s}+2 \xMW^2) \BoWW -2 \xMW^2 (2 \hat{s}+\xMW^2) \CoWoW \Big) \bigg] 
\nl
&& {} + \frac{(I^3_{\mathrm{w},f}-Q_f \sw^2)^2}{\hat{s} \cw^2 \sw^2} \bigg[-4 \hat{s} +
  4 (2 \hat{s}+\xMZ^2) \BoooZ -2 (3 \hat{s} + 2 \xMZ^2)  \Bosoo \nl
&& {}- 4 (\hat{s}+\xMZ^2)^2  \CoZo\bigg] 
\Biggr\} \,.
\end{eqnarray}
Due to the heavy mass of the top quark the weak corrections for
incoming \Pb-quarks differ from those for partonic processes with
incoming \Pd/\Ps - quarks by $\delta F^\sigma_{\Pb\Pb \ga/\PZ,\weak} =
{}F^\sigma_{\Pb\Pb\ga/\PZ,\weak} - F^\sigma_{\Pd\Pd/\Ps\Ps\,\ga/\PZ,\weak}$, where
\begin{eqnarray}
\delta F^+_{\Pb\Pb \ga,\weak}(\hat s) &=&
\delta F^+_{\Pb\Pb\PZ,\weak}(\hat s) = 0 ,\\
\delta F^-_{\Pb\Pb \ga,\weak}(\hat s) &=& 
+\frac{\al}{8 \pi} \Biggl\{ -\frac{2}{\hat{s} \sw^2}  
\bigg[ (2 \hat{s}+\xMW^2)(\BoooW-\BootW) 
\nl
&& {} +(3 \hat{s}+2\xMW^2)(\Bosoo-\Bostt)\nl
&& {} + 2 (\hat{s}+\xMW^2)^2 (\CoWo-\CotWt) \nl
&& {} +3 \xMW^2 (2 \hat{s}+\xMW^2) (\CoWoW-\CoWtW) \bigg]
\nl 
&& {} -\frac{\xMT^2}{\hat{s} \sw^2 \xMW^2} 
\bigg[(\xMT^2+\xMW^2)(\BootW+2 \Bostt-3 \BoWW)
\nl
&& {} + \hat{s} \big(\Bostt-\frac{3}{2}\BoWW-\frac{5}{2}\big) + (\xMW^2 (2 \hat{s}+3 \xMW^2)-\xMT^2(\xMT^2+\hat{s}))  
\nl
&&  {}\times (2 \CotWt+3 \CoWtW)
\bigg]\Biggr\}\,, 
\\ 
\delta F^-_{\Pb\Pb \PZ,\weak}(\hat s) &=& 
+\frac{\al}{8 \pi } \Biggl\{ - \frac{1}{(2 \sw^2-3) \hat{s}\sw^2} 
\bigg[2(2 \hat{s}+\xMW^2) (2 \sw^2-3) (\BoooW-\BootW) 
\nl
&& {}      +(3 \hat{s}+2 \xMW^2) (4\sw^2-3)(\Bosoo-\Bostt) 
\nl
&& {}      -12 \cw^2 \xMW^2 (2 \hat{s}+\xMW^2)(\CoWoW-\CoWtW) 
\nl
&& {}      + 2(4 \sw^2-3)(\hat{s}+\xMW^2)^2(\CoWo-\CotWt)\bigg] 
\nl
&& {} -\frac{\xMT^2}{(2 \sw^2-3) \hat{s}\sw^2 \xMW^2} 
\bigg[ (\xMT^2+\xMW^2) (4 \Bostt \sw^2-3 \BoWW (\sw^2-\cw^2)
\nl
&& {} +\BootW (2 \sw^2-3)) +(2 \hat{s} \sw^2-6 \xMW^2)\Bostt  
+\hat{s} \bigg(\frac{3}{2}-5 \sw^2\bigg)
\nl 
&& {}    +\bigg(6 \xMW^2-\frac{3}{2} \hat{s} (\sw^2-\cw^2)\bigg)\BoWW -12 \hat{s} \xMW^2 \CotWt 
\nl
&& {}
-3 \left(4 \xMW^4-2 \xMT^2 \xMW^2-\xMT^2 \hat{s}\right) (\CotWt+\CoWtW)
\nl
&& {}    -3 \left(3 \xMW^4-\xMT^4\right)\CoWtW 
+2 \sw^2(\xMW^2 (2 \hat{s}+3 \xMW^2)-\xMT^2 (\xMT^2+\hat{s}))
\nl 
&& {} \times    (2 \CotWt+3 \CoWtW)  \bigg]\Biggr\}\,.
\end{eqnarray}

As mentioned in \refse{sec:surv-radi-corr}, the calculation of the
box diagrams leads to additional Dirac chains. However, these can be
reduced to the Dirac structure \refeq{eq:sme}
appearing in the LO matrix element,
\begin{equation}
\mathcal{A}^{\sigma \tau} = \left[\bar{v}_{q} \ga^\mu \omega_\sigma u_{q}  \right] \left[\bar{u}_{l} \ga_\mu \omega_\tau v_{l} \right] \,.
\end{equation}
We used the identities~\cite{Denner:2005fg}
\begin{eqnarray}
\big[\bar{v}_{q} \ga^\al \ga^\beta \ga^\delta \omega_{\pm} u_{q}\big] 
\big[\bar{u}_{l} \ga_\al \ga^{\beta'} \ga_\delta \omega_{\pm} v_{l} \big]
&=& 4 g^{\beta \beta'} \mathcal{A}^{\pm\pm} \,,
\nl
\big[\bar{v}_{q} \ga^\al \ga^\beta \ga^\delta \omega_{\pm} u_{q}\big] 
\big[\bar{u}_{l} \ga_\al \ga^{\beta'} \ga_\delta \omega_{\mp} v_{l} \big]
&=& 
4\big[\bar{v}_{q}  \ga^{\beta'} \omega_\pm u_{q}\big]  \big[\bar{u}_{l} \ga^\beta \omega_{\mp} v_{l} \big] \,,
\nl
\big[\bar{v}_{q} \, \ks_1\, \omega_\pm  u_{q}\big]  
\big[\bar{u}_{l} \,\ps_2\, \omega_\pm v_{l} \big] 
&=& + \hat u \,\mathcal{A}^{\pm\pm} / 2 \,,
\nl
\big[\bar{v}_{q} \, \ks_1\, \omega_\pm u_{q}\big]  
\big[\bar{u}_{l} \,\ps_2\, \omega_\mp    v_{l} \big] 
&=& - \hat t \,\mathcal{A}^{\pm\mp} / 2 \,,
\end{eqnarray}
which are valid in four space-time dimensions, to reduce the Dirac
structures of the box diagrams. The electroweak
corrections due to box diagrams can be written in terms of the functions
\begin{eqnarray}
\lefteqn{
b^{\pm\mp}_t(M_V,M_{V'},m_{Q},m_q,m_l) = - 2 \Big\{\CoslVVpQ + \CosqVVpQ } &&
\nl
&& {}- ( \hat t - m_Q^2 ) \DotlqVQVpo  \Big\} \,,
\\
\lefteqn{
b^{\pm\pm}_t(M_V,M_{V'},m_{Q},m_q,m_l) =
\frac{1}{\hat u^2}\Big\{ 2 \hat u \big( B_0(\hat s,M_V,M_{V'})-B_0(\hat t,m_{Q},0) \big) } &&
\nl
&& {} - \hat t \big(m_Q^2 - M_V^2 - M_{V'}^2 - \hat t + \hat u \big)  \big(\CotlqVQ+\CotlqVpQ \big) 
\nl
&& {}- \big(\hat t^2 + \hat u^2 + \hat s ( m_Q^2-M_V^2-M_{V'}^2) \big) 
\big(\CoslVVpQ 
\nl
&& {} + \CosqVVpQ \big) 
+  \Big(\hat t (M_V^2+M_{V'}^2-m_Q^2-2 \hat s)^2 
\nl
&& {} + (\hat u (2 \hat u - m_Q^2) -2 \hat s^2)(M_V^2+M_{V'}^2-m_Q^2-2 \hat s)
- 2 \hat u (\hat u^2 - M_V^2M_{V'}^2) + \hat u \hat s (\hat s - m_Q^2 ) - \hat s^3 \Big) 
\nl
&& {} \times\DotlqVQVpo
 \Big\} \,,
\\
\lefteqn{
b^{\pm\pm}_u(M_V,M_{V'},m_{Q},m_q,m_l) = -b^{\pm\mp}_t(M_V,M_{V'},m_{Q},m_q,m_l)\big|_{\hat t \leftrightarrow \hat u} \,, } && \\
\lefteqn{
b^{\pm\mp}_u(M_V,M_{V'},m_{Q},m_q,m_l) = -b^{\pm\pm}_t(M_V,M_{V'},m_{Q},m_q,m_l)\big|_{\hat t \leftrightarrow \hat u} \,. } &&
\end{eqnarray}
Here the subscripts $q$ and $l$ refer to light quarks and leptons. The
heavy mass $m_Q$ of the weak isospin partner of down-type quarks is
present only in the case of \PW\PW~box diagrams for incoming
\Pb-quarks. In terms of these functions the electroweak box correction
factors read
\begin{eqnarray}
f_{\qqb}^{\PZ\PZ,\, \sigma \tau}(\hat s,\hat t)&=& 
\alpha^2  (\gqqZ^\sigma \gllZ^\tau)^2 \big(
b^{\sigma\tau}_t(\xMZ,\xMZ,0,0,0) + b^{\sigma\tau}_u(\xMZ,\xMZ,0,0,0) \big) \,
\end{eqnarray}
for the \PZ\PZ~box-diagrams, and
\begin{eqnarray}
f_{\qqb}^{\PW\PW,\,++}(\hat s,\hat t) &=& f_{\qqb}^{\PW\PW,\,\pm\mp}(\hat s,\hat t) = 0 \,,\\
f_{\Pu \bar \Pu}^{\PW\PW,\,--}(\hat s,\hat t) &=& \frac{\alpha^2}{4 \sw^4} b^{--}_u(\xMW,\xMW,0,0,0) \,,\\ 
f_{\Pd \bar \Pd}^{\PW\PW,\,--}(\hat s,\hat t) &=& \frac{\alpha^2}{4 \sw^4} b^{--}_t(\xMW,\xMW,0,0,0) \,,\\ 
f_{\Pb \bar \Pb}^{\PW\PW,\,--}(\hat s,\hat t) &=& \frac{\alpha^2}{4 \sw^4} b^{--}_t(\xMW,\xMW,\xMT,0,0)
\end{eqnarray}
for the \PW\PW~box-diagrams. The photonic box corrections are given by
\begin{eqnarray}
f_{\qqb}^{\ga\ga,\, \sigma \tau} &=& \alpha^2\, Q_q^2 Q_l^2\; 
\big(
b^{\sigma \tau}_t(m_\ga,m_\ga,0,m_q,m_l) + b^{\sigma \tau}_u(m_\ga,m_\ga,0,m_q,m_l) \big) \,,\\
f_{\qqb}^{\PZ\ga,\,\sigma \tau} &=& 2\alpha^2\, Q_q Q_l\, \gqqZ^\sigma \gllZ^\tau \;\big(
b^{\sigma \tau}_t(\xMZ,m_\ga,0,m_q,m_l) + b^{\sigma \tau}_u(\xMZ,m_\ga,0,m_q,m_l) \big) \,.
\end{eqnarray}

\section{SPS benchmark scenarios}
\label{app:SPS}

{}For the SPS benchmark~\cite{Allanach:2002nj} scenarios discussed in
this work we use the low-energy input specified in
Table~\ref{ta:SUSY_input}.  The input variables are the ratio $\tanb$
of the vacuum expectation values of the Higgs bosons giving rise to
up- and down-type fermion masses, the mass of the CP-odd Higgs boson,
$\MA$, the supersymmetric Higgs mass parameter $\mu$, the electroweak
gaugino mass parameters $M_{1,2}$, the gluino mass $m_{\tilde{g}}$,
the trilinear couplings $A_{\tau,\Pt,\Pb}$, the scale at which the
\DRbar-input values are defined, $\mu_{\mathrm{R}} ($\DRbar$)$, the soft
SUSY-breaking parameters in the diagonal entries of the squark and
slepton mass matrices of the first and second generations $M_{fi}$
(where $i=L,R$ refers to the left- and right-handed sfermions, $\Pf=\Pq,\Pl$
to quarks and leptons, and $\Pf=\Pu,\Pd,\Pe$ to up and down quarks and
electrons, respectively), and the analogous soft SUSY-breaking
parameters for the third generation $M^{3G}_{fi}$. 
\begin{table}
\begin{center}
\vspace{18cm}
\begin{rotate}{90}
\begin{tabular}{|r|r|r|r|r|r|r|r|r|r|r|}
\hline
\mbox{} \hspace{1cm} \mbox{} & SPS~1a     & SPS~1b    & SPS~2    & SPS~3         & SPS~4      & SPS~5       & SPS~6    & SPS~7        & SPS~8      & SPS~9   \\ \hline
$\tanb$                      & $     10 $ & $    30 $ & $     10 $ & $      10 $ & $     50 $ & $       5 $ & $     10 $ & $     15 $ & $     15 $ & $     10 $  \\
$\MA$[GeV]                   & $  393.6 $ & $ 525.5 $ & $ 1443.0 $ & $   572.4 $ & $  404.4 $ & $   693.9 $ & $  463.0 $ & $  377.9 $ & $  514.5 $ & $  911.7 $  \\
$\mu$[GeV]                   & $  352.4 $ & $ 495.6 $ & $  124.8 $ & $   508.6 $ & $  377.0 $ & $   639.8 $ & $  393.9 $ & $  300.0 $ & $  398.3 $ & $  869.9 $  \\
$M_1$[GeV]                   & $   99.1 $ & $ 162.8 $ & $  120.4 $ & $   162.8 $ & $  120.8 $ & $   121.4 $ & $  195.9 $ & $  168.6 $ & $  140.0 $ & $ -550.6 $  \\
$M_2$[GeV]                   & $  192.7 $ & $ 310.9 $ & $  234.1 $ & $   311.4 $ & $  233.2 $ & $   234.6 $ & $  232.1 $ & $  326.8 $ & $  271.8 $ & $ -175.5 $  \\
$m_{\tilde{g}}$[GeV]         & $  595.2 $ & $ 916.1 $ & $  784.4 $ & $   914.3 $ & $  721.0 $ & $   710.3 $ & $  708.5 $ & $  926.0 $ & $  820.5 $ & $ 1275.2 $  \\
$A_\tau$[GeV]                & $ -254.2 $ & $-195.8 $ & $ -187.8 $ & $  -246.1 $ & $ -102.3 $ & $ -1179.3 $ & $ -213.4 $ & $  -39.0 $ & $  -36.7 $ & $ 1162.4 $  \\
$A_\Pt$[GeV]                 & $ -510.0 $ & $-729.3 $ & $ -563.7 $ & $  -733.5 $ & $ -552.2 $ & $  -905.6 $ & $ -570.0 $ & $ -319.4 $ & $ -296.7 $ & $ -350.3 $  \\
$A_\Pb$[GeV]                 & $ -772.7 $ & $-987.4 $ & $ -797.2 $ & $ -1042.2 $ & $ -729.5 $ & $ -1671.4 $ & $ -811.3 $ & $ -350.5 $ & $ -330.3 $ & $  216.4 $  \\
$\mu_{\mathrm{R}} ($\DRbar$)$[GeV]      & $  454.7 $ & $ 706.9 $ & $ 1077.1 $ & $   703.8 $ & $  571.3 $ & $   449.8 $ & $  548.3 $ & $  839.6 $ & $  987.8 $ & $ 1076.1 $  \\ 
$M_{\Pq L}$[GeV]             & $  539.9 $ & $ 836.2 $ & $ 1533.6 $ & $   818.3 $ & $  732.2 $ & $   643.9 $ & $  641.3 $ & $  861.3 $ & $ 1081.6 $ & $ 1219.2 $  \\
$M_{\Pd R}$[GeV]             & $  519.5 $ & $ 803.9 $ & $ 1530.3 $ & $   788.9 $ & $  713.9 $ & $   622.9 $ & $  621.8 $ & $  828.6 $ & $ 1029.0 $ & $ 1237.6 $  \\
$M_{\Pu R}$[GeV]             & $  521.7 $ & $ 807.5 $ & $ 1530.5 $ & $   792.6 $ & $  716.0 $ & $   625.4 $ & $  629.3 $ & $  831.3 $ & $ 1033.8 $ & $ 1227.9 $  \\
$M_{\Pl L}$[GeV]             & $  196.6 $ & $ 334.0 $ & $ 1455.6 $ & $   283.3 $ & $  445.9 $ & $   252.2 $ & $  260.7 $ & $  257.2 $ & $  353.5 $ & $  316.2 $  \\
$M_{eR}$[GeV]                & $  136.2 $ & $ 248.3 $ & $ 1451.0 $ & $   173.0 $ & $  414.2 $ & $   186.8 $ & $  232.8 $ & $  119.7 $ & $  170.4 $ & $  300.0 $  \\
$M^{3G}_{\Pq L}$[GeV]        & $  495.9 $ & $ 762.5 $ & $ 1295.3 $ & $   760.7 $ & $  640.1 $ & $   535.2 $ & $  591.2 $ & $  836.3 $ & $ 1042.7 $ & $ 1111.6 $  \\
$M^{3G}_{\Pd R}$[GeV]        & $  516.9 $ & $ 780.3 $ & $ 1519.9 $ & $   785.6 $ & $  673.4 $ & $   620.5 $ & $  619.0 $ & $  826.9 $ & $ 1025.5 $ & $ 1231.7 $  \\
$M^{3G}_{\Pu R}$[GeV]        & $  424.8 $ & $ 670.7 $ & $  998.5 $ & $   661.2 $ & $  556.8 $ & $   360.5 $ & $  517.0 $ & $  780.1 $ & $  952.7 $ & $ 1003.2 $  \\
$M^{3G}_{\Pl L}$[GeV]        & $  195.8 $ & $ 323.8 $ & $ 1449.6 $ & $   282.4 $ & $  394.7 $ & $   250.1 $ & $  259.7 $ & $  256.8 $ & $  352.8 $ & $  307.4 $  \\
$M^{3G}_{\Pe R}$[GeV]        & $  133.6 $ & $ 218.6 $ & $ 1438.9 $ & $   170.0 $ & $  289.5 $ & $   180.9 $ & $  230.5 $ & $  117.6 $ & $  167.2 $ & $  281.2 $  \\ \hline
\end{tabular}
\end{rotate}
\end{center}
\caption{ The low-energy input for the SPS scenarios. See text for details.}
\label{ta:SUSY_input}
\end{table}

\end{appendix}


\begin{thebibliography}{99}
\frenchspacing
\newcommand{\epj}[3]{{\sl Eur. Phys. J.} {\bf #1} (19#2) #3}
\newcommand{\zp}[3]{{\sl Z. Phys.} {\bf #1} (19#2) #3}
\newcommand{\np}[3]{{\sl Nucl. Phys.} {\bf #1} (19#2) #3}
\newcommand{\phm}[3]{{\sl Phil. Mag.} {\bf #1} (19#2) #3}
\newcommand{\pl}[3]{{\sl Phys. Lett.} {\bf #1} (19#2) #3}
\newcommand{\pr}[3]{{\sl Phys. Rev.} {\bf #1} (19#2) #3}
\newcommand{\prep}[3]{{\sl Phys.\ Rep.} {\bf #1} (19#2) #3}
\newcommand{\prl}[3]{{\sl Phys. Rev. Lett.} {\bf #1} (19#2) #3}
\newcommand{\prs}[3]{{\sl Proc. Roy. Soc.} {\bf #1} (19#2) #3}
\newcommand{\fp}[3]{{\sl Fortschr. Phys.} {\bf #1} (19#2) #3}
\newcommand{\cpc}[3]{{\sl Comput. Phys. Commun.} {\bf #1} (19#2) #3}
\newcommand{\ijmp}[3]{{\sl Int. J. Mod. Phys.} {\bf #1} (19#2) #3}
\newcommand{\nim}[3]{{\sl Nucl. Instr. Meth.} {\bf #1} (19#2) #3}
\newcommand{\nc}[3]{{\sl Nuovo Cimento} {\bf #1} (19#2) #3}
\newcommand{\vj}[4]{{\sl #1} {\bf #2} (19#3) #4}
\newcommand{\jcp}[3]{{\sl J. Comp. Phys.} {\bf #1} (19#2) #3}

\bibitem{Gerber:2007xk}
  C.~E.~Gerber {\it et al.}  [TeV4LHC Top and Electroweak Working Group],
  ``Tevatron-for-LHC report: Top and electroweak physics,''
  arXiv:0705.3251 [hep-ph].

\bibitem{Buescher:2006jm}
  V.~B\"uscher  {\it et al.} [TeV4LHC Landscape Working Group],
  ``Tevatron-for-LHC report: Preparations for discoveries,''
  hep-ph/0608322.

\bibitem{Haywood:1999qg}
  S.~Haywood {\it et al.},
  hep-ph/0003275.

\bibitem{Hamberg:1990np}
  R.~Hamberg, W.~L.~van Neerven and T.~Matsuura,
  Nucl.\ Phys.\  B {\bf 359} (1991) 343
  [Erratum-ibid.\  B {\bf 644} (2002) 403];\\
  W.~L.~van Neerven and E.~B.~Zijlstra,
  Nucl.\ Phys.\  B {\bf 382} (1992) 11
  [Erratum-ibid.\  B {\bf 680} (2004) 513];\\
%
  R.~V.~Harlander and W.~B.~Kilgore,
  Phys.\ Rev.\ Lett.\  {\bf 88} (2002) 201801
  [hep-ph/0201206].

\bibitem{Anastasiou:2003yy}
  C.~Anastasiou, L.~J.~Dixon, K.~Melnikov and F.~Petriello,
  Phys.\ Rev.\ Lett.\  {\bf 91} (2003) 182002
  [hep-ph/0306192] and
%
  Phys.\ Rev.\  D {\bf 69} (2004) 094008
  [hep-ph/0312266];\\
%
  K.~Melnikov and F.~Petriello,
  Phys.\ Rev.\ Lett.\  {\bf 96} (2006) 231803
  [hep-ph/0603182] and
  Phys.\ Rev.\  D {\bf 74} (2006) 114017
  [hep-ph/0609070];\\
%
  S.~Catani, L.~Cieri, G.~Ferrera, D.~de Florian and M.~Grazzini,
  Phys.\ Rev.\ Lett.\  {\bf 103} (2009) 082001
  [arXiv:0903.2120 [hep-ph]].


\bibitem{Moch:2005ky}
  S.~Moch and A.~Vogt,
  Phys.\ Lett.\  B {\bf 631} (2005) 48
  [hep-ph/0508265];\\
  E.~Laenen and L.~Magnea,
  Phys.\ Lett.\  B {\bf 632}, 270 (2006)
  [hep-ph/0508284];\\
  A.~Idilbi, X.~d.~Ji, J.~P.~Ma and F.~Yuan,
  Phys.\ Rev.\  D {\bf 73}, 077501 (2006)
  [hep-ph/0509294];\\
  V.~Ravindran and J.~Smith,
  Phys.\ Rev.\  D {\bf 76}, 114004 (2007)
  [arXiv:0708.1689 [hep-ph]].

\bibitem{Frixione:2006gn}
  S.~Frixione and B.~R.~Webber,
  hep-ph/0612272.

\bibitem{Arnold:1990yk}
  P.~B.~Arnold and R.~P.~Kauffman,
  Nucl.\ Phys.\  B {\bf 349} (1991) 381;\\
%
  C.~Balazs, J.~w.~Qiu and C.~P.~Yuan,
  Phys.\ Lett.\  B {\bf 355} (1995) 548
  [hep-ph/9505203];\\
%
  C.~Balazs and C.~P.~Yuan,
  Phys.\ Rev.\  D {\bf 56} (1997) 5558
  [hep-ph/9704258];\\
%
  R.~K.~Ellis, D.~A.~Ross and S.~Veseli,
  Nucl.\ Phys.\  B {\bf 503} (1997) 309
  [hep-ph/9704239];\\
%
  R.~K.~Ellis and S.~Veseli,
  Nucl.\ Phys.\  B {\bf 511} (1998) 649
  [hep-ph/9706526];\\
%
  J.~w.~Qiu and X.~f.~Zhang,
  Phys.\ Rev.\ Lett.\  {\bf 86} (2001) 2724
  [hep-ph/0012058] and
%
  Phys.\ Rev.\  D {\bf 63} (2001) 114011
  [hep-ph/0012348];\\
%
  A.~Kulesza and W.~J.~Stirling,
  Eur.\ Phys.\ J.\  C {\bf 20} (2001) 349
  [hep-ph/0103089];\\
%
  A.~Kulesza, G.~Sterman and W.~Vogelsang,
  Phys.\ Rev.\  D {\bf 66} (2002) 014011
  [hep-ph/0202251];\\
%
  F.~Landry, R.~Brock, P.~M.~Nadolsky and C.~P.~Yuan,
  Phys.\ Rev.\  D {\bf 67} (2003) 073016
  [hep-ph/0212159];\\
%
  S.~Berge, P.~M.~Nadolsky and F.~I.~Olness,
  Phys.\ Rev.\  D {\bf 73} (2006) 013002
  [hep-ph/0509023];\\
%
  G.~Bozzi {\it et al.},
  arXiv:0812.2862 [hep-ph].

\bibitem{Baur:1998kt}
  U.~Baur, S.~Keller and D.~Wackeroth,
  Phys.\ Rev.\  D {\bf 59} (1999) 013002
  [hep-ph/9807417].

\bibitem{Zykunov:2001mn}
  V.~A.~Zykunov,
  Eur.\ Phys.\ J.\ direct C {\bf 3} (2001) 9
  [hep-ph/0107059];\\
%
  U.~Baur and D.~Wackeroth,
  Phys.\ Rev.\  D {\bf 70} (2004) 073015
  [hep-ph/0405191];\\
%
  A.~Arbuzov {\it et al.},
  Eur.\ Phys.\ J.\  C {\bf 46} (2006) 407
  [Erratum-ibid.\  C {\bf 50} (2007) 505]
  [hep-ph/0506110].

\bibitem{Dittmaier:2001ay}
  S.~Dittmaier and M.~Kr\"amer,
  Phys.\ Rev.\  D {\bf 65} (2002) 073007
  [hep-ph/0109062].

\bibitem{CarloniCalame:2006zq}
  C.~M.~Carloni Calame, G.~Montagna, O.~Nicrosini and A.~Vicini,
  JHEP {\bf 0612} (2006) 016
  [hep-ph/0609170].

\bibitem{Baur:1997wa}
  U.~Baur, S.~Keller and W.~K.~Sakumoto,
  Phys.\ Rev.\  D {\bf 57} (1998) 199
  [hep-ph/9707301].

\bibitem{Baur:2001ze}
  U.~Baur {\it et al.},
  Phys.\ Rev.\  D {\bf 65} (2002) 033007
  [hep-ph/0108274].

\bibitem{Zykunov:2005tc}
  V.~A.~Zykunov,
  Phys.\ Rev.\  D {\bf 75} (2007) 073019
  [hep-ph/0509315].

\bibitem{CarloniCalame:2007cd}
  C.~M.~Carloni Calame, G.~Montagna, O.~Nicrosini and A.~Vicini,
  JHEP {\bf 0710} (2007) 109
  [arXiv:0710.1722 [hep-ph]].

\bibitem{Arbuzov:2007db}
  A.~Arbuzov {\it et al.},
  Eur.\ Phys.\ J.\  C {\bf 54} (2008) 451
  [arXiv:0711.0625 [hep-ph]].

\bibitem{Buttar:2006zd}
  C.~Buttar {\it et al.},
  ``Les Houches physics at TeV colliders 2005, standard model, QCD, EW, and
  Higgs working group: Summary report,''
  hep-ph/0604120.

\bibitem{Buttar:2008jx}
  C.~Buttar {\it et al.},
  arXiv:0803.0678 [hep-ph].

\bibitem{DK_LH} S.~Dittmaier and M.~Kr\"amer, in \citere{Buttar:2006zd}.

\bibitem{Arbuzov:2007kp}
  A.~B.~Arbuzov and R.~R.~Sadykov,
  J.\ Exp.\ Theor.\ Phys.\  {\bf 106} (2008) 488
  [arXiv:0707.0423 [hep-ph]].

\bibitem{Brensing:2007qm}
  S.~Brensing, S.~Dittmaier, M.~Kr\"amer and A.~M\"uck,
  Phys.\ Rev.\  D {\bf 77} (2008) 073006
  [arXiv:0710.3309 [hep-ph]].

\bibitem{Placzek:2003zg}
  W.~Placzek and S.~Jadach,
  Eur.\ Phys.\ J.\  C {\bf 29} (2003) 325
  [hep-ph/0302065];\\
%
  C.~M.~Carloni Calame, G.~Montagna, O.~Nicrosini and M.~Treccani,
  Phys.\ Rev.\  D {\bf 69} (2004) 037301
  [hep-ph/0303102];\\
%
  C.~M.~Carloni Calame {\it et al.},
  Acta Phys.\ Polon.\  B {\bf 35} (2004) 1643
  [hep-ph/0402235].

\bibitem{CarloniCalame:2005vc}
  C.~M.~Carloni Calame, G.~Montagna, O.~Nicrosini and M.~Treccani,
  JHEP {\bf 0505} (2005) 019
  [hep-ph/0502218].

\bibitem{Balossini:2008cs}
  G.~Balossini {\it et al.},
  Acta Phys.\ Polon.\  B {\bf 39} (2008) 1675
  [arXiv:0805.1129 [hep-ph]] and
%
  arXiv:0907.0276 [hep-ph].

\bibitem{Kotikov:2007vr}
  A.~Kotikov, J.~H.~K\"uhn and O.~Veretin,
  Nucl.\ Phys.\  B {\bf 788} (2008) 47
  [hep-ph/0703013].


\bibitem{Denner:1999gp}
A.~Denner, S.~Dittmaier, M.~Roth and D.~Wackeroth,
Nucl.\ Phys.\ B {\bf 560} (1999) 33
[hep-ph/9904472].

\bibitem{Denner:2005fg}
  A.~Denner, S.~Dittmaier, M.~Roth and L.~H.~Wieders,
  Nucl.\ Phys.\  B {\bf 724} (2005) 247
  [hep-ph/0505042].

\bibitem{Stuart:1991xk}
  R.~G.~Stuart,
  Phys.\ Lett.\  B {\bf 262}, 113 (1991).

\bibitem{Aeppli:1993cb}
  A.~Aeppli, F.~Cuypers and G.~J.~van Oldenborgh,
  Phys.\ Lett.\ B {\bf 314} (1993) 413
  [hep-ph/9303236];\\
%
  A.~Aeppli, G.~J.~van Oldenborgh and D.~Wyler,
  Nucl.\ Phys.\ B {\bf 428} (1994) 126
  [hep-ph/9312212];\\
%
  H.~G.~J.~Veltman,
  Z.\ Phys.\  C {\bf 62} (1994) 35.
%

\bibitem{Denner:1998tb}
  A.~Denner and S.~Dittmaier,
  Eur.\ Phys.\ J.\  C {\bf 9} (1999) 425
  [hep-ph/9812411].

\bibitem{Kuraev:1985hb}
  E.~A.~Kuraev and V.~S.~Fadin,
  Sov.\ J.\ Nucl.\ Phys.\ {\bf 41} (1985) 466
  [Yad.\ Fiz.\  {\bf 41} (1985) 733];\\
  G.~Altarelli and G.~Martinelli,
  {\it  In Ellis, J. ( Ed.), Peccei, R.d. ( Ed.): Physics At Lep, Vol. 1,
  47-57;}\\
  O.~Nicrosini and L.~Trentadue,
  Phys.\ Lett.\ B {\bf 196} (1987) 551 and
  Z.\ Phys.\ C {\bf 39} (1988) 479;\\
  F.~A.~Berends, W.~L.~van Neerven and G.~J.~H.~Burgers,
  Nucl.\ Phys.\ B {\bf 297} (1988) 429
  [Erratum-ibid.\ B {\bf 304} (1988) 921];\\
  A.~B.~Arbuzov,
  Phys.\ Lett.\ B {\bf 470} (1999) 252
  [hep-ph/9908361].

\bibitem{Denner:2006ic}
  A.~Denner and S.~Dittmaier,
  Nucl.\ Phys.\ Proc.\ Suppl.\  {\bf 160} (2006) 22
  [hep-ph/0605312].

\bibitem{Grunewald:2000ju}
M.~W.~Gr\"unewald {\it et al.},
in {\it Reports of the Working Groups on Precision Calculations
for LEP2 Physics}, eds.\ S.~Jadach, G.~Passarino and R.~Pittau
(CERN 2000-009, Geneva, 2000), p.~1
[hep-ph/0005309].

\bibitem{Dittmaier:2002nd}
  S.~Dittmaier and A.~Kaiser,
  Phys.\ Rev.\  D {\bf 65} (2002) 113003
  [hep-ph/0203120].

\bibitem{Denner:1993kt}
A.~Denner,
{}Fortsch.\ Phys.\  {\bf 41} (1993) 307.

\bibitem{Bardin:1988xt}
D.~Y.~Bardin, A.~Leike, T.~Riemann and M.~Sachwitz,
Phys.\ Lett.\ B {\bf 206} (1988) 539.

\bibitem{Beenakker:1996kn}
  W.~Beenakker {\it et al.},
  Nucl.\ Phys.\ B {\bf 500} (1997) 255
  [hep-ph/9612260].

\bibitem{Sirlin:1991fd}
  A.~Sirlin,
  Phys.\ Rev.\ Lett.\  {\bf 67} (1991) 2127 and
%
  Phys.\ Lett.\ B {\bf 267} (1991) 240;\\
%
  R.~G.~Stuart,
  Phys.\ Rev.\ Lett.\  {\bf 70} (1993) 3193;\\
%
  P.~Gambino and P.~A.~Grassi,
  Phys.\ Rev.\ D {\bf 62} (2000) 076002
  [hep-ph/9907254];\\
%
  P.~A.~Grassi, B.~A.~Kniehl and A.~Sirlin,
  Phys.\ Rev.\ D {\bf 65} (2002) 085001
  [hep-ph/0109228].

\bibitem{Sirlin:1980nh}
A.~Sirlin,
Phys.\ Rev.\ D {\bf 22} (1980) 971;\\
W.~J.~Marciano and A.~Sirlin,
Phys.\ Rev.\ D {\bf 22} (1980) 2695
[Erratum-ibid.\ D {\bf 31} (1980) 213] and
Nucl.\ Phys.\ B {\bf 189} (1981) 442.

\bibitem{LEP1}
{\sl Z Physics at LEP1}, eds.\ G.~Altarelli, R.~Kleiss and
C.~Verzegnassi (CERN 89-08, Geneva, 1989), Vol.~1.

\bibitem{Bardin:1999gt}
  D.~Y.~Bardin, M.~Grunewald and G.~Passarino,
  hep-ph/9902452.

\bibitem{Kublbeck:1990xc}
J.~K\"ublbeck, M.~B\"ohm and A.~Denner,
Comput.\ Phys.\ Commun.\  {\bf 60} (1990) 165;
H.~Eck and J.~K\"ublbeck, {\it Guide to FeynArts 1.0\/},
University of W\"urzburg, 1992.

\bibitem{Hahn:2000kx}
T.~Hahn,
Comput.\ Phys.\ Commun.\  {\bf 140} (2001) 418
[hep-ph/0012260].

\bibitem{Hahn:1998yk}
T.~Hahn and M.~P\'erez-Victoria,
Comput.\ Phys.\ Commun.\  {\bf 118} (1999) 153
[hep-ph/9807565];\\
%
T.~Hahn,
Nucl.\ Phys.\ Proc.\ Suppl.\  {\bf 89} (2000) 231
[hep-ph/0005029].

\bibitem{Mertig:1991an}
R.~Mertig, M.~B\"ohm and A.~Denner,
Comput.\ Phys.\ Commun.\  {\bf 64} (1991) 345;\\
R.~Mertig, {\it Guide to FeynCalc 1.0\/}, University of W\"urzburg, 1992.

\bibitem{Passarino:1979jh}
G.~Passarino and M.~Veltman,
Nucl.\ Phys.\ B {\bf 160} (1979) 151.

\bibitem{'tHooft:1979xw}
G.~'t Hooft and M.~Veltman,
Nucl.\ Phys.\ B {\bf 153} (1979) 365.

\bibitem{Beenakker:1990jr}
W.~Beenakker and A.~Denner,
Nucl.\ Phys.\ B {\bf 338} (1990) 349.

\bibitem{Denner:1991qq}
A.~Denner, U.~Nierste and R.~Scharf,
Nucl.\ Phys.\ B {\bf 367} (1991) 637.

\bibitem{Denner:2005nn}
  A.~Denner and S.~Dittmaier,
  Nucl.\ Phys.\ B {\bf 734} (2006) 62
  [hep-ph/0509141].

\bibitem{scalints}
  A.~Denner and S.~Dittmaier, in preparation.

\bibitem{Dittmaier:2003bc}
  S.~Dittmaier,
  Nucl.\ Phys.\  B {\bf 675} (2003) 447
  [hep-ph/0308246].

\bibitem{Bredenstein:2008zb}
  A.~Bredenstein, A.~Denner, S.~Dittmaier and S.~Pozzorini,
  JHEP {\bf 0808} (2008) 108
  [arXiv:0807.1248 [hep-ph]].

\bibitem{Dittmaier:1999nn}
S.~Dittmaier,
Phys.\ Rev.\ D {\bf 59} (1999) 016007
[hep-ph/9805445].

\bibitem{Dittmaier:1999mb}
S.~Dittmaier,
Nucl.\ Phys.\ B {\bf 565} (2000) 69
[hep-ph/9904440].

\bibitem{Dittmaier:2008md}
  S.~Dittmaier, A.~Kabelschacht and T.~Kasprzik,
  Nucl.\ Phys.\  B {\bf 800} (2008) 146
  [arXiv:0802.1405 [hep-ph]].
  
\bibitem{Diener:2005me}
  K.~P.~Diener, S.~Dittmaier and W.~Hollik,
  Phys.\ Rev.\ D {\bf 72} (2005) 093002
  [hep-ph/0509084].

\bibitem{Martin:2004dh}
  A.~D.~Martin, R.~G.~Roberts, W.~J.~Stirling and R.~S.~Thorne,
  Eur.\ Phys.\ J.\ C {\bf 39} (2005) 155
  [hep-ph/0411040].

\bibitem{Burkhardt:1995tt}
H.~Burkhardt and B.~Pietrzyk,
Phys.\ Lett.\ B {\bf 356} (1995) 398;\\
S.~Eidelman and F.~Jegerlehner,
Z.\ Phys.\ C {\bf 67} (1995) 585
[hep-ph/9502298].

\bibitem{Consoli:1989pc}
  M.~Consoli, W.~Hollik and F.~Jegerlehner,
  Presented at Workshop on Z Physics at LEP, CERN-TH-5527-89.

\bibitem{Consoli:1989fg}
  M.~Consoli, W.~Hollik and F.~Jegerlehner,
  Phys.\ Lett.\  B {\bf 227} (1989) 167.

\bibitem{Fleischer:1993ub}
J.~Fleischer, O.~V.~Tarasov and F.~Jegerlehner,
Phys.\ Lett.\ B {\bf 319} (1993) 249.

\bibitem{Fadin:1999bq}
  V.~S.~Fadin, L.~N.~Lipatov, A.~D.~Martin and M.~Melles,
  Phys.\ Rev.\  D {\bf 61} (2000) 094002
  [hep-ph/9910338].

\bibitem{Ciafaloni:2000df}
  M.~Ciafaloni, P.~Ciafaloni and D.~Comelli,
  Phys.\ Rev.\ Lett.\  {\bf 84} (2000) 4810
  [hep-ph/0001142].

\bibitem{Hori:2000tm}
  M.~Hori, H.~Kawamura and J.~Kodaira,
  Phys.\ Lett.\  B {\bf 491} (2000) 275
  [hep-ph/0007329].

\bibitem{Melles:2001dh}
  M.~Melles,
  Eur.\ Phys.\ J.\  C {\bf 24} (2002) 193
  [hep-ph/0108221].

\bibitem{Beenakker:2001kf}
  W.~Beenakker and A.~Werthenbach,
  Nucl.\ Phys.\  B {\bf 630} (2002) 3
  [hep-ph/0112030].

\bibitem{Denner:2003wi}
  A.~Denner, M.~Melles and S.~Pozzorini,
  Nucl.\ Phys.\  B {\bf 662} (2003) 299
  [hep-ph/0301241].

\bibitem{Jantzen:2005xi}
  B.~Jantzen, J.~H.~K\"uhn, A.~A.~Penin and V.~A.~Smirnov,
  Phys.\ Rev.\  D {\bf 72} (2005) 051301
  [Erratum-ibid.\  D {\bf 74} (2006) 019901]
  [hep-ph/0504111] and
%
  Nucl.\ Phys.\  B {\bf 731} (2005) 188
  [Erratum-ibid.\  B {\bf 752} (2006) 327]
  [hep-ph/0509157].

\bibitem{Denner:2006jr}
  A.~Denner, B.~Jantzen and S.~Pozzorini,
  Nucl.\ Phys.\  B {\bf 761} (2007) 1
  [hep-ph/0608326].

\bibitem{Ciafaloni:2006qu}
  P.~Ciafaloni and D.~Comelli,
  JHEP {\bf 0609} (2006) 055
  [hep-ph/0604070].

\bibitem{Baur:2006sn}
  U.~Baur,
  Phys.\ Rev.\  D {\bf 75} (2007) 013005
  [hep-ph/0611241].

\bibitem{Kinoshita:1962ur}
  T.~Kinoshita,
  J.\ Math.\ Phys.\  {\bf 3} (1962) 650;
%
  T.~D.~Lee and M.~Nauenberg,
  Phys.\ Rev.\  {\bf 133} (1964) B1549.

\bibitem{Yao:2006px}
  W.~M.~Yao {\it et al.}  [Particle Data Group],
  J.\ Phys.\ G {\bf 33} (2006) 1.

\bibitem{Jegerlehner:2001ca}
  F.~Jegerlehner,
  DESY 01-029, LC-TH-2001-035 [hep-ph/0105283].

\bibitem{Allanach:2002nj}
  B.~C.~Allanach {\it et al.},
  Eur.\ Phys.\ J.\ C {\bf 25} (2002) 113
  [eConf {\bf C010630} (2001) P125]
  [hep-ph/0202233].

\bibitem{SPShomepage} see: \url{http://www.cpt.dur.ac.uk/~georg/sps/}

\bibitem{Catani:1996vz}
  S.~Catani and M.~H.~Seymour,
  Nucl.\ Phys.\  B {\bf 485} (1997) 291
  [Erratum-ibid.\  B {\bf 510} (1998) 503]
  [arXiv:hep-ph/9605323].


\end{thebibliography}
\end{document}